%% file: main.tex
\documentclass[fleqn,10pt]{wlscirep}
\usepackage[utf8]{inputenc}
\usepackage[T1]{fontenc}

\usepackage[english]{babel}
\usepackage{amsmath}
\usepackage{dsfont}

\usepackage[colorlinks=true, allcolors=blue]{hyperref}
\usepackage{comment}
\usepackage{appendix}
\usepackage{pdfpages}
\usepackage{subcaption}
\usepackage{booktabs}
\usepackage{authblk}
\usepackage{cite} 
\usepackage{makecell}
\usepackage{placeins}
\usepackage{lineno}

\title{Interplay between social contact and media exposure in the overestimation of racial diversity in the U.S.}

\author[1]{Clara Eminente}
\author[2]{Henrik Olsson}
\author[2]{Ljubica Nedelkoska}
\author[2]{Rafael Prieto-Curiel}
\author[2]{Mirta Galesic}
\author[1,2,*]{Elisa Omodei}

\affil[1]{Department of Network and Data Science, Central European University, Vienna, Austria}
\affil[2]{Complexity Science Hub, Vienna, Austria}
\affil[*]{Email: omodeie@ceu.edu}

\begin{abstract}
The general population systematically overestimates the size of minority groups, yet how these misperceptions vary across racial groups and geographical scales remains poorly understood. Using a purpose-built survey of the U.S. population, we examine overestimation of people of color (PoC) communities across four nested geographical scales: neighborhood, city, state, and nation. Our results demonstrate that overestimation is both scale- and group-dependent: the probability of overestimation increases progressively from local to national levels, and people of color overestimate their own group size more frequently than white people do at both the neighborhood and national levels. Among white respondents, we identify a scale-dependent divide in exposure mechanisms: direct interethnic social contact is the primary correlate of overestimation at local levels, whereas perceived frequency of coverage of people of color in news dominates at the national level. Furthermore, across both groups, frequent news consumption is associated with reduced rates of overestimation, while frequent social media use is associated with higher rates. These findings suggest that overestimation is real and present across scales and groups. This in turn can foster an `illusion of diversity', potentially undermining support for equity-promoting policies by creating the erroneous belief that representation goals have already been achieved.
\end{abstract}

\begin{document}

\makeatletter
\@input{main.aux}
\makeatother

\maketitle

\section*{Introduction}
A striking and widely documented phenomenon is that members of the general population tend to substantially overestimate the size of racial minority groups and immigrant minority groups, both in Europe ~\cite{herda2010many, sides2007european, duffy2014perceptions} and the U.S. ~\cite{nadeau1993innumeracy, sigelman2001innumeracy, alba_distorted_2005, citrin2008immigration, guay_quirks_2025}. 
For instance, in a recent survey, U.S. respondents overestimated the share of black people in the population by more than 3 times (41\% estimated vs 12\% actual) and that of Asian people by nearly 5 times (29\% vs 6\%)~\cite{orth2022}.
Misperceptions of the prevalence of specific racial groups have been shown to positively correlate with inflated perceptions of crime and threat~\cite{dixon2008crime, chiricos1997racial}, which in turn can shape intergroup attitudes and pose serious challenges to social integration~\cite{lawrence_consequences_2014, bursztyn2022misperceptions}.
Misperceptions of group size are not confined to majority group members, though, as minority group members also tend to overestimate their own size~\cite{sigelman2001innumeracy, alba_distorted_2005}. This can have consequences: when individuals overestimate the proportion of minorities in their environment, it can create an `illusion of diversity', whereby the erroneous belief that equity and representation have already been achieved leads to decreased support for diversity-promoting policies~\cite{kardosh_minority_2022}.
Prior work has established that the tendency to overestimate the size of racial minority and immigrant populations varies systematically with sociodemographic characteristics, as younger individuals, women, and those with lower levels of education are more likely to overestimate the size of minority groups~\cite{nadeau1993innumeracy, herda2010many}. Both direct and indirect exposure to a group has also been found to correlate positively with overestimation~\cite{nadeau1993innumeracy, sigelman2001innumeracy, alba_distorted_2005, kunovich_perceptions_2017, herda2010many}.

Despite extensive evidence that individuals systematically overestimate the size of racial and immigrant minority groups, important limitations remain in our understanding of these misperceptions. Most prior studies have examined such misperceptions at the national level and among the general population, without distinguishing between the perceptions of the white majority and those of minority groups themselves~\cite{nadeau1993innumeracy,gallagher2003miscounting}. Sigelman and Niemi were among the first to compare estimates by racial groups in the U.S. at the national level, finding that black people were more likely than white people to overestimate their own group size~\cite{sigelman2001innumeracy}. Subsequent work extended this comparison to local community estimates and showed that black and Hispanic people estimated their own group size to be substantially larger than that estimated by white people, and that overestimation tended to be higher for national than for local estimates~\cite{alba_distorted_2005,wong2012bringing}. However, these comparisons have not been systematically examined across multiple geographical scales. In addition, although previous studies have linked misperceptions to both direct and indirect exposure to minority groups~\cite{nadeau1993innumeracy,sigelman2001innumeracy,alba_distorted_2005,kunovich_perceptions_2017,herda2010many}, the mechanisms underlying these relationships remain poorly understood. In particular, most existing studies rely on secondary survey data that do not measure the composition of individuals' social circles directly, or their exposure to information about other groups in the media, making it difficult to test theories that attribute population misperceptions to the structure of individuals' social environments.

In this study, we address these limitations by examining three aspects of misperceptions about the size of the minority population: variation across geographical scales and racial groups of those making the estimate, the role of direct and indirect exposure to minority groups, and the influence of broader information environments. In a purpose-built survey of U.S. participants, we measure perceptions of the size of different populations across four nested geographical levels---neighborhood, city, state, and nation---among both white people and people of color. While we acknowledge that each racial group in the United States has its own distinct identity, we follow an established body of literature supporting the use of the term ``people of color'' (PoC) as a bridging identity term, particularly in contexts where the experiences of non-white groups are compared to those of the white population, which is the focus of the present study~\cite{perez2021diversity, starr2023re, starr2024people}. 

We first examine whether the likelihood of overestimating PoC group sizes differs across geographical scales. We ask white and PoC respondents to estimate the proportion of people of color in their social circle of family members, friends, colleagues, and acquaintances; as well as in their neighborhood, city, state, and the whole country. 
We then investigate the mechanisms underlying these misperceptions. Drawing on the social sampling model of social judgment~\cite{galesic2018sampling,galesic2021integrating}, we first consider how the composition of individuals’ social circles relates to their perceived size of the PoC population at different geographical scales. According to this model, people estimate the prevalence of attributes in broader populations by drawing on their knowledge of the distribution of those attributes among their social contacts. However, they do not naively rely on their social circles, but only to the extent they believe those circles represent the target population. For example, if an individual perceives their social circle as quite representative of their neighborhood but not of their overall state, we would expect their neighborhood estimates to be more affected by the composition of their social circle than their state estimates. The latter are likely to be more influenced by indirect information sources such as traditional news media and social media. We therefore investigate how misperceptions at different geographical scales relate to respondents' overall news consumption and social media use, and specifically the frequency with which they encounter news coverage about people of color and the perceived tone of that coverage. 

Media portrayals of minorities have long been shown to influence public perceptions and attitudes. For example, exposure to local television news that overrepresents black individuals as criminals has been associated with heightened perceptions of violence~\cite{dixon2008crime}, and disproportionate media coverage of immigrant populations has been linked to higher rates of overestimation of their population size and more negative attitudes toward immigrants~\cite{herda2010many,schemer2012influence,benesch2019media}. More generally, prior research has shown that exposure to political information and news coverage can influence perceptions of immigration and minority populations. Specifically, providing reliable statistics has been shown to correct misperceptions about the sizes of these groups~\cite{hopkins2019muted}, while exposure to positive narratives has been found to generate more favorable attitudes toward their integration~\cite{grigorieff2020does}. However, the role of social media in driving overestimation of minority group size remains understudied. Unlike traditional news media, social media combines news diffusion with direct interpersonal contact, creating an online public sphere that is potentially more ethnically diverse than one's social circle or traditional news environment. Gelovani and colleagues, for instance, found that greater time spent on X is correlated with lower intergroup ethnocentrism~\cite{gelovani_intergroup_2025}. At the same time, other studies have shown that individuals who rely heavily on social media for news are more likely to believe misinformation \cite{ahmed2022social, ahmed2023examining}, as excessive social media use may erode the quality of information users are exposed to and reduce critical engagement with it.

Our analyses reveal several consistent patterns. Overestimation of the size of the PoC population is both group- and scale-dependent: people of color overestimate their own group size more often than white people do at both the neighborhood and national levels, and the probability of overestimation increases progressively from local to national estimates. Consistent with predictions from the social sampling model, among the white population, the composition of individuals’ social circles is most strongly associated with overestimation of group size at more local geographical scales, whereas the frequency of news coverage of people of color is most strongly associated with overestimation at the national level. Finally, we find that frequent news consumption is associated with less overestimation, while frequent social media use with more.

\section*{Results}

\subsection*{Misperceptions vary across racial groups and geographical scales}

\begin{figure}[ht!]
\centering
\includegraphics[width=0.65\textwidth]{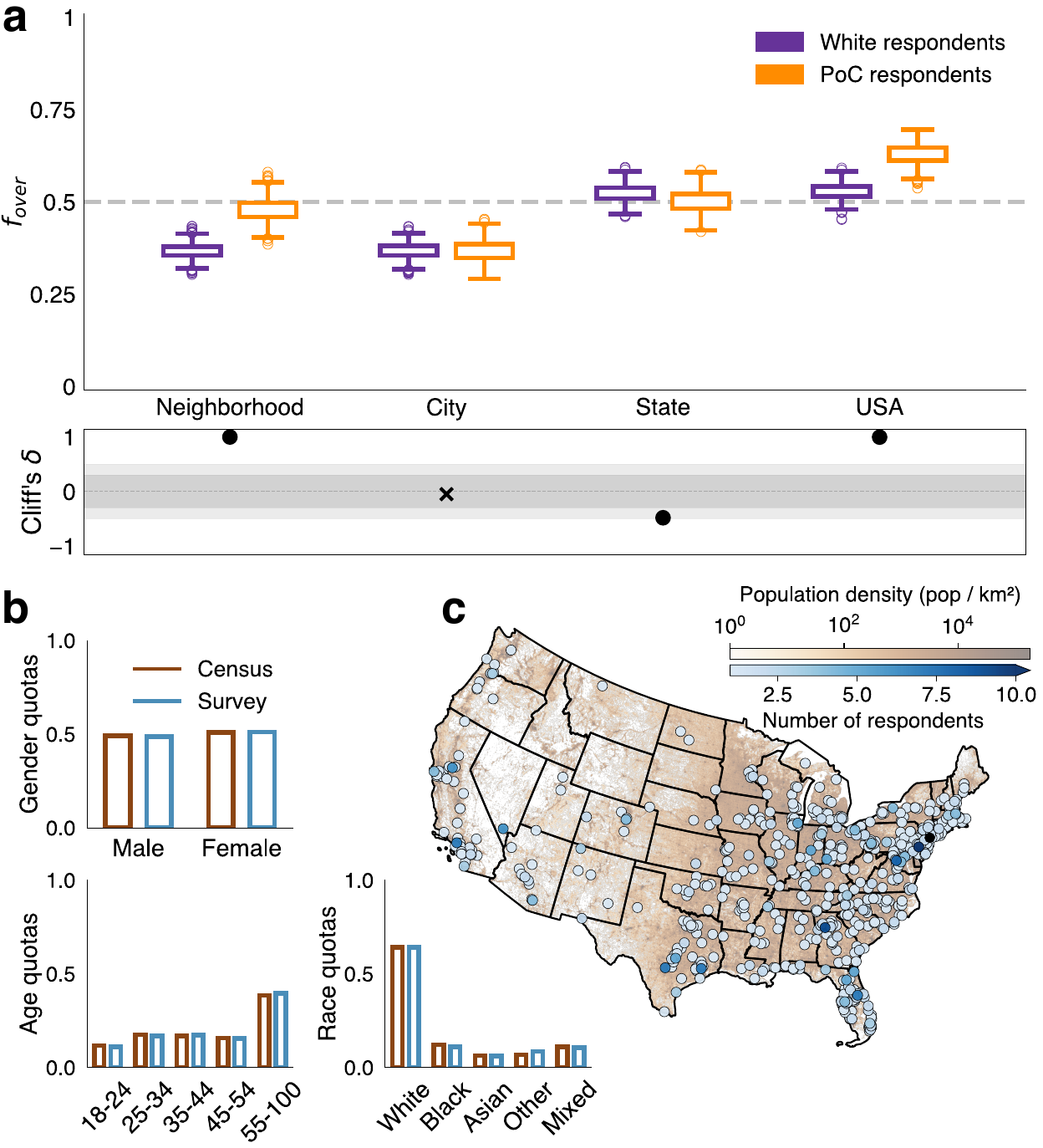}
\caption{\textbf{Geographical patterns in the overestimation of the size of the PoC population and sample representativeness}. Panel \textbf{a}: Bootstrapped fraction of respondents who overestimate the size of the PoC population at each geographical resolution, split by racial group. Each boxplot is obtained by creating $1000$ bootstrapped samples of the labels assigned to respondents at that geographical level (neighborhood, town or city, state, and country) and, for each sample, computing the fraction of respondents who overestimate. The lower plot shows the magnitude of the difference between the distributions being compared, using Cliff's delta, a non-parametric effect size measure ranging from $-1$ to $1$. In this case, a positive value indicates that the white population tends to overestimate more often than the PoC population does, whereas a negative value indicates the opposite. Absolute values of $\delta$ between $0.5$ and $1$ indicate strong differences between the compared distributions (white background), values between $0.3$ and $0.5$ indicate moderate difference (light grey background), and values $\delta<0.3$ indicate negligible differences (dark grey background), often corresponding to non-statistically significant Mann-Whitney U tests (indicated by a cross sign in the plot). The dashed line at $f_{over}=0.5$ marks the threshold above which the majority of respondents are overestimating the size of PoC communities. Panels \textbf{b-c}: Representativeness of our sample in terms of gender, age, race of the respondents, and in terms of their geographical distribution. In panel \textbf{b}, we compare the quotas of each group in our sample and in the U.S. census, finding good alignment. Only binary gender identity is included due to sample size limitations, as further detailed in Section~\ref{sec:supmat:datacleaning} of the Supplementary Information. In panel \textbf{c}, we overlay the population density distribution of the U.S. with that of our respondents. We exclude non-continental areas and Alaska for visualization purposes. }
\label{fig:methods_geores}
\end{figure}

We conducted a survey of a sample representative of the U.S. population in terms of demographic characteristics and geographical location ($N=866$, Figure~\ref{fig:methods_geores}b-c). We asked respondents to estimate the size of PoC and white communities in their social circle (defined as family members, friends, colleagues, and acquaintances with whom respondents had interacted in the previous month), neighborhood, town or city, state, and the U.S., as well as for their race, geographical location, social media and information consumption habits, and perceived frequency of coverage and tone of news about people of color (see Methods for more details on survey design and administration).

We first analyze the fraction of respondents who overestimate the proportion of people of color at each geographical resolution, separately for white and PoC respondents. Given the extensive evidence documenting systematic overestimation of the size of racial minority communities, our primary focus is on overestimation patterns. However, in Section~\ref{sec:supmat:undercorrect} we also present patterns of correct estimation and underestimation to provide a more complete picture.
Figure~\ref{fig:methods_geores}a illustrates overestimation patterns of the size of PoC communities across different geographical resolutions, disaggregated by the racial group of the respondents. A respondent is considered to have overestimated if their guess about the size of a given racial group at a given geographic level exceeds the actual census value by more than 10\% of the census value itself.

The upper plot of Figure~\ref{fig:methods_geores}a shows the bootstrapped distribution of the fraction of people who overestimated the size of the PoC population at each geographical level, $f_{over}$. For each geographical level, we obtain two distributions, based on the racial group of the respondents. The central tendencies of the two distributions at each geographical level are compared using two-sided Mann–Whitney U tests (\textit{p} < 0.0001 for all geographical levels except city, detailed results for all tests are reported in Table~\ref{tab:mwu_geores}). The bottom plot of Figure~\ref{fig:methods_geores}a reports the effect sizes of the tests, measured by the non-parametric effect size measure Cliff’s delta~\cite{cliff1993dominance}, to quantify the magnitude of the difference between the two distributions beyond the test's statistical significance.

We find that at the state and national level, more than 50\% of respondents in both our white and PoC samples overestimate the size of the PoC population. At the national level, this is a well-established result for perceptions of racial minorities in general~\cite{nadeau1993innumeracy} and for the black population in particular~\cite{sigelman2001innumeracy}. We also find that, for both white and PoC respondents, the likelihood of overestimation, while always above 40\%, is lower at more local geographical scales (neighborhood and city) than at broader scales (state and country). This result is consistent with the two earlier studies comparing local community and national estimates~\cite{alba_distorted_2005, wong2012bringing}, though our findings confirm and extend these results, as this pattern has not been systematically studied across multiple scales before. Finally, we find that people of color overestimate their own group size more often than white people do at the neighborhood and national levels. This is broadly consistent with previous studies~\cite{sigelman2001innumeracy, alba_distorted_2005, kunovich_perceptions_2017}. Interestingly, at the city level, there is no significant difference, and at the state level, the relation is reversed. 

In sum, overestimation of the size of the PoC population is both group- and scale-dependent: the probability of overestimation increases as geographical resolution expands from the neighborhood to the national level, and people of color overestimate their own group size more frequently than white people do at both scales. These patterns suggest that the information sources informing these estimates may shift across scales, moving from experience-based inference at local levels toward a greater reliance on indirect information for broader populations. We investigate these potential mechanisms next.

\subsection*{The role of direct and indirect exposure varies across geographical scales}

We evaluate how perceptions of population composition are associated with direct exposure through social contact and indirect exposure through news. To test whether, as expected from the social sampling model, people form their perceptions of racial group composition based on direct exposure, and if this effect is stronger at the local level, we investigate whether exposure to people of color through respondents’ social circles correlates with their likelihood of overestimating the PoC population size. 

\begin{figure}[ht!]
\centering
\includegraphics[width=0.7\textwidth]{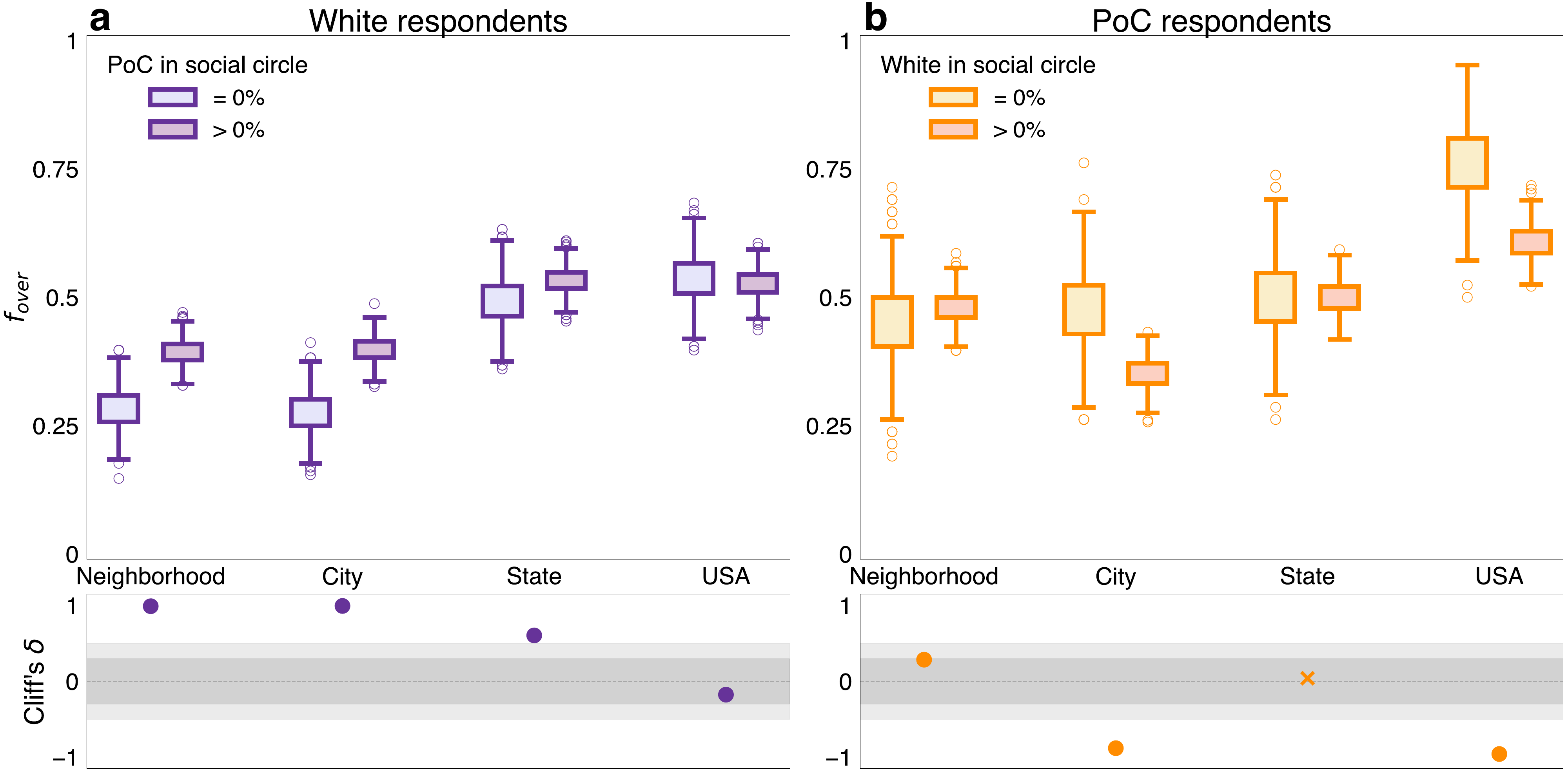}
\caption{\textbf{Probability of overestimation against exposure through social contact.} Upper plots: distributions of the fraction of overestimating respondents (white on the left, PoC on the right) when estimating the size of the PoC population at different geographical levels. Respondents are split according to the composition of their social circle. Lower plots: magnitude of the difference between the compared distributions using Cliff's delta.}
\label{fig:soccirc}
\end{figure}

At the neighborhood and city levels, white people who are exposed to people of color through their social circle consistently overestimate the size of the PoC population more often than those who are not exposed (Figure~\ref{fig:soccirc}a, \textit{p} < 0.0001 for all tests, see details in Table~\ref{tab:mwu_socciclewhite}). The effect size weakens at the state level, and the difference becomes almost non-significant at the national level. This is consistent with our expectation that information gathered through direct contact influences local estimates, whereas for national estimates, respondents also rely on indirect information beyond social circle composition.

All people of color reported having other people of color in their social circle, but not all reported having white people (a symmetric pattern to that observed for white respondents, who all reported having whites in their social circle, but not all reported having people of color). Comparing people of color who have at least one white person in their social circle with those who do not (Figure~\ref{fig:soccirc}b), we find no noticeable effect on overestimation at the neighborhood and state level. However, at the city and national level, we find that people of color who do not have any white people in their social circle (i.e., who are directly exposed to people of color only) are more prone to overestimating their group size, as compared to those exposed to white people (see test details in Table~\ref{tab:mwu_socciclepoc}).

\begin{figure}[ht!]
\centering
\includegraphics[width=0.7\textwidth]{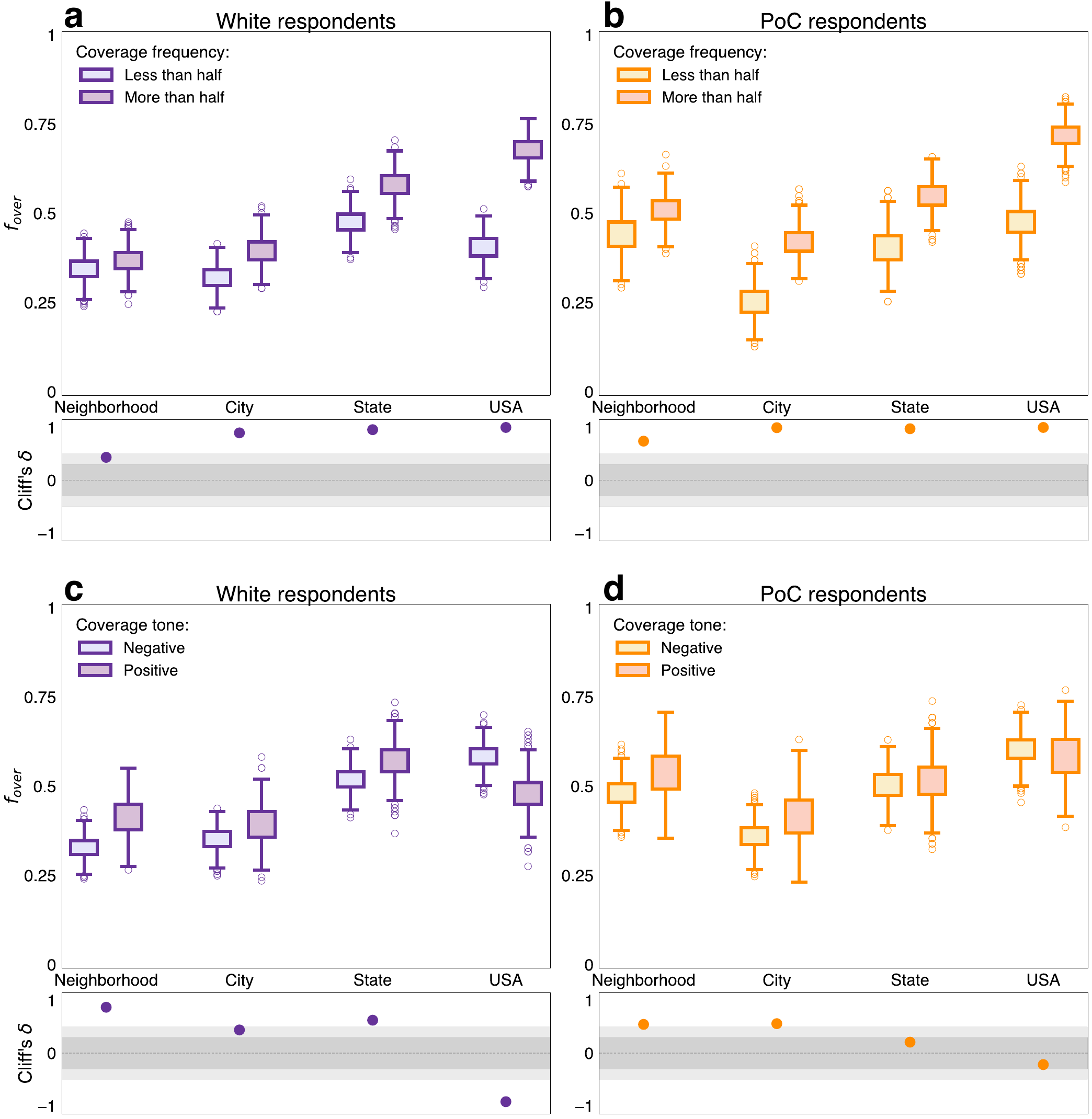}
\caption{\textbf{Probability of overestimation against news coverage about people of color.} Panels \textbf{a-b}: distributions of the fraction of respondents who overestimate the size of the PoC population at different geographical levels, split according to the perceived frequency of news coverage about people of color. Panels \textbf{c-d}: distributions split according to the perceived tone of news coverage about people of color. Respondents are separated by racial group (white on the left, PoC on the right). The lower plots show the magnitude of the difference between the compared distributions using Cliff's delta.}
\label{fig:coverage}
\end{figure}

Moving on to indirect exposure, we examine the portion of news that people consumed that focused on people of color. 
Among people of color (Figure~\ref{fig:coverage}b), those who perceived more news coverage about people of color consistently show higher overestimation rates across all geographical levels (\textit{p} < 0.0001 for all tests, details in Table~\ref{tab:mwu_coveragepoc}), with the effect slightly stronger at the national level than at the neighborhood level. The same trend is present for white people (Figure~\ref{fig:coverage}a), but it is more pronounced at higher geographical scales, with weaker or moderate effects at local levels (\textit{p} < 0.0001 for all tests, details in Table~\ref{tab:mwu_coveragewhite}).

We thus find that, for the white population, direct exposure through social circles mainly correlates with overestimation at local geographical levels, whereas indirect exposure through news correlates more strongly with overestimation at state and national levels.

Turning to the tone of such news, we observe that when estimating at local levels, white people who reported a generally positive tone of the news toward people of color are more likely to overestimate than those who reported negative tones (Figure~\ref{fig:coverage}c, \textit{p} < 0.0001 for all tests, details in Table~\ref{tab:mwu_tonewhite}). At the national level, however, the pattern reverses: negative tone corresponds to higher rates of overestimation compared to positive tone.

For people of color (Figure~\ref{fig:coverage}d), we find a similar pattern (\textit{p} < 0.0001 for all tests, details in Table~\ref{tab:mwu_tonepoc}), although the effects are generally weaker.

In summary, perceptions of population composition are associated with the interplay between direct and indirect exposure in ways that vary by scale. In the white population, while social circle composition is the primary correlate of overestimation at local levels, the frequency of news coverage becomes more influential as the scale of estimation expands. Additionally, the association with news tone reverses across scales for both white and PoC respondents. Positive tone correlates with higher overestimation at the local level, whereas negative tone correlates with higher overestimation at the national level. These results suggest that direct contact and specific media representations could drive overestimation at different scales. However, specific representations are only one facet of indirect exposure. We next investigate whether general habits of information consumption could further shape these perceptions of population composition.

\subsection*{Information environments are further associated with overestimation}
We evaluate how perceptions of population composition are further associated with broader information environments, specifically the overall frequency of news consumption and social media use. Beyond the specific news content and social contact examined in the previous section, we test whether these general habits could serve as mechanisms that either counteract or reinforce misperceptions of PoC group size. This analysis allows us to assess whether the frequency of information engagement, which represents a broader measure of indirect exposure, could influence these perceptions independently of the specific representations encountered in news coverage. Consistent with the idea that general habits affect accuracy, we examine whether engaging with news leaves citizens better informed about actual population statistics.

\begin{figure}[ht!]
    \centering
    \includegraphics[width=0.7\textwidth]{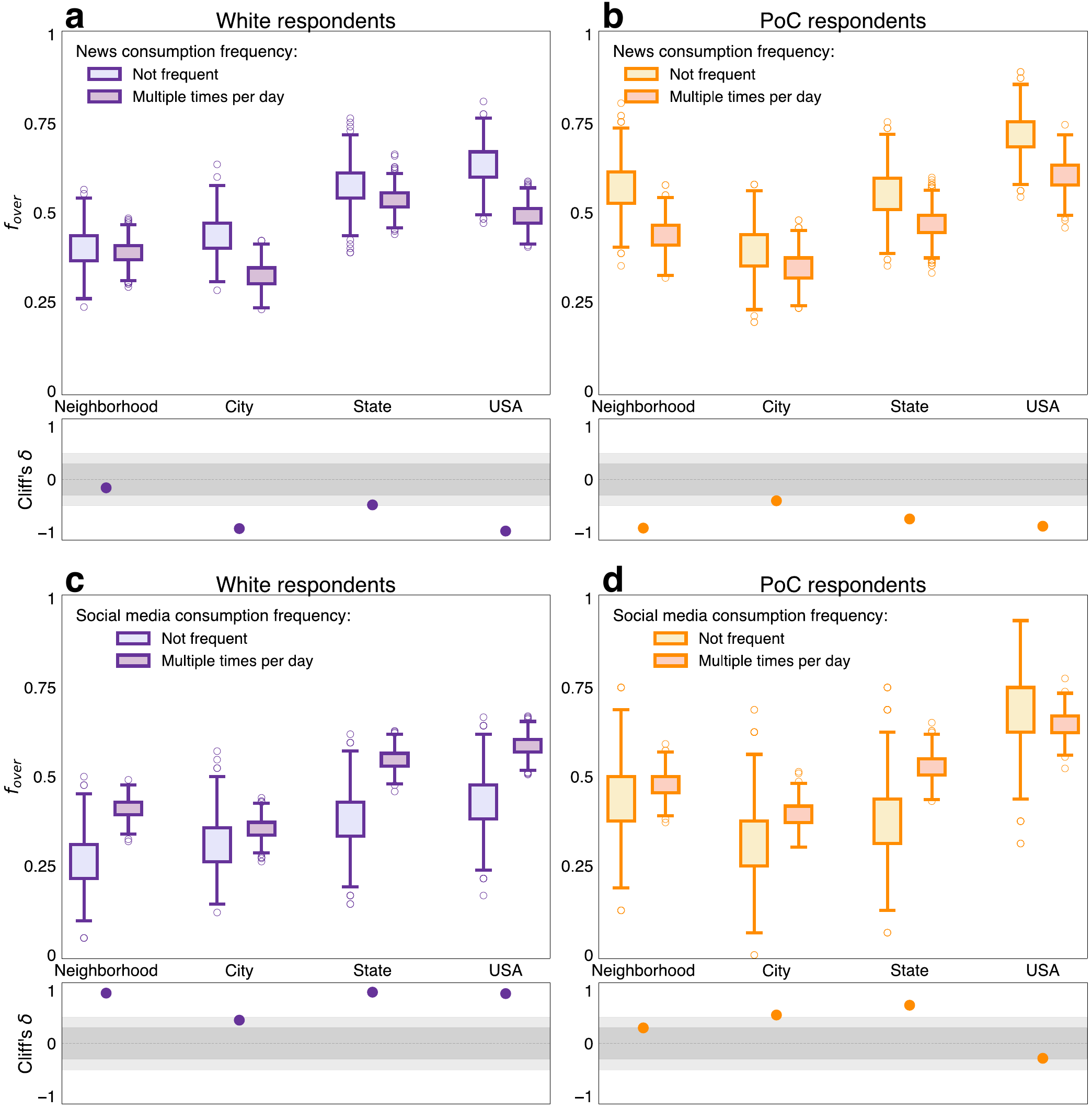}
    \caption{\textbf{Probability of overestimation against news consumption frequency and social media use.} The upper plots in each panel show the distributions of the fraction of respondents who overestimate the size of the PoC population at different geographical levels. Respondents are first split by race (white on the left ---Panels~\textbf{a, c}--- and PoC on the right ---Panels~\textbf{b, d}---), and then further split according to how often they consume news (Panels~\textbf{a, b}) and how often they use social media (Panels~\textbf{c, d}). The lower plots show the magnitude of the difference between the compared distributions using Cliff's delta.}
    \label{fig:frequency}
\end{figure}

We find that for both people of color and white people, higher news consumption frequency negatively correlates with the likelihood of overestimating the size of the PoC community (Figure~\ref{fig:frequency}a-b, \textit{p} < 0.0001 for all tests, details in Tables~\ref{tab:mwu_newswhite} and ~\ref{tab:mwu_newspoc}). That is, respondents who rarely consume news are significantly more likely to overestimate the size of PoC communities than those who consume news multiple times per day. This is consistent with our expectation that less informed people hold more inaccurate perceptions. As put forward by Herda in his study of overestimation of immigrant population size across over twenty European countries~\cite{herda2010many}, engaging with news leaves citizens better informed about actual statistics and encourages more thoughtful analysis and critical thinking.

While the influence of traditional media on racial and immigration misperceptions and attitudes has been widely studied, comparatively little is known about social media usage and racial minority overestimation. Therefore, we next explore the association between the frequency of social media use and the likelihood of overestimation (Figure~\ref{fig:frequency}c-d).
We find that white respondents who use social media multiple times per day are significantly more likely to overestimate the size of the PoC population than those who use it infrequently (\textit{p} < 0.0001 for all tests, details in Tables~\ref{tab:mwu_smwhite} and ~\ref{tab:mwu_smpoc}). The same pattern, although weaker, holds for PoC respondents.

In summary, our results suggest how general information habits could play a role in driving or correcting misperceptions of population composition. While frequent news consumption may counteract overestimation by providing a more accurate baseline of population statistics, frequent social media use could reinforce the illusion of diversity by increasing the likelihood of overestimation. These habits represent a broader form of indirect exposure that operates independently of the specific news content individuals encounter. However, having identified several distinct mechanisms that may influence these perceptions, it remains unclear which of these factors is most consequential overall. We conclude our analysis by evaluating the relative importance of these different predictors across all geographical scales.

\subsection*{Exposure effects persist when considered together while controlling for socio-demographic factors}

We next evaluate the relative importance of direct and indirect exposure mechanisms in the overestimation of the size of PoC communities across different geographical scales. This analysis integrates the three primary factors examined in our research: the composition of immediate social circles, the frequency and tone of news coverage about people of color, and general habits of news and social media consumption. Moreover, we take into account respondents' sociodemographic characteristics and the confidence they associated with their estimates. We train two sets of binary random forest classifiers to predict whether a respondent overestimated the size of the PoC population. We use random forest models because they are able to capture non-linear relationships and complex interactions between variables~\cite{breiman2001random}. Specifically, we train one classifier for each geographical resolution, resulting in four models for each respondent's group (i.e., white and PoC). All models are evaluated on a held-out test set and found to outperform a random baseline, as indicated by precision-recall curves exceeding the positive class prevalence, with the exception of one model (PoC respondents, neighborhood level).

\begin{figure}[ht!]
    \centering
    \includegraphics[width=0.95\textwidth]{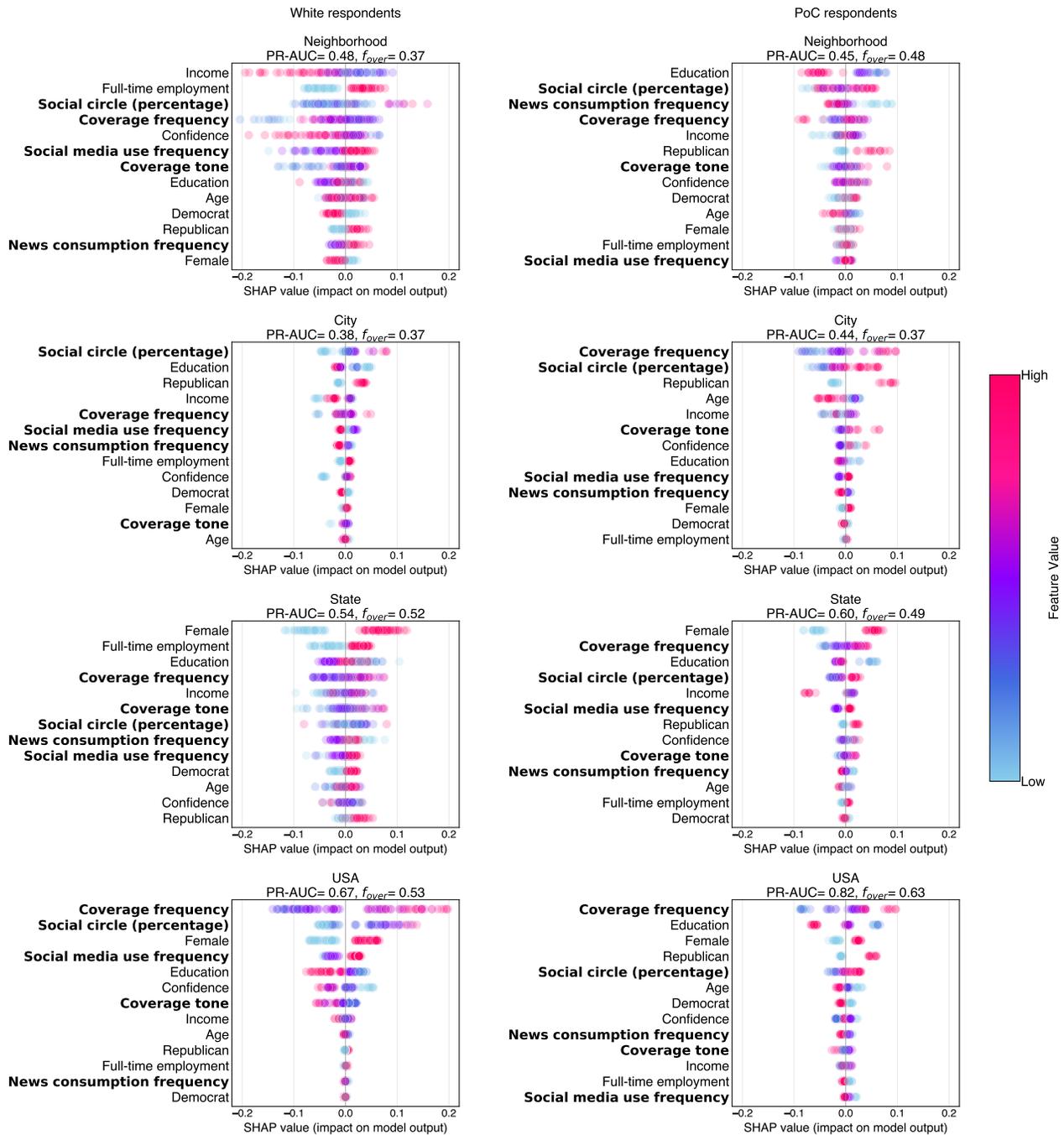}
    \caption{\textbf{Variable importance in the random forest models.} A separate random forest classifier is trained for each combination of the respondent's race and geographical level. Each plot displays the relative importance of the model's predictor variables, quantified using SHAP values. Variables in bold are those examined individually in previous sections. Each point represents a single observation in the test set (i.e., a respondent's estimate at a given geographical level), where the position along the horizontal axis indicates the corresponding SHAP value, reflecting the magnitude and direction of that feature's contribution to classifying the observation as an overestimation. Point color encodes the feature value: blue denotes low values and red denotes high values of the respective variable.
As an illustrative example, in the upper right panel (PoC respondents at the neighborhood level), respondents with higher levels of education (red points) are associated with negative SHAP values, whereas those with lower levels of education (blue points) exhibit positive SHAP values. This indicates that higher educational attainment is negatively associated with overestimation, while lower educational attainment shows the opposite relationship. Education ranks first among all predictor variables in this panel, indicating that it exerts the greatest overall influence (measured as the mean absolute SHAP value) on the model's prediction of overestimation, regardless of direction. Plot titles report the Precision-Recall Area Under the Curve (PR-AUC) and the positive class prevalence. Models outperform the baselines when PR-AUC $> f_{over}$.}
    \label{fig:shap}
\end{figure}

Figure~\ref{fig:shap} shows to what extent each independent variable impacts the probability of overestimating, with the variables ordered from the most to the least important. Variable importance is quantified using SHAP values, which capture how strongly each feature contributes to predicting overestimation~\cite{lundberg2017unified}.
For both white and PoC respondents, exposure-related variables, such as the frequency with which people of color appear in the news, the tone of such news, and the composition of the respondents' social circles, measured as the percentage of contacts who are people of color, appear consistently as important features in the model. This further supports the results from the previous sections and confirms that these exposure variables are among the most influential features for classifying respondents as overestimators. Moreover, frequent news consumption and social media use are also important features in most models and, even in the models when they are lower in the ranking, they nevertheless remain significant (i.e., in none of the models their SHAP values are zero, indicating that they contribute to the prediction).
SHAP values also provide information about the direction in which each feature value influences the prediction, that is, whether it pushes the model output closer to or further from predicting overestimation. We find that the direction of these effects aligns with the patterns identified in the previous sections. 

In addition to the analyzed variables, income and education emerge as the most important socio-demographic predictors, in the direction one would expect: lower income and less educated respondents are more likely to hold more inaccurate demographic perceptions, as also found in previous studies~\cite{nadeau1993innumeracy, herda2010many}. The inclusion of these control variables does not, however, attenuate the effect of exposure variables, confirming their independent contribution to predicting overestimation.

Using a random forest classifier, we aimed at providing a more comprehensive overview of possible determinants of overestimation of the size of PoC communities. Once again, we find that while social circle composition is a key predictor of overestimation at the local level, indirect exposure through news and social media emerges as one of the dominant variables as the scale expands. These results demonstrate that in a classification setting, the predictive power of direct experience is superseded by broader information environments at the national level. By evaluating these mechanisms within an integrated model, we establish that the effects highlighted in the previous analyses not only persist when all variables are considered together, but are also present even when mediated by additional control variables such as sociodemographic characteristics.

\clearpage

\section*{Discussion}

This study examined misperceptions about PoC's group size across four geographical scales (neighborhood, city, state, and nation) among both the white and PoC population in the U.S., and investigated how social contact, news media exposure, and social media use relate to these misperceptions. Our findings extend and deepen the existing literature in several ways.

We replicate the well-established finding that overestimation of minority groups' sizes is widespread, and additionally show that it is both group- and scale-dependent. PoC respondents overestimate their own group size more consistently than white respondents do at both the neighborhood and national levels. This finding aligns with earlier work at the national level, which this study extends to finer geographical resolutions. Crucially, we show that the shift toward higher overestimation rates at broader geographical scales could reflect a transition from experience-based to media-based inference: at local scales, where direct observation is possible, respondents appear to anchor their estimates in their immediate social environment, whereas at the national level, where direct experience is necessarily incomplete, they rely more heavily on indirect sources such as news and social media. This reliance on indirect sources at broader scales appears to be a primary driver of the overestimation of diversity, as national-level perceptions become increasingly detached from actual demographic realities. This interpretation is supported by our finding that the composition of people's social circles is most strongly associated with overestimation rates at local levels, while news coverage frequency is most strongly associated with overestimation rates at the national level.

These results are consistent with social sampling theory, which holds that individuals draw on their social circles and media environments to form judgments about population-level distributions. Our purpose-built survey allowed us to operationalize social circle composition more precisely than prior work relying on secondary data, and to test, for the first time, how its role varies across four geographical scales.

With respect to media exposure, we find that both the frequency and the tone of news coverage about people of color are associated with overestimation, but in ways that vary by geographical scale. Negative news tone is associated with higher overestimation at the national level, consistent with the hypothesis that the hyper-visibility of minorities in negative news stories inflates perceived group size. At the local level, however, positive tone is associated with higher overestimation, suggesting that positive coverage may reinforce the salience of PoC in ways that amplify the effect of direct social contact. These associations suggest that different types of media representations can contribute to the illusion through distinct pathways, either by the hyper-visibility of minorities in negative news or through the heightened salience of their presence in positive coverage. These scale-dependent effects of news tone are, to our knowledge, novel findings that warrant further investigation.

Finally, we find that overall news consumption frequency is associated with reduced overestimation rates, while social media use is associated with increased rates, a pattern that holds for both white and PoC respondents. This divergence between traditional and social media mirrors broader findings on information quality and misinformation, and raises the possibility that social media's distinctive combination of news diffusion and interpersonal contact creates a perceived social environment that is more ethnically diverse than the population at large. This perceived diversity reinforces the illusion by fostering the erroneous belief that equity and representation have already been achieved. As established in our introduction, such a belief can significantly reduce public support for the policies designed to promote these goals.

Several limitations of this study should be acknowledged. First, and most importantly, our design is observational and cross-sectional, which precludes causal claims. The associations we report between social circle, media exposure, and overestimation are consistent with theoretically motivated predictions, but alternative explanations, including reverse causation and unmeasured confounding factors, cannot be ruled out. Our findings can, however, serve as a foundation for future experimental work, including survey experiments that manipulate media framing and exposure through social media, and longitudinal designs that track how changes in social circle composition or media habits relate to changes in perceptions over time.

Second, we treat people of color as a single analytical category, aggregating across groups with distinct demographic profiles. This choice was motivated in part by sample size constraints, as obtaining sufficiently large and representative samples for each racial group separately is a substantial practical challenge, and in part by the theoretical focus on white - nonwhite comparisons, for which the PoC category is well established in the literature. Future work should disaggregate this category to examine whether the patterns we identify hold across specific groups, such as black, Hispanic, and Asian respondents, and whether the mechanisms operate differently across groups.

Third, our outcome variable is binary, distinguishing only between respondents who overestimate and those who do not. A continuous measure of the magnitude of overestimation would provide richer information. However, the magnitude of overestimation is inherently bounded by and dependent on the true census value: a respondent estimating the share of people of color in a neighborhood where they represent 5\% of the population has far more room to overestimate than one living where people of color represent 40\% of the population. While it would be possible to perform this comparison at the country level, where all respondents are estimating the same quantities, the asymmetry in the baselines (i.e., census values) at the other geographical levels makes comparisons of overestimation magnitudes at the neighborhood, city and state level difficult to interpret without careful adjustment, and we opted for the binary classification to ensure comparability. Future work could develop bounded or relative measures of overestimation that account for this constraint.

Finally, the rich individual-level data collected in our purpose-built survey, including multi-scale geographical estimates, social circle composition, media habits, and demographic characteristics, provide a natural foundation for future cognitive modeling work. Fitting formal models of judgment under uncertainty to our data could shed light on the cognitive processes through which U.S. residents form beliefs about the racial composition of society at different geographical scales, and could help disentangle the contributions of prior beliefs, environmental sampling, and information exposure to the misperceptions we measured. Ultimately, identifying these drivers is essential for understanding how to mitigate the illusion of racial diversity and ensure that demographic perceptions are grounded in accurate information, thereby sustaining informed public support for equity initiatives.

\section*{Methods}

\subsection*{Survey design and administration}

We designed an original questionnaire and administered it between May and June 2024 through the online platform \textit{Prolific} on $1000$ respondents aged $18$ or older in the U.S., nationally representative by gender, age, and race. In the analyses, we excluded $72$ respondents due to serious inconsistencies in the answers they provided (e.g., zip code, city and state of residence not matching), and further $62$ respondents were discarded because their answers to some sociodemographic questions were either uninformative or too rare to allow us to draw reliable conclusions (e.g., respondents who identified as a third or non-binary gender, or selected `Other' where possible). Details on the data cleaning procedure can be found in the Supplementary Material~\ref{sec:supmat:datacleaning}. The final sample, composed of $N = 866$ respondents, remains representative by gender, age and race (Figure\ref{fig:methods_geores}a). Furthermore, the geographical distribution of respondents closely matches the population density of the U.S. (Figure\ref{fig:methods_geores}b).

First, respondents answered standard sociodemographic questions, reported the zip code of their neighborhood of residence, and indicated their town or city, and state of residence. They were also asked whether they identify as of Hispanic or Latino origin and to report their race, following the structure and wording used by the U.S. Census Bureau~\cite{acsquest2024}.

Respondents then completed a training section, where they were asked to translate fractions to proportions in simple scenarios (such as having a certain number of apples in one color, a certain number of apples in another color, and translating this into percentages of apples in each color). To move to the next section, respondents had to correctly complete these scenarios. This allowed us to familiarize them with the upcoming task (estimating the percentage of different races in the population) and to disqualify respondents who were not at all comfortable with thinking in percentages. These respondents were terminated from the survey and were therefore not included in our initial sample of $1000$ respondents, which only include respondents who completed the full survey according to \textit{Prolific}.

Respondents were then asked to estimate the proportions (in percentage points) of major racial groups in the U.S. (white, Hispanic, black, Asian, and Other). They provided these estimates for their social circle (defined as friends, family, colleagues, and acquaintances with whom they had communicated in the previous month, either face-to-face or otherwise), as well as for their neighborhood (based on their zip code), their town or city, their state, and the country as a whole. After each estimate, they reported their confidence on a scale from $1$ (not confident at all) to $6$ (very confident). At each geographical level, we aggregated the estimates made of the sizes of black, Asian, Hispanic, and Other communities to obtain the estimate of the PoC population. We give an in-depth description of the reasons behind this choice and the way it was implemented in Section~\ref{sec:supmat:poc} of the Supplementary Materials.

Finally, we collected information on social media use and information consumption. Respondents indicated how often they use various social media platforms, where they primarily obtain news from (from TV and radio, to podcasts and social media), and their perceptions of media coverage of different racial groups in terms of both frequency and tone.

\subsection*{Reference data for population composition}
The U.S. Census Bureau provides publicly available information about several statistics about the U.S. population. For our analysis, we used the information provided by the American Community Survey~\cite{acs}~\cite{censusdataweb}, an ongoing yearly survey that collects several demographic characteristics of the U.S. population. As the ground truth against which to compare the estimates of our respondents and the representativeness of our sample, we used the 5-year detailed tables, as they are the only ones with resolution down to the zip code level (which we used as a proxy for the participants' neighborhood). In particular, we use the table `B03002: Hispanic or Latino Origin by Race' for the year 2024 as the ground-truth values to compare respondents' estimates to in order to measure overestimation. 

\subsection*{Measuring estimation bias}\label{sec:bias}
The first step in our analysis was to identify bias in the estimates made by our respondents. In particular, we were interested in assessing whether they over, under or correctly estimated the prevalence of people of color at a certain geographical level. Let us denote by $s_i^{\ell}$ the estimate provided by respondent $i$ for the size of the PoC community at geographical level $\ell$ ($\ell$ = Neighborhood, City, State, USA). Using the information collected in the first section of the survey regarding the geographical location of each respondent, we matched each estimate to the corresponding census value, $c_i^{\ell}$. For each respondent $i$ and geographical level $\ell$, we then computed the difference between $s_i^{\ell}$ and $c_i^{\ell}$ and classified the response as an overestimate, a correct estimate, or an underestimate according to:

\begin{equation}
    \lambda_i^{\ell} =
    \begin{cases}
        \text{over } & \text{if } \quad s_i^{\ell} - c_i^{\ell} > \beta, \\
        \text{correct } & \text{if} \quad |s_i^{\ell} - c_i^{\ell}| < \beta \text{, and}\\
        \text{under } & \text{if} \quad s_i^{\ell} - c_i^{\ell} < -\beta.
    \end{cases}
\end{equation}

Here, $\lambda_i^{\ell}$ is the label assigned to respondent $i$ according to their estimate of the PoC population size at geographical level $\ell$, and $\beta$ is a buffer. The buffer accounts for the fact that an exact match between the survey estimate and the census value is unlikely. Supplementary Material~\ref{sec:supmat:buffer} discusses possible definitions of the buffer and shows that our results remain stable across different choices. In the results reported in the main section, we adopted a 10\% proportional buffer.

Once each estimate was labeled, we computed the fraction of respondents who overestimated, correctly estimated, or underestimated the size of the PoC community at geographical level $\ell$. In all of our analyses, respondents were split by the race they identified with, hence the fraction of respondents of race $r$ ($r=$White or PoC) that overestimated the size of the PoC community at geographical level $\ell$ was computed as:
\begin{equation}
    f_{\text{over}}^{r,\ell} = \frac{1}{N_r}\sum_{i=1}^{N_r}\mathbf{1}_{\{\lambda_i^{\ell} = \text{over}\}}.
\end{equation}

An identical formula provides an analogous value for the labels `correct' and `under'. By construction,
\begin{equation}\label{eq:fover}
    \sum_{\text{label} \in {\text{{over, under, correct}}}} f_{\text{label}}^{r,\ell} = 1 \quad \forall {r, \ell}.
\end{equation}

Besides splitting respondents by race, in some parts of our analysis $f_{over}$ was computed within subgroups of respondents, for instance based on the racial composition of their social circle or on their social media consumption frequency. In Supplementary Materials~\ref{sec:supmat:undercorrect}, we extend our analysis to patterns of correct and underestimation, to provide a more complete picture of the phenomenon. 

To ensure robustness, instead of relying on a single value of $f_{\text{over}}$, we generated $B=1000$ bootstrapped samples for the specific subset of respondents we were interested in, at each geographical resolution, across the analysis. The bootstrapped samples were obtained by sampling with replacement the estimation labels of the respondents $B$ times and computing the fraction of labels corresponding to overestimation for each of the $B$ samples.

 The first part of our results is obtained by comparing the distributions resulting from the bootstrapping procedure at each level of geographical resolution while accounting for different characteristics of our respondents (either just their race as in Figure~\ref{fig:methods_geores}a or a combination of their race and other variables as in Figures~\ref{fig:soccirc}-\ref{fig:frequency}). To compare the two distributions, we first perform two-sided Mann-Whitney U tests, to assess the statistical significance of the difference. Since we are also interested in establishing the magnitude of this difference, for each pair of compared distributions, we additionally compute the effect size by means of the Cliff's delta. When comparing two distributions $P_A$ and $P_B$, the Cliff's delta varies between $-1$ (when $x < y \quad \forall x\in P_A, \quad \forall y \in P_B$) and $1$ (opposite scenario), with values near zero corresponding to very weak effects, sometimes coinciding with non-statistically significant tests.

\subsection*{Random forest classifiers}

Once we established the relationship between the probability of overestimating the size of the PoC population and single variables of interest, we moved on to studying all the variables considered together and mediated by sociodemographic characteristics and other control variables (See Table~\ref{tab:rfvars} for the full list of variables and their definition). We used Random Forest (RF) Classifiers, with the particular goal of predicting whether a respondent will be an overestimator or not given their answers to the other questions in the survey.
The choice of Random Forest Classifiers instead of other statistical methods (such as logistic regression) was due to their non-parametric nature and their ability to capture non-linear relationships~\cite{breiman2001random}. We trained eight RF classifiers in total: one for each racial group of respondents at each of four geographical levels of population estimation (neighborhood, city, state, and country). The training of each model was preceded by the fine-tuning of its hyperparameters using a grid search (see Table~\ref{tab:hyperpars} for further details). Model performance was evaluated on a held-out test set using the Precision-Recall Area Under the Curve (PR-AUC), a metric that summarizes the trade-off between precision (the proportion of predicted overestimators who truly overestimate) and recall (the proportion of true overestimators correctly identified) across all classification thresholds. We favored PR-AUC over more conventional metrics such as accuracy or ROC-AUC because the proportion of overestimators varies meaningfully across subgroups, ranging from approximately 40\% among white respondents at the neighborhood and city levels to more than 60\% among respondents of color at the country level, making class imbalance a possible concern. In imbalanced settings, PR-AUC is more sensitive to model performance on the minority class and therefore provides a more informative evaluation. For each model, the baseline is the fraction of overestimators in the test set ($f_{over}$), which is equivalent to the PR-AUC a naïve classifier would achieve by always predicting the positive class. We thus assess the performance of our models by checking whether PR-AUC $> f_{over}$.

To move beyond black-box predictions and interpret how each predictor contributes to individual classifications, we employed SHAP (SHapley Additive exPlanations) values, a game-theoretic framework that assigns each feature a contribution score for a given prediction~\cite{lundberg2017unified}. Intuitively, the SHAP value for a given variable quantifies how much that variable pushes a prediction toward or away from the positive class (here, overestimation) relative to the average model output. For a single prediction, a positive SHAP value indicates that the feature increases the probability of an estimate to be classified as an overestimation at that geographical scale, while a negative SHAP value indicates the opposite. Aggregated across all respondents, the mean absolute SHAP value for each feature serves as a measure of global importance, reflecting how much that variable influences predictions on average across the sample. We use the mean absolute SHAP value of each feature to rank them from most to least important, and use this ranking to order the features in each of the plots in Figure~\ref{fig:shap}. As an illustrative example, in the upper right panel of Figure~\ref{fig:shap} (PoC respondents at the neighborhood level), respondents with higher levels of education (red points) are associated with negative SHAP values, whereas those with lower levels of education (blue points) exhibit positive SHAP values. This indicates that higher educational attainment is negatively associated with overestimation, while lower educational attainment shows the opposite relationship. Education ranks first among all predictor variables in this panel, indicating that it exerts the greatest overall influence (measured as the mean absolute SHAP value) on the model's prediction of overestimation, regardless of direction. In the same Figure, the features are ordered within each panel by decreasing mean absolute SHAP value, such that those appearing at the top exert greater average influence on the model.

\bibliography{references}

\subsection*{Data and code availability}

The data supporting the findings of this study will be made openly available upon publication.

\subsection*{Acknowledgments}
The authors wish to thank Fariba Karimi and Petra Kralj Novak for the fruitful discussions during the conception of the project.

\subsection*{Funding}
This work was supported by CEU Research Support Scheme. H.O. and M.G. were partially supported by the Austrian Research Promotion Agency grant number 873927 and the ERC Advanced Grant COLLADAPT (101140741).

\newpage
\nolinenumbers

\setcounter{section}{0}          
\setcounter{figure}{0}           
\setcounter{table}{0}            

\renewcommand{\thesection}{S\arabic{section}}   
\renewcommand{\thefigure}{S\arabic{figure}}     
\renewcommand{\thetable}{S\arabic{table}}       

\setcounter{page}{1}

\input{supplementary}
\end{document}

%% file: supplementary.tex
\section*{{\Large \textbf{Supplementary Information}}}
\vspace{0.5cm}

\section{Data cleaning}\label{sec:supmat:datacleaning}

Of the initial sample of $1000$ respondents, the present analysis is based on a subsample of $866$. Below, we describe the data cleaning procedure used to filter out unusable responses.

In the first stage, responses were discarded primarily due to serious internal inconsistencies. The main inconsistencies identified concerned self-reported social media and news consumption habits and reported place of residence. Inconsistencies in the latter arose from mismatches between the reported postal code, city, and state of residence. For a subset of these cases, we were able to contact the respondents directly and resolve the inconsistencies manually. Responses for which no reply was received or no consistent answer could be established were discarded. A small number of additional responses were excluded because the reported location could not be matched to available census data. This initial filtering procedure yielded $928$ valid responses.

Due to the insufficient sample size, which would not allow us to draw statistically robust conclusions, we excluded respondents who identified as non-binary ($\sim2\%$ of the sample), and who did not disclose their political orientation ($\sim3\%$) or their income ($\sim1.7\%$).

At the end of the filtering procedure, we re-assessed the representativeness of our sample in terms of gender, age, race, and geographical location, finding a good alignment between our sample and census values (see Figure~\ref{fig:methods_geores}b-c).

\section{Respondents race distribution}
When collecting information about the race of our respondents, we use the same phrasing used by the Census Bureau. Consequently, our respondents were initially asked whether they had Hispanic or Latino origin, and subsequently, which race they identified with (with the option of selecting more than one). Following previous works~\cite{alba_distorted_2005, kunovich_perceptions_2017}, we classify as ``Hispanic'' all respondents who reported having Hispanic or Latino origin, regardless of the race they reported. All other respondents who did not report Hispanic or Latino origins were classified as ``Black'', ``Asian'', ``White'' and of ``Other'' race in accordance with their specified race. Respondents who reported more than one race were classified as ``'Mixed''. The final breakdown of the ethnicity and race of our respondents is as follows:
\begin{table}[h]
\centering
\begin{tabular}{lcccccc}
\hline
Race & White & Hispanic & Black & Asian & Mixed & Other \\
\hline
Percentage & 63\% & 12\% & 12\% & 6\% & 5\% & 2\% \\
\hline
\end{tabular}
\end{table}

\section{PoC-level aggregations}\label{sec:supmat:poc}
Information about the respondents' race allows us to determine whether they will be estimating the size of the racial group they belong to. To have perfect matching between the estimating and estimated groups, we split our respondents into the same groups we ask for estimates for, as mentioned in the previous paragraph. However, due to the fact that our sample is nationally representative in terms of race, if we have enough statistics about White respondents ($544$ respondents), the same is not true for other races. For instance, only $52$ respondents identified as Asian, and $108$ and $100$ identified as Hispanic and Black, respectively. Since our analysis involves further splitting these groups (e.g., based on their social media consumption habits), the reduced sample size among non-White respondents poses serious limitations on the statistical power of our analysis. For this reason, we decided to adopt a known concept in the literature and aggregate all non-White respondents under the description of \textit{people of color} (\textit{PoC}). From now on, we consider only two groups, both among the estimators (respondents) and estimated: \textit{ White} and \textit{PoC}.
In terms of the races of the respondents, those who reported a race different from White (including Mixed and Other) were labeled as PoC.

Since the questions in our survey were phrased in terms of disaggregated races, we employ the following strategy to aggregate some of the variables across races to form the corresponding variable in terms of PoC:
\begin{itemize}
    \item Estimates of population sizes were summed, both at the survey and at the census level. So for both census and survey, given the estimate of the population size of race $r$ at geographical level $\ell$ by respondent $i$, $e_i^{r,\ell}$:
    \begin{equation*}
        e_i^{PoC,\ell} = \sum_{r }e_i^{r,\ell},
    \end{equation*}
    for $r$ corresponding to black, Hispanic, Asian, Other. In the case of estimates (and corresponding census values), the category ``Other'' includes all possible alternatives to the four main races, including mixed race.
    \item The perceived tone of the news involving people of color was obtained by averaging the perceived tone of the news involving black, Hispanic, and Asian people.
    \item The perceived coverage of people of color was obtained by summing the perceived coverage of stories involving black, Hispanic, and Asian people.
\end{itemize}

\FloatBarrier

\section{Mann-Whitney U test results}\label{sec:supmat:mwu}

All tests are performed by comparing distributions obtained from $B=1000$ bootstrapped sample points.

\begin{table}[!ht]
\begin{center}
\caption{Mann-Whitney U Test results at each geographical level. The test compares the central tendencies of the distributions of $f_{over}$ White vs PoC respondents (when estimating the size of PoC communities), as in Figure~\ref{fig:methods_geores}a.}
\label{tab:mwu_geores}
\begin{tabular}{lrrrrr}
\toprule
 & Median (White) & Median (PoC) & U-stat & p-value & Cliff's delta \\
Geographical level &  &  &  &  &  \\
\midrule
Neighborhood & 0.37 & 0.48 & 662 & <0.0001 & 1.00 \\
City & 0.37 & 0.37 & 524398 & 0.059 & -0.05 \\
State & 0.52 & 0.50 & 740821 & <0.0001 & -0.48 \\
USA & 0.53 & 0.63 & 1861 & <0.0001 & 1.00 \\
\bottomrule
\end{tabular}
\end{center}
\end{table}

\begin{table}[!ht]
\begin{center}
\caption{Mann-Whitney U Test results at each geographical level for white respondents. Each test compares the central tendencies of the distributions of  $f_{over}$. Respondents are split according to the fraction of people of color in their social circle (> 0\% vs 0\%), as in Figure~\ref{fig:soccirc}a.}
\label{tab:mwu_socciclewhite}
\begin{tabular}{lrrrrrr}
\toprule
 & Median (> 0\%) & Median (= 0\%) & U-stat & p-value & Cliff's delta \\
Geographical level &  &  &  &  &  &  \\
\midrule
Neighborhood & 0.39 & 0.28 & 994770 & <0.0001 & 0.99 \\
City & 0.40 & 0.27 & 996571 & <0.0001 & 0.99 \\
State & 0.53 & 0.49 & 802326 & <0.0001 & 0.60 \\
USA & 0.52 & 0.54 & 411935 & <0.0001 & -0.18 \\
\bottomrule
\end{tabular}
\end{center}
\end{table}

\begin{table}[!ht]
\begin{center}
\caption{Mann-Whitney U Test results at each geographical level for PoC respondents. Each test compares the central tendencies of the distributions of  $f_{over}$. Respondents are split according to the fraction of white people in their social circle (> 0\% vs 0\%), as in Figure~\ref{fig:soccirc}b.}
\label{tab:mwu_socciclepoc}
\begin{tabular}{lrrrrrr}
\toprule
 & Median (> 0\%) & Median (= 0\%) & U-stat & p-value & Cliff's delta \\
Geographical level &  &  &  &  &  &  \\
\midrule
Neighborhood & 0.48 & 0.45 & 642366 & <0.0001 & 0.28 \\
City & 0.35 & 0.48 & 59327 & <0.0001 & -0.88 \\
State & 0.50 & 0.50 & 520211 & 0.117 & 0.04 \\
USA & 0.61 & 0.76 & 20894 & <0.0001 & -0.96 \\
\bottomrule
\end{tabular}
\end{center}
\end{table}

\begin{table}[!ht]
\begin{center}
\caption{Mann-Whitney U Test results at each geographical level for white respondents. Each test compares the central tendencies of the distributions of $f_{over}$. Respondents are split according to their perceived amount of news coverage dedicated to people of color(More than half vs Less than half), as in Figure~\ref{fig:coverage}a.}
\label{tab:mwu_coveragewhite}
\begin{tabular}{lrrrrrr}
\toprule
 & Median (More than half) & Median (Less than half) & U-stat & p-value & Cliff's delta \\
Geographical level &  &  &  &  &  &  \\
\midrule
Neighborhood & 0.37 & 0.34 & 716726 & <0.0001 & 0.43 \\
City & 0.40 & 0.32 & 948420 & <0.0001 & 0.90 \\
State & 0.58 & 0.48 & 978830 & <0.0001 & 0.96 \\
USA & 0.68 & 0.40 & 1000000 & <0.0001 & 1.00 \\
\bottomrule
\end{tabular}
\end{center}
\end{table}

\begin{table}[!ht]
\begin{center}
\caption{Mann-Whitney U Test results at each geographical level for PoC respondents. Each test compares the central tendencies of the distributions of  $f_{over}$. Respondents are split according to their perceived amount of news coverage dedicated to people of color(More than half vs Less than half), as in Figure~\ref{fig:coverage}b.}
\label{tab:mwu_coveragepoc}
\begin{tabular}{lrrrrrr}
\toprule
 & Median (More than half) & Median (Less than half) & U-stat & p-value & Cliff's delta \\
Geographical level &  &  &  &  &  &  \\
\midrule
Neighborhood & 0.51 & 0.44 & 870168 & <0.0001 & 0.74 \\
City & 0.42 & 0.25 & 997625 & <0.0001 & 1.00 \\
State & 0.55 & 0.41 & 987822 & <0.0001 & 0.98 \\
USA & 0.72 & 0.48 & 999972 & <0.0001 & 1.00 \\
\bottomrule
\end{tabular}
\end{center}
\end{table}

\begin{table}[!ht]
\begin{center}
\caption{Mann-Whitney U Test results at each geographical level for white respondents. Each test compares the central tendencies of the distributions of  $f_{over}$. Respondents are split according to their perceived tone of the news concerning people of color (Positive vs Negative), as in Figure~\ref{fig:coverage}c.}
\label{tab:mwu_tonewhite}
\begin{tabular}{lrrrrrr}
\toprule
 & Median (Positive) & Median (Negative) & U-stat & p-value & Cliff's delta \\
Geographical level &  &  &  &  &  &  \\
\midrule
Neighborhood & 0.42 & 0.33 & 934545 & <0.0001 & 0.87 \\
City & 0.39 & 0.35 & 719421 & <0.0001 & 0.44 \\
State & 0.57 & 0.52 & 811744 & <0.0001 & 0.62 \\
USA & 0.48 & 0.58 & 39327 & <0.0001 & -0.92 \\
\bottomrule
\end{tabular}
\end{center}
\end{table}

\begin{table}
\begin{center}
\caption{Mann-Whitney U Test results at each geographical level for PoC respondents. Each test compares the central tendencies of the distributions of  $f_{over}$. Respondents are split according to their perceived tone of the news concerning people of color (Positive vs Negative), as in Figure~\ref{fig:coverage}d.}
\label{tab:mwu_tonepoc}
\begin{tabular}{lrrrrrr}
\toprule
 & Median (Positive) & Median (Negative) & U-stat & p-value & Cliff's delta \\
Geographical level &  &  &  &  &  &  \\
\midrule
Neighborhood & 0.54 & 0.48 & 772159 & <0.0001 & 0.54 \\
City & 0.42 & 0.36 & 779094 & <0.0001 & 0.56 \\
State & 0.52 & 0.50 & 603666 & <0.0001 & 0.21 \\
USA & 0.58 & 0.60 & 390858 & <0.0001 & -0.22 \\
\bottomrule
\end{tabular}
\end{center}
\end{table}

\begin{table}
\begin{center}
\caption{Mann-Whitney U Test results at each geographical level for white respondents. Each test compares the central tendencies of the distributions of  $f_{over}$. Respondents are split according to how often they consume news (Multiple times per day vs Not frequently), as in Figure~\ref{fig:frequency}a.}
\label{tab:mwu_newswhite}
\begin{tabular}{lrrrrrr}
\toprule
 & \makecell{Median \\(Multiple times per day)} & \makecell{Median \\(Not frequent)} & U-stat & p-value & Cliff's delta \\
Geographical level &  &  &  &  &  &  \\
\midrule
Neighborhood & 0.39 & 0.40 & 420398 & <0.0001 & -0.16 \\
City & 0.32 & 0.44 & 33914 & <0.0001 & -0.93 \\
State & 0.54 & 0.58 & 258521 & <0.0001 & -0.48 \\
USA & 0.49 & 0.64 & 10444 & <0.0001 & -0.98 \\
\bottomrule
\end{tabular}
\end{center}
\end{table}

\begin{table}
\begin{center}
\caption{Mann-Whitney U Test results at each geographical level for PoC respondents. Each test compares the central tendencies of the distributions of  $f_{over}$. Respondents are split according to how often they consume news (Multiple times per day vs Not frequently), as in Figure~\ref{fig:frequency}b.}
\label{tab:mwu_newspoc}
\begin{tabular}{lrrrrrr}
\toprule
 & \makecell{Median \\(Multiple times per day)} & \makecell{Median \\(Not frequent)} & U-stat & p-value & Cliff's delta \\
Geographical level &  &  &  &  &  &  \\
\midrule
Neighborhood & 0.44 & 0.58 & 38518 & <0.0001 & -0.92 \\
City & 0.35 & 0.39 & 297420 & <0.0001 & -0.41 \\
State & 0.46 & 0.56 & 125116 & <0.0001 & -0.75 \\
USA & 0.61 & 0.72 & 56447 & <0.0001 & -0.89 \\
\bottomrule
\end{tabular}
\end{center}
\end{table}

\begin{table}
\begin{center}
\caption{Mann-Whitney U Test results at each geographical level for white respondents. Each test compares the central tendencies of the distributions of  $f_{over}$. Respondents are split according to how often they use social media (Multiple times per day vs Not frequently), as in Figure~\ref{fig:frequency}c.}
\label{tab:mwu_smwhite}
\begin{tabular}{lrrrrrr}
\toprule
 & \makecell{Median \\(Multiple times per day)} & \makecell{Median \\(Not frequent)} & U-stat & p-value & Cliff's delta \\
Geographical level &  &  &  &  &  &  \\
\midrule
Neighborhood & 0.41 & 0.26 & 976992 & <0.0001 & 0.95 \\
City & 0.35 & 0.31 & 720356 & <0.0001 & 0.44 \\
State & 0.55 & 0.38 & 985906 & <0.0001 & 0.97 \\
USA & 0.59 & 0.43 & 972268 & <0.0001 & 0.94 \\
\bottomrule
\end{tabular}
\end{center}
\end{table}

\begin{table}
\begin{center}
\caption{Mann-Whitney U Test results at each geographical level for PoC respondents. Each test compares the central tendencies of the distributions of  $f_{over}$. Respondents are split according to how often they use social media (Multiple times per day vs Not frequently), as in Figure~\ref{fig:frequency}d.}
\label{tab:mwu_smpoc}
\begin{tabular}{lrrrrrr}
\toprule
 & \makecell{Median \\(Multiple times per day)} & \makecell{Median \\(Not frequent)} & U-stat & p-value & Cliff's delta \\
Geographical level &  &  &  &  &  &  \\
\midrule
Neighborhood & 0.48 & 0.44 & 647163 & <0.0001 & 0.29 \\
City & 0.39 & 0.31 & 769402 & <0.0001 & 0.54 \\
State & 0.53 & 0.38 & 862156 & <0.0001 & 0.72 \\
USA & 0.65 & 0.69 & 358721 & <0.0001 & -0.28 \\
\bottomrule
\end{tabular}
\end{center}
\end{table}

\FloatBarrier

\section{Buffer robustness checks}\label{sec:supmat:buffer}
We mentioned in the main text that, when assigning a label to a respondent's estimate, we consider an estimate correct if it is within a certain buffer around the census value. The buffer we use throughout the main text is what we call a \textit{proportional} buffer of magnitude 10\%. This means that the survey estimate is considered correct if it is inside the interval defined as:
\begin{equation*}
    s^{r, \ell}_i \in c^{r, \ell}_i(1 \pm 10\%),
\end{equation*}
or more generically, given a buffer value $b$ (given as a percentage), if:
\begin{equation*}
    s^{r, \ell}_i \in c^{r, \ell}_i(1 \pm b).
\end{equation*}
This definition of the buffer penalizes small errors when the quantity to be guessed is small, but allows for bigger errors when the census value is larger. It corresponds to the intuition that when the quantity to be guessed is very small, even a small variation should be considered as an error, whereas if the quantity to be guessed is larger, more room for error can be allowed.

We also define a fixed buffer, meaning that we mark an estimate as correct if, regardless of the value of the census, it falls in a fixed range around the census value:
\begin{equation*}
    s^{r, \ell}_i \in c^{r, \ell}_i \pm b.
\end{equation*}

In both cases, the buffer should be expressed in percentage points, but the meaning of the percentage is different. In the case of a proportional buffer, its magnitude corresponds to the proportion of the census value we allow as an error. In the case of a fixed buffer, the magnitude of $b$ should be expressed in the same fashion as the census value, hence as a fraction or in percentage points (as the census value of the size of the population of a racial group is usually expressed as a fraction or a percentage of the population). Our results are robust against different buffer specifications and magnitudes.

\begin{figure}[ht!]
    \centering
    \begin{subfigure}{0.45\textwidth}
        \includegraphics[width=\linewidth]{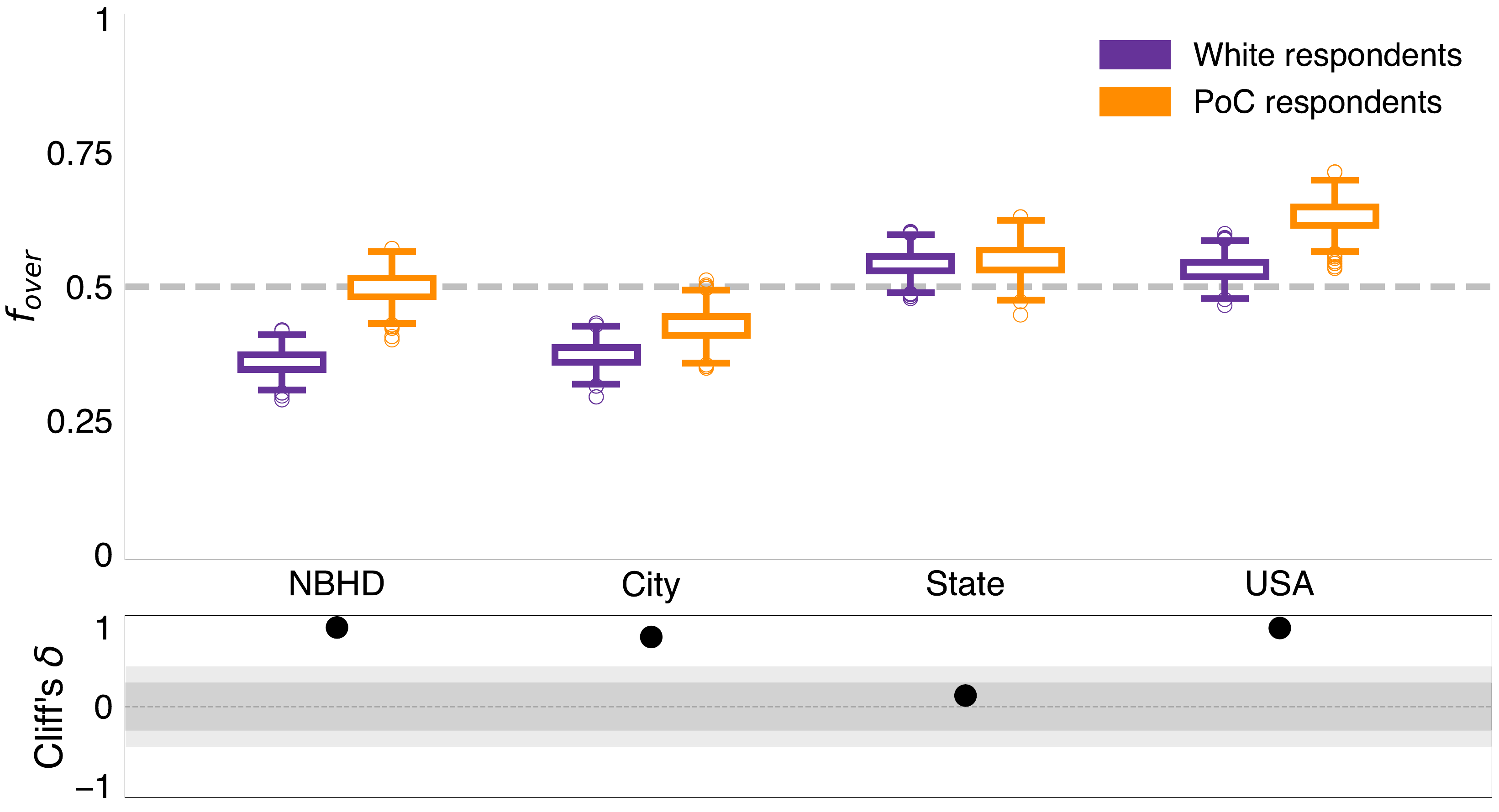}
        \caption{Fixed buffer at 3\%.}
    \end{subfigure}\hfill
    \begin{subfigure}{0.45\textwidth}
        \includegraphics[width=\linewidth]{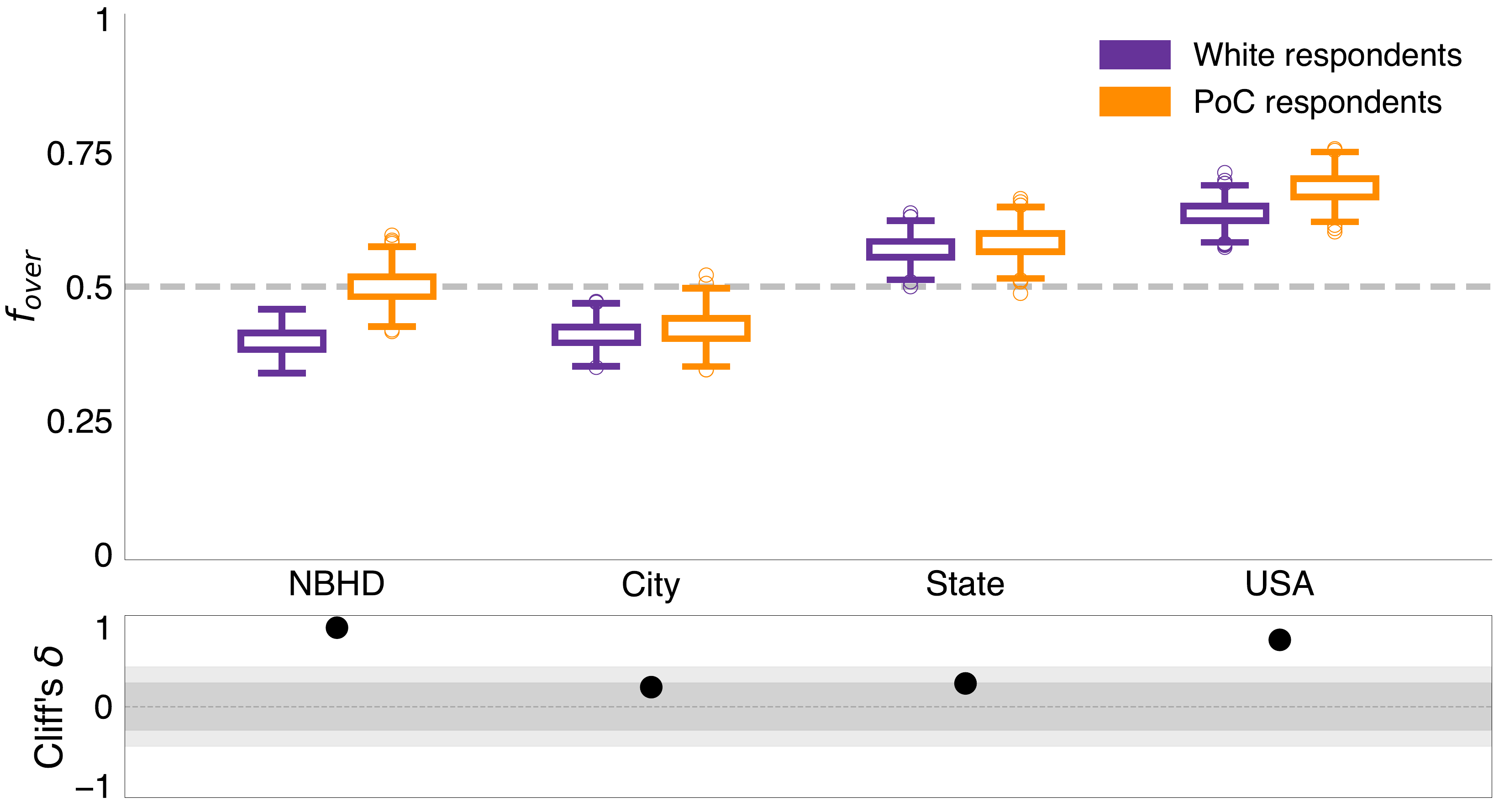}
        \caption{Proportional buffer at 5\%.}
    \end{subfigure}
    \vspace{0.5em}
    \begin{subfigure}{0.45\textwidth}
        \includegraphics[width=\linewidth]{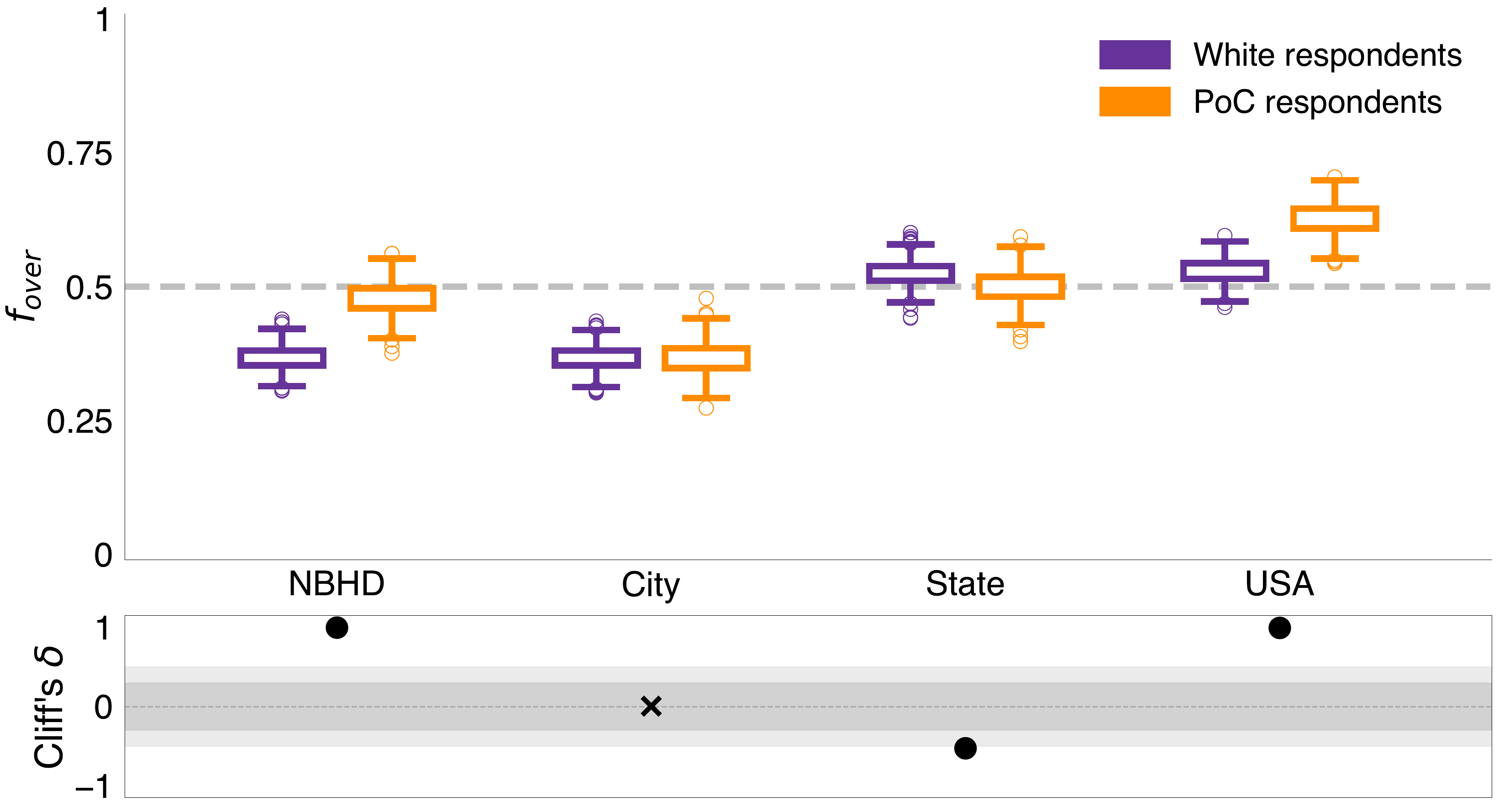}
        \caption{Proportional buffer at 10\%(as in the main text).}
    \end{subfigure}\hfill
    \begin{subfigure}{0.45\textwidth}
        \includegraphics[width=\linewidth]{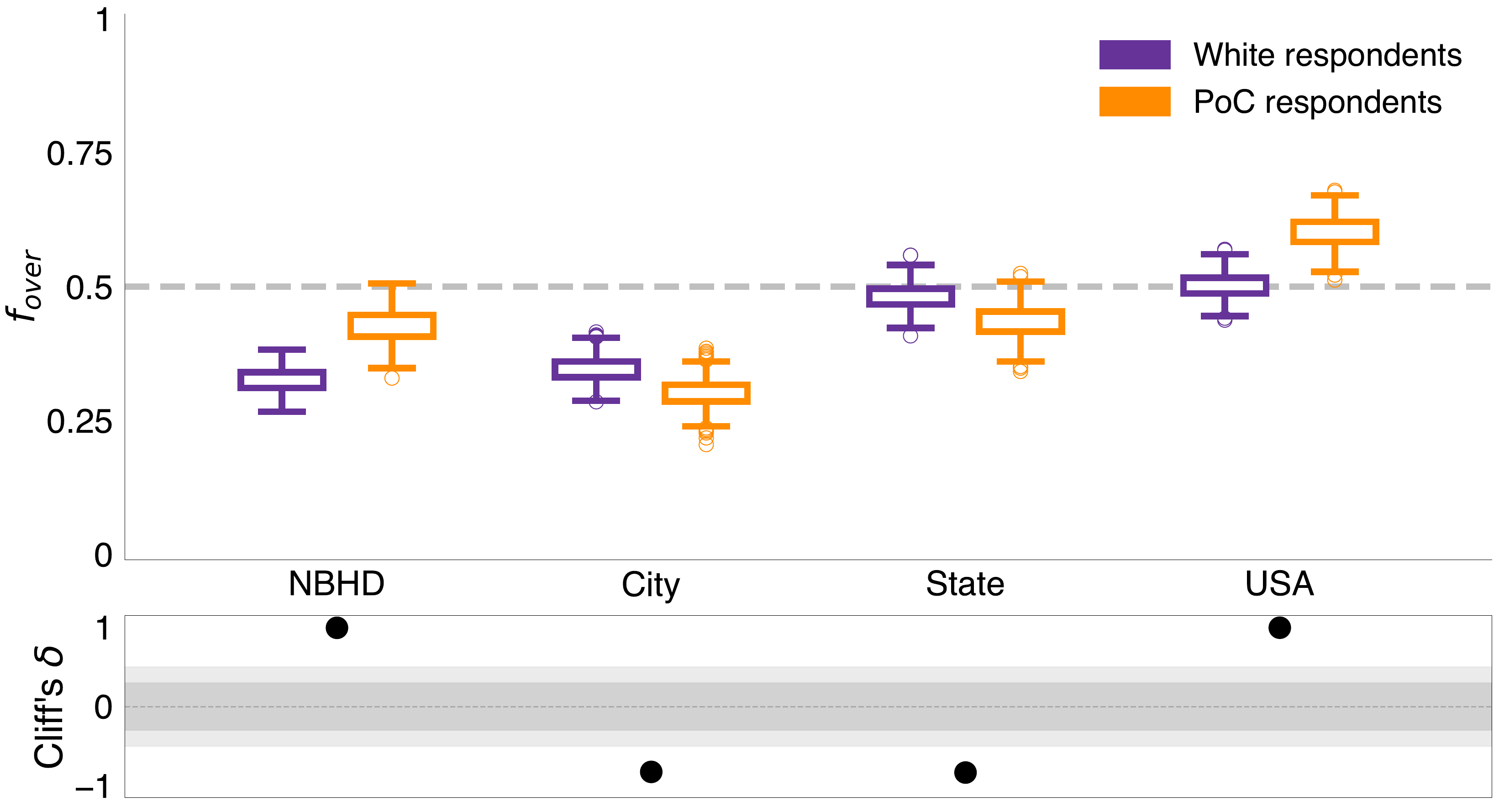}
        \caption{Proportional buffer at 15\%.}
    \end{subfigure}
    \caption{Figure~\ref{fig:methods_geores}a reproduced for different types and values of the buffer.}
\end{figure}

\begin{figure}[ht!]
    \centering
    \begin{subfigure}{0.45\textwidth}
        \includegraphics[width=\linewidth]{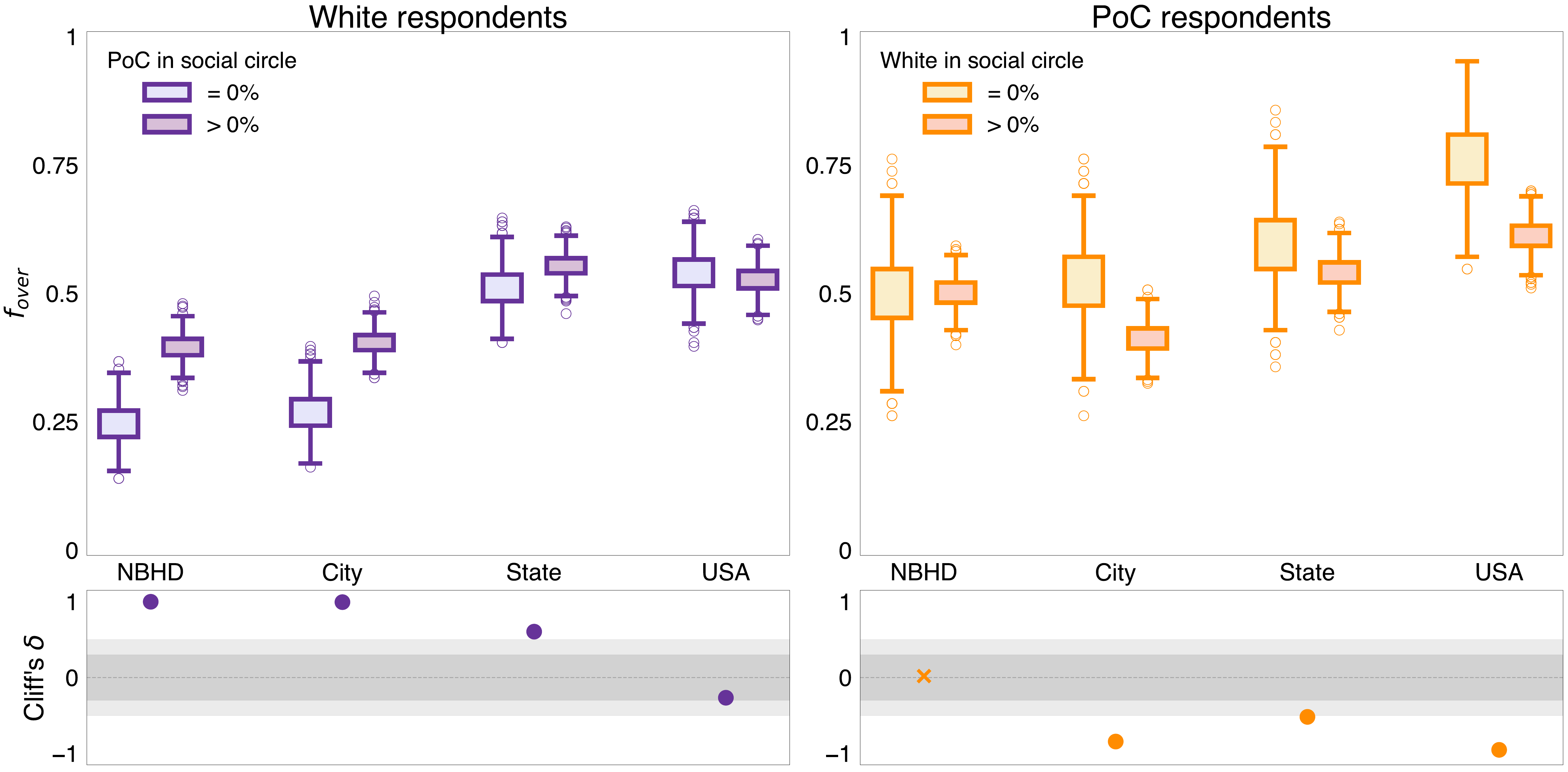}
        \caption{Fixed buffer at 3\%.}
    \end{subfigure}\hfill
    \begin{subfigure}{0.45\textwidth}
        \includegraphics[width=\linewidth]{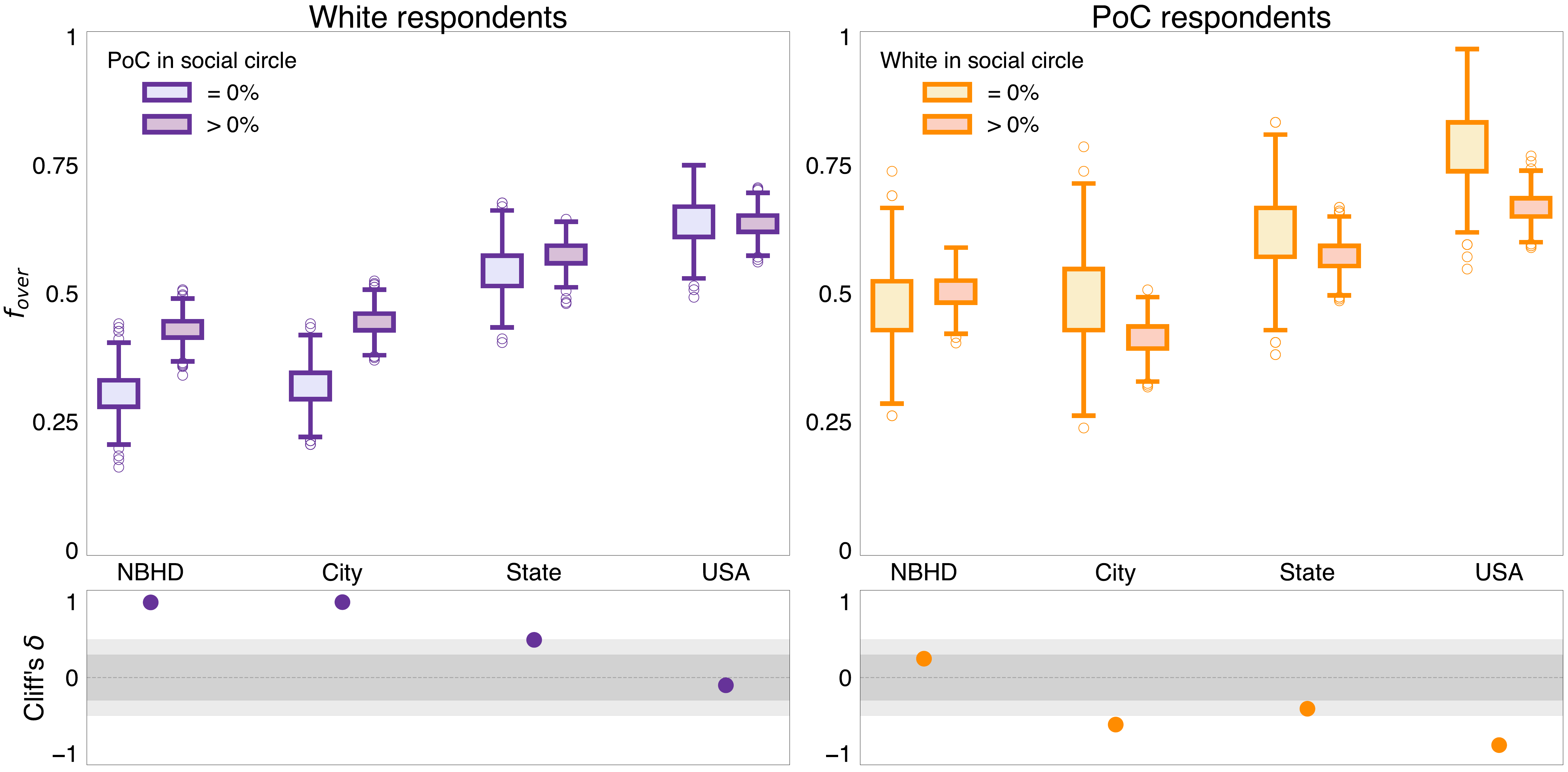}
        \caption{Proportional buffer at 5\%.}
    \end{subfigure}
    \vspace{0.5em}
    \begin{subfigure}{0.45\textwidth}
        \includegraphics[width=\linewidth]{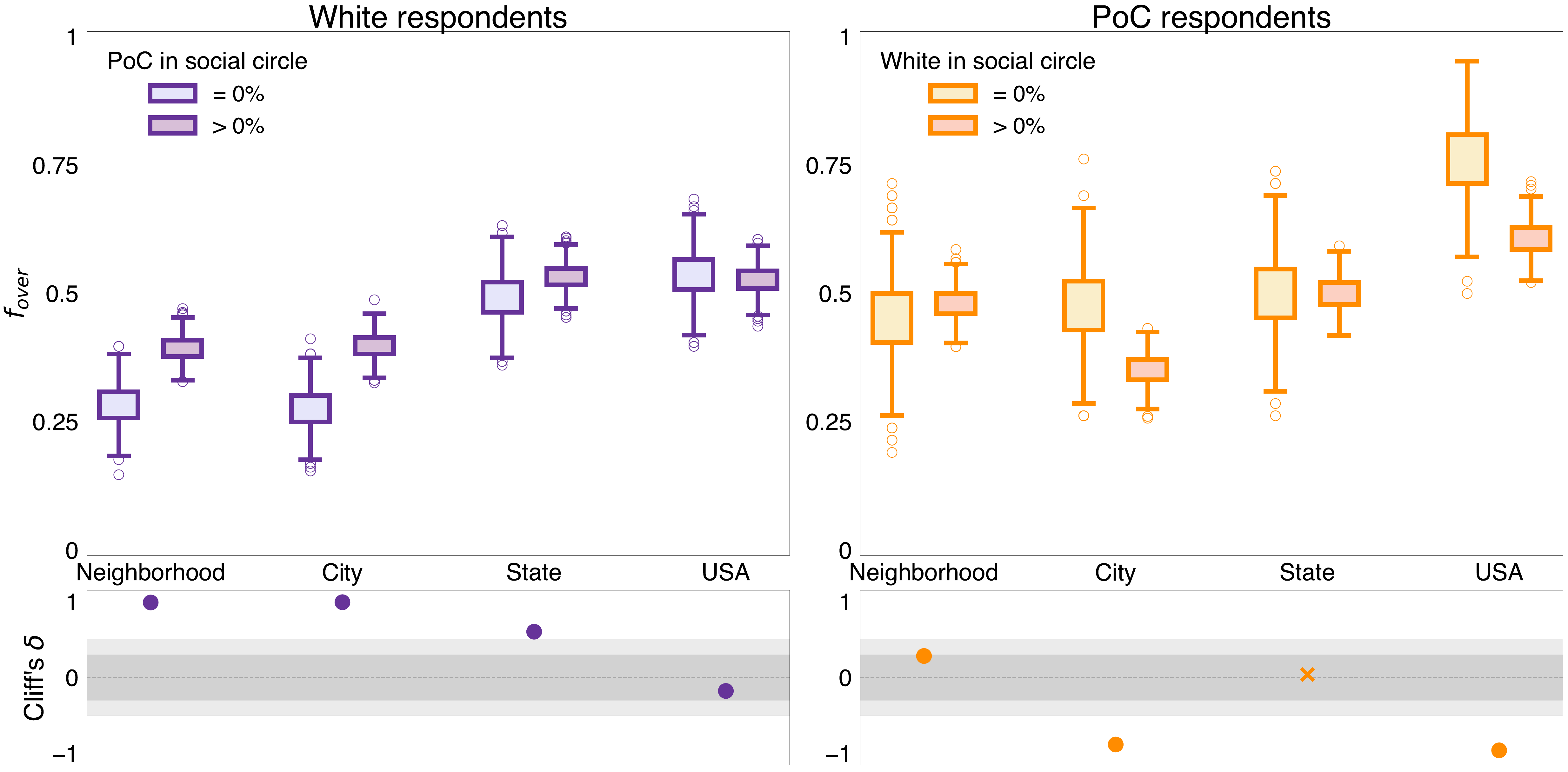}
        \caption{Proportional buffer at 10\%(as in the main text).}
    \end{subfigure}\hfill
    \begin{subfigure}{0.45\textwidth}
        \includegraphics[width=\linewidth]{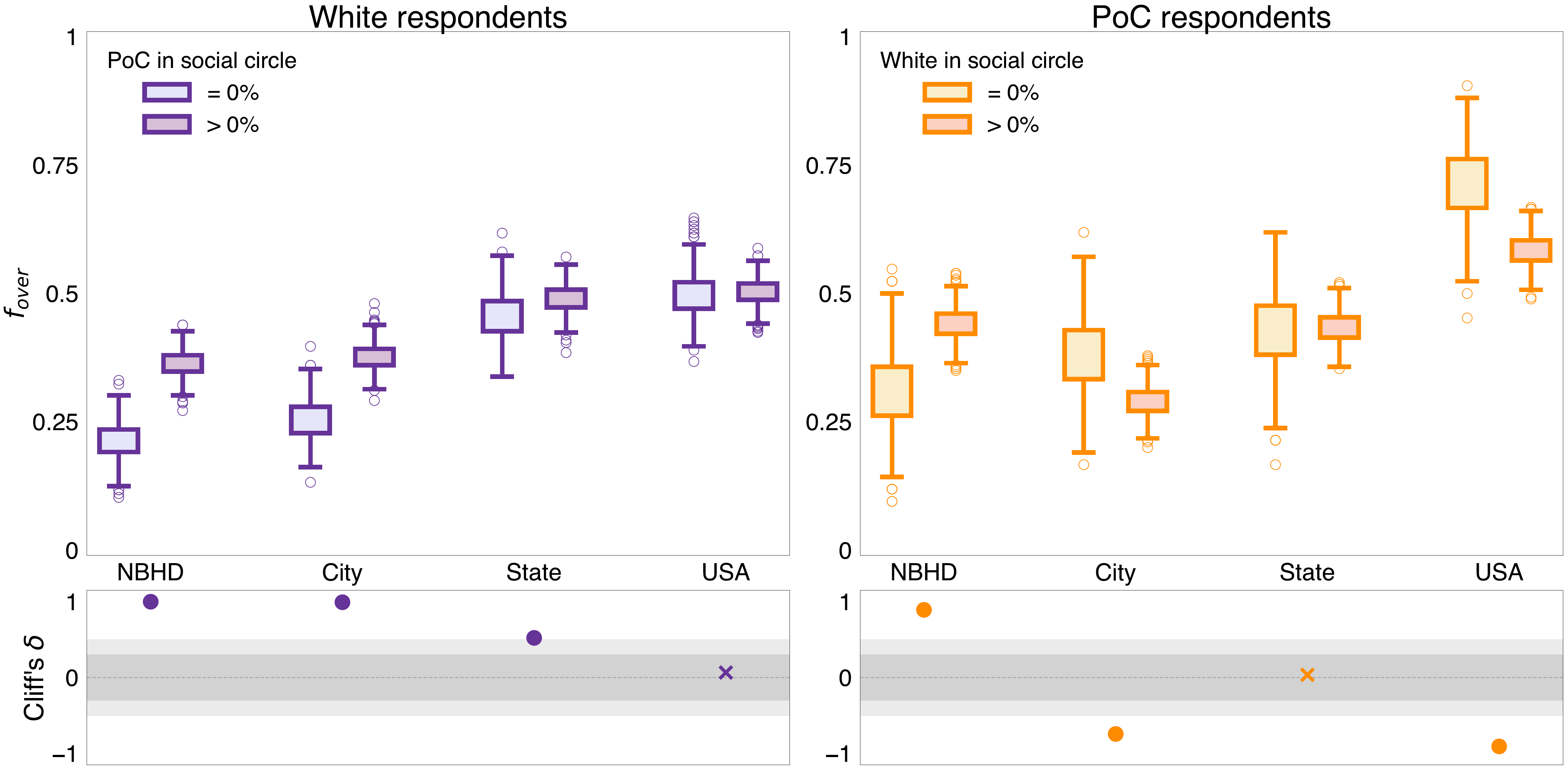}
        \caption{Proportional buffer at 15\%.}
    \end{subfigure}
    \caption{Figure~\ref{fig:soccirc} reproduced for different types and values of the buffer.}
\end{figure}

\begin{figure}[ht!]
    \centering
    \begin{subfigure}{0.45\textwidth}
        \includegraphics[width=\linewidth]{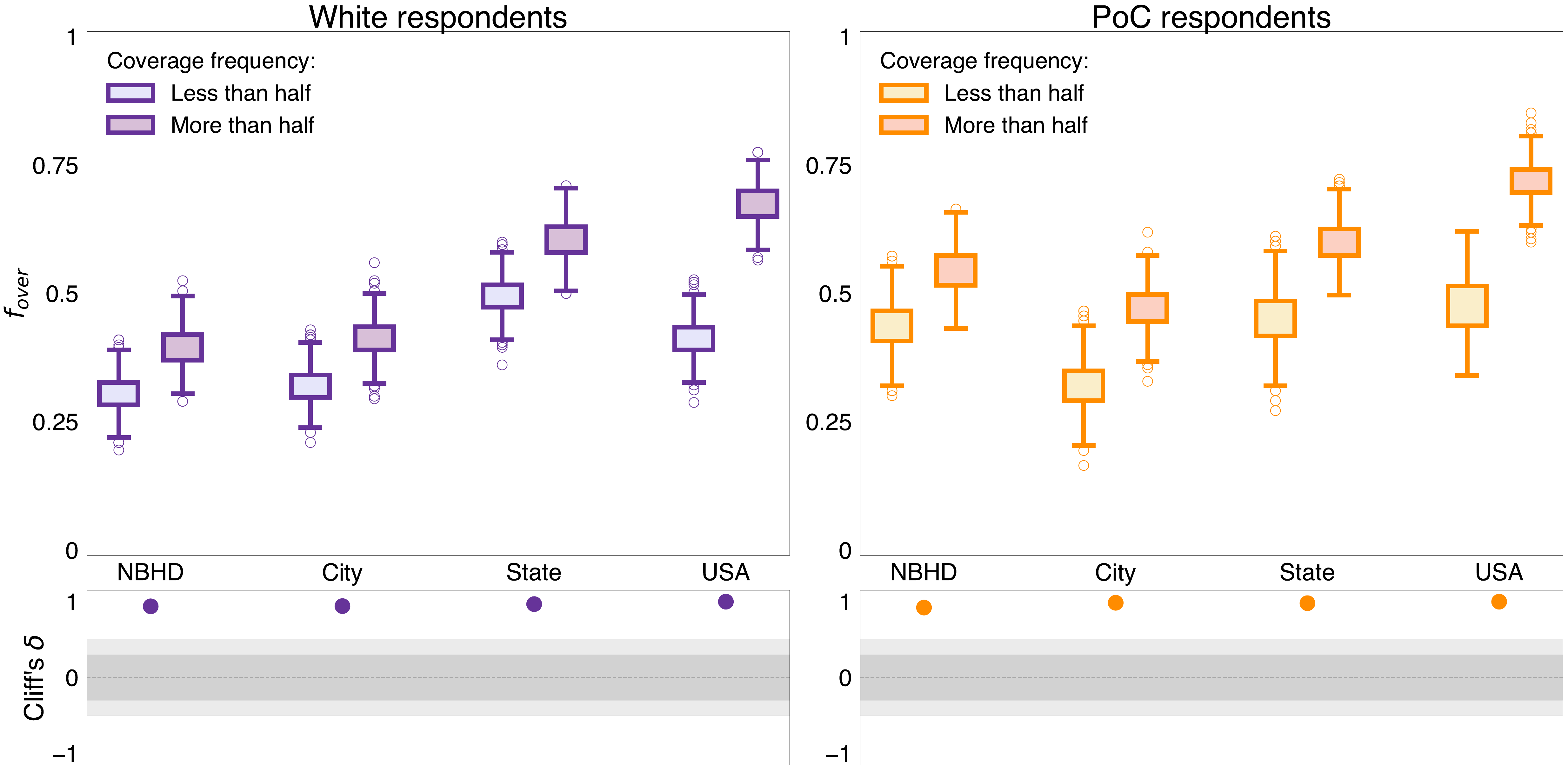}
        \caption{Fixed buffer at 3\%.}
    \end{subfigure}\hfill
    \begin{subfigure}{0.45\textwidth}
        \includegraphics[width=\linewidth]{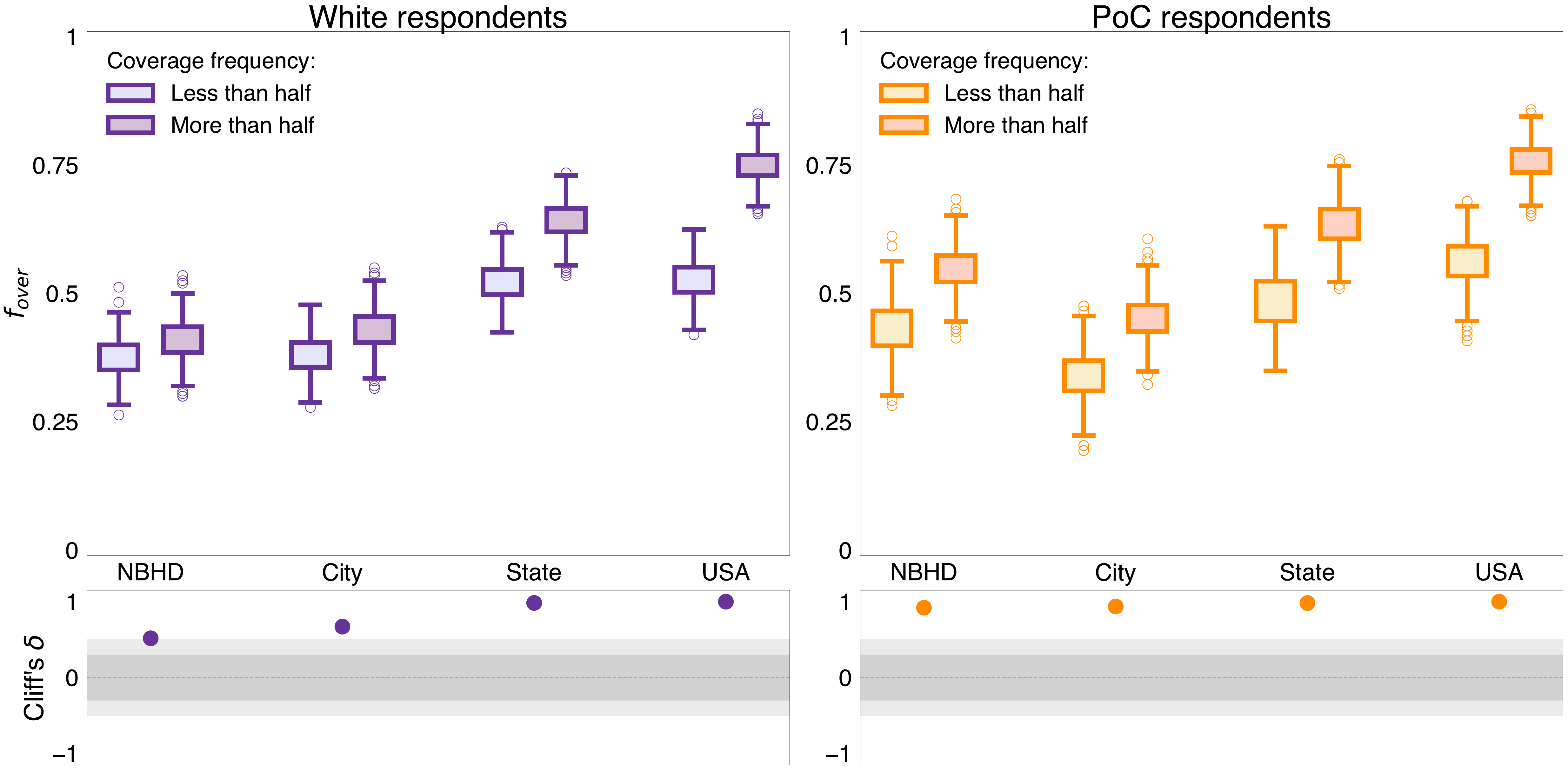}
        \caption{Proportional buffer at 5\%.}
    \end{subfigure}
    \vspace{0.5em}
    \begin{subfigure}{0.45\textwidth}
        \includegraphics[width=\linewidth]{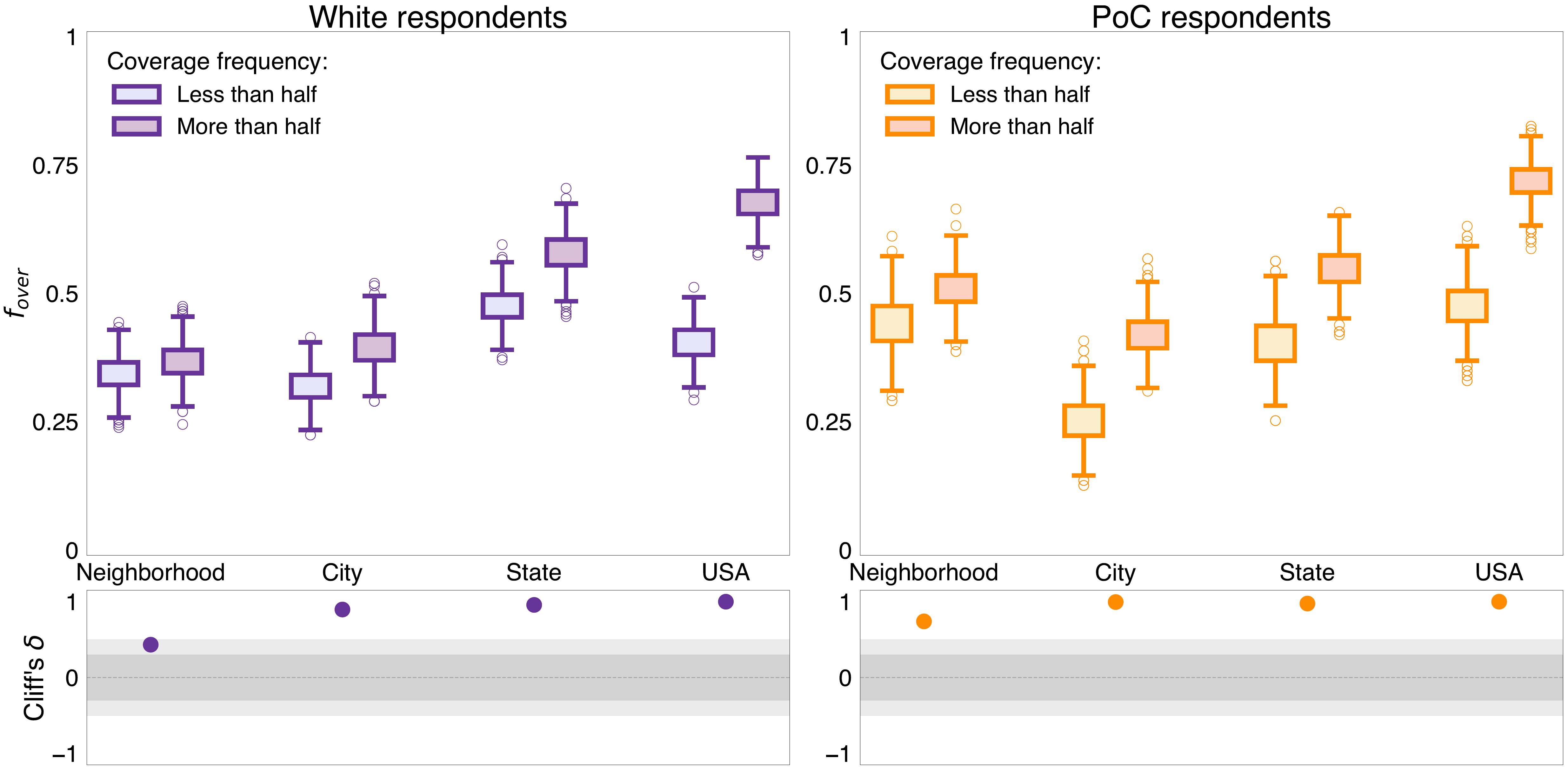}
        \caption{Proportional buffer at 10\%(as in the main text).}
    \end{subfigure}\hfill
    \begin{subfigure}{0.45\textwidth}
        \includegraphics[width=\linewidth]{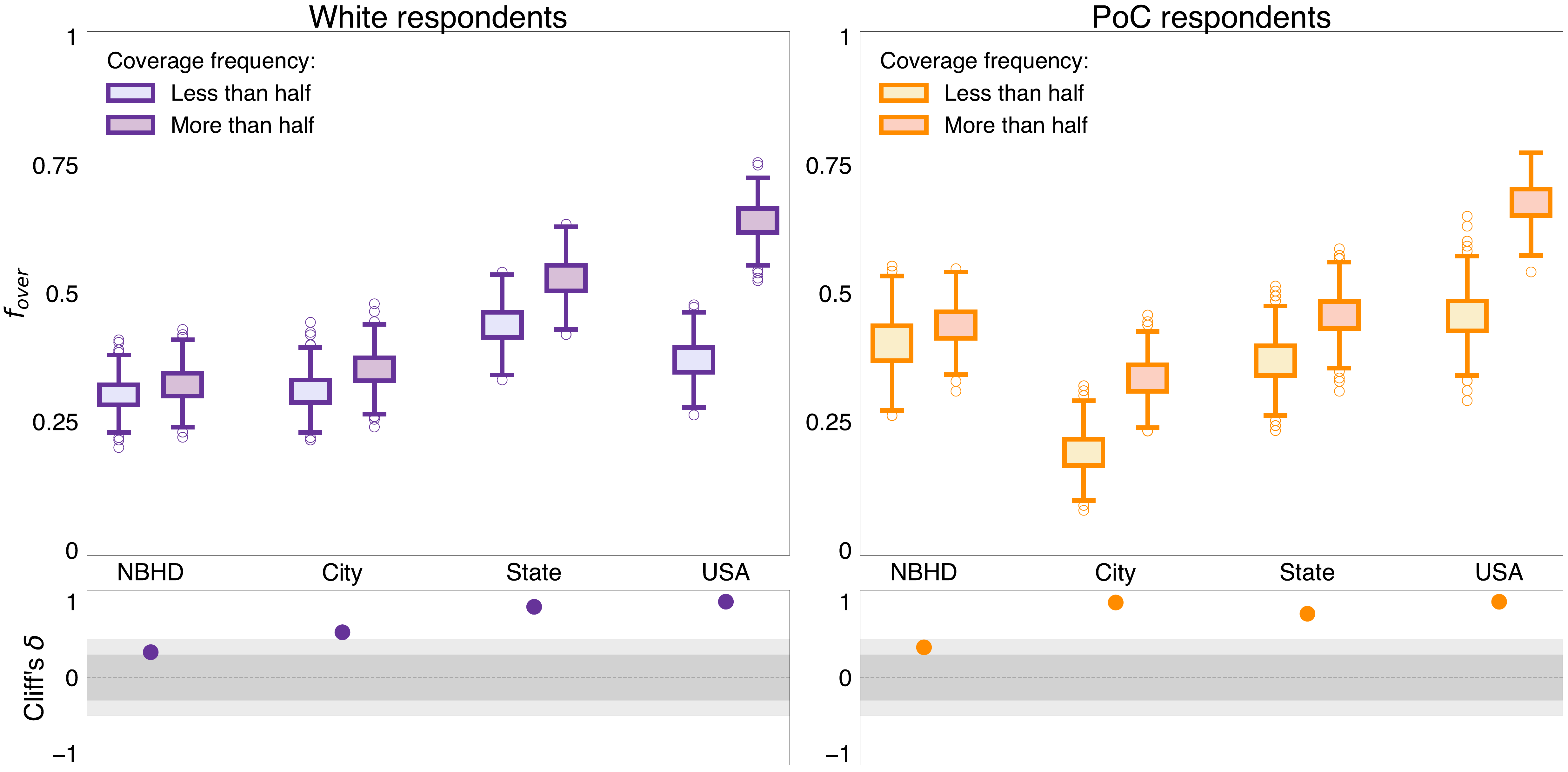}
        \caption{Proportional buffer at 15\%.}
    \end{subfigure}
    \caption{Figure~\ref{fig:coverage}a-b reproduced for different types and values of the buffer.}
\end{figure}

\begin{figure}[ht!]
    \centering
    \begin{subfigure}{0.45\textwidth}
        \includegraphics[width=\linewidth]{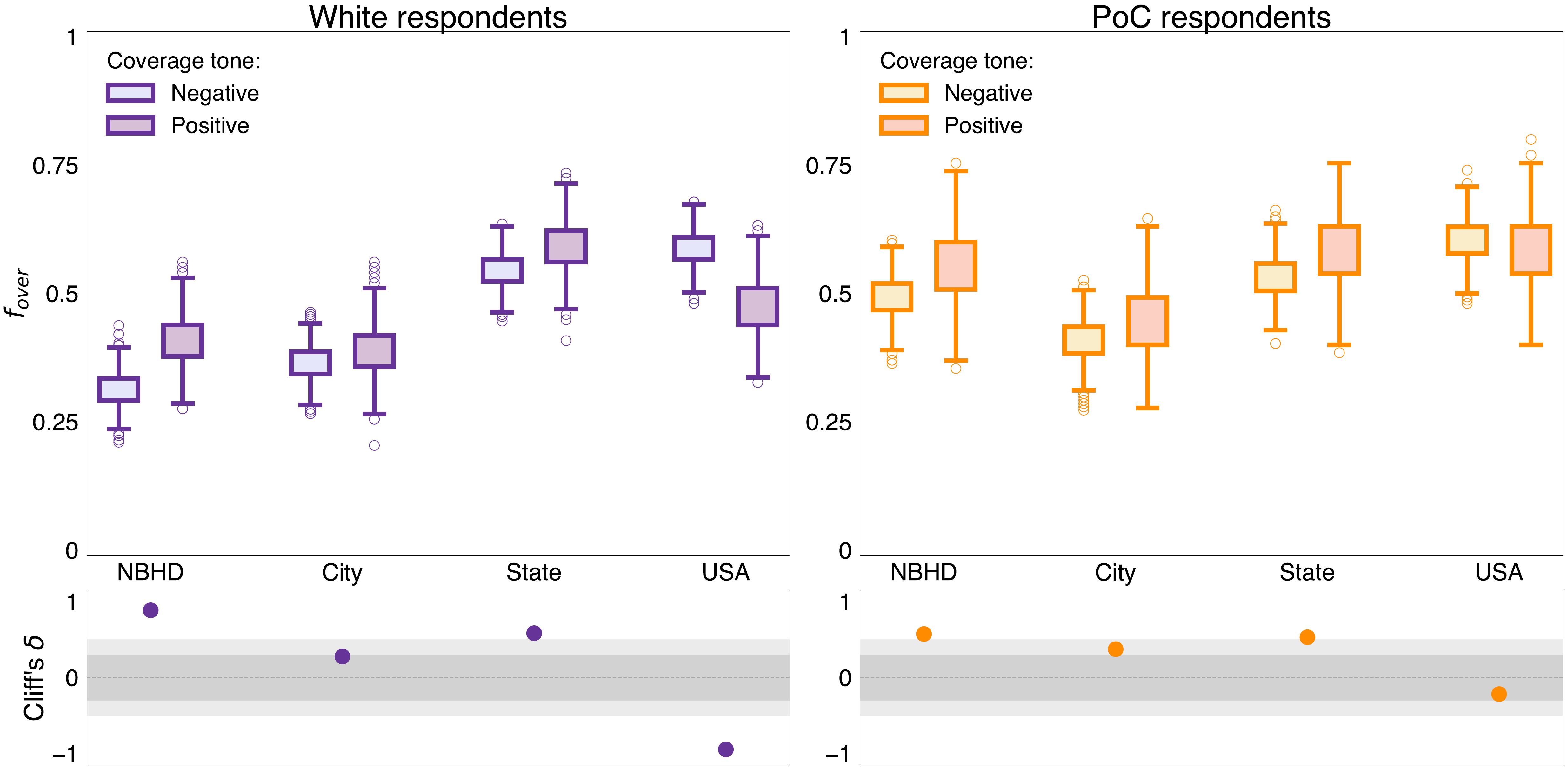}
        \caption{Fixed buffer at 3\% around.}
    \end{subfigure}\hfill
    \begin{subfigure}{0.45\textwidth}
        \includegraphics[width=\linewidth]{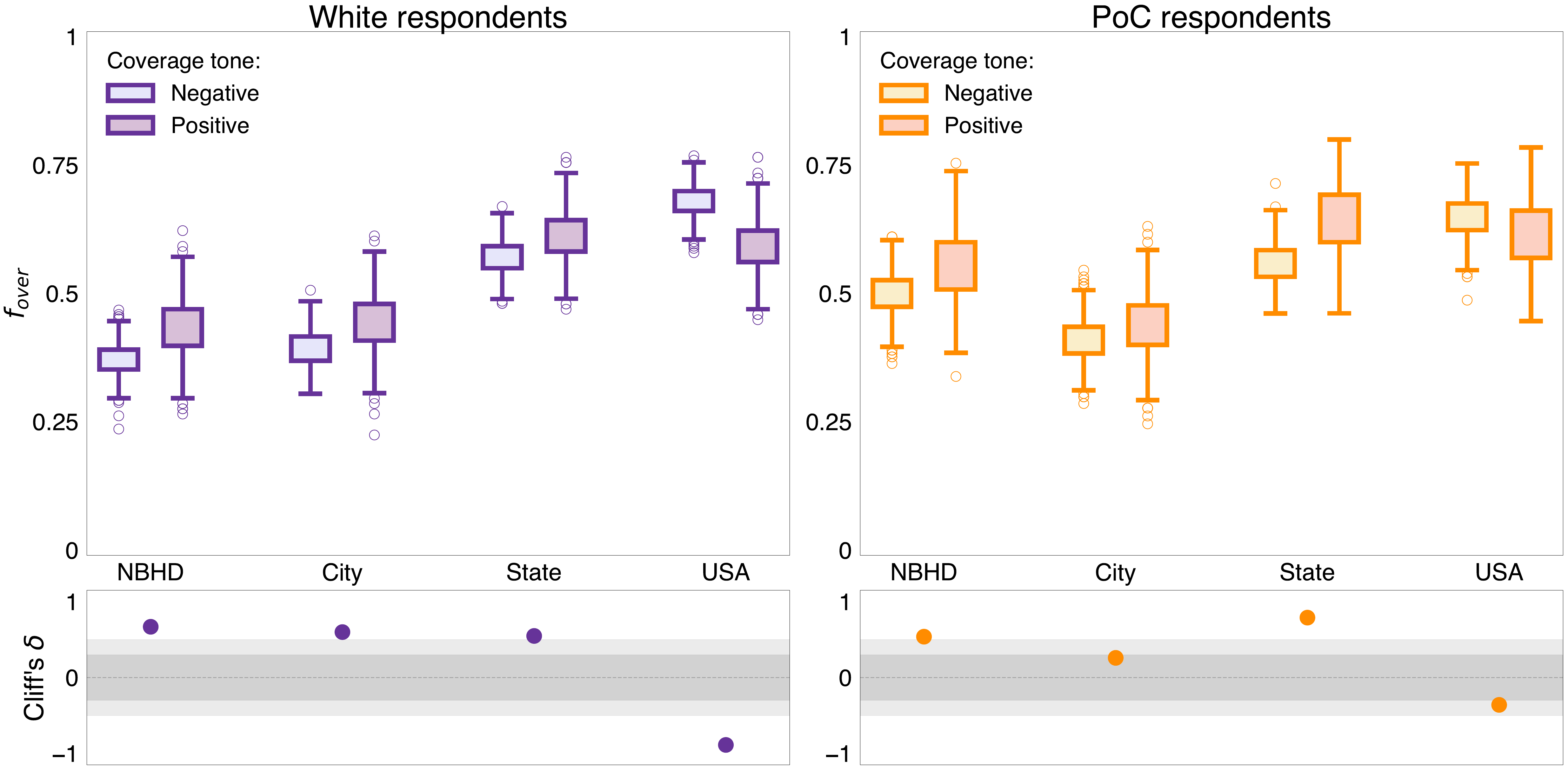}
        \caption{Proportional buffer at 5\%.}
    \end{subfigure}
    \vspace{0.5em}
    \begin{subfigure}{0.45\textwidth}
        \includegraphics[width=\linewidth]{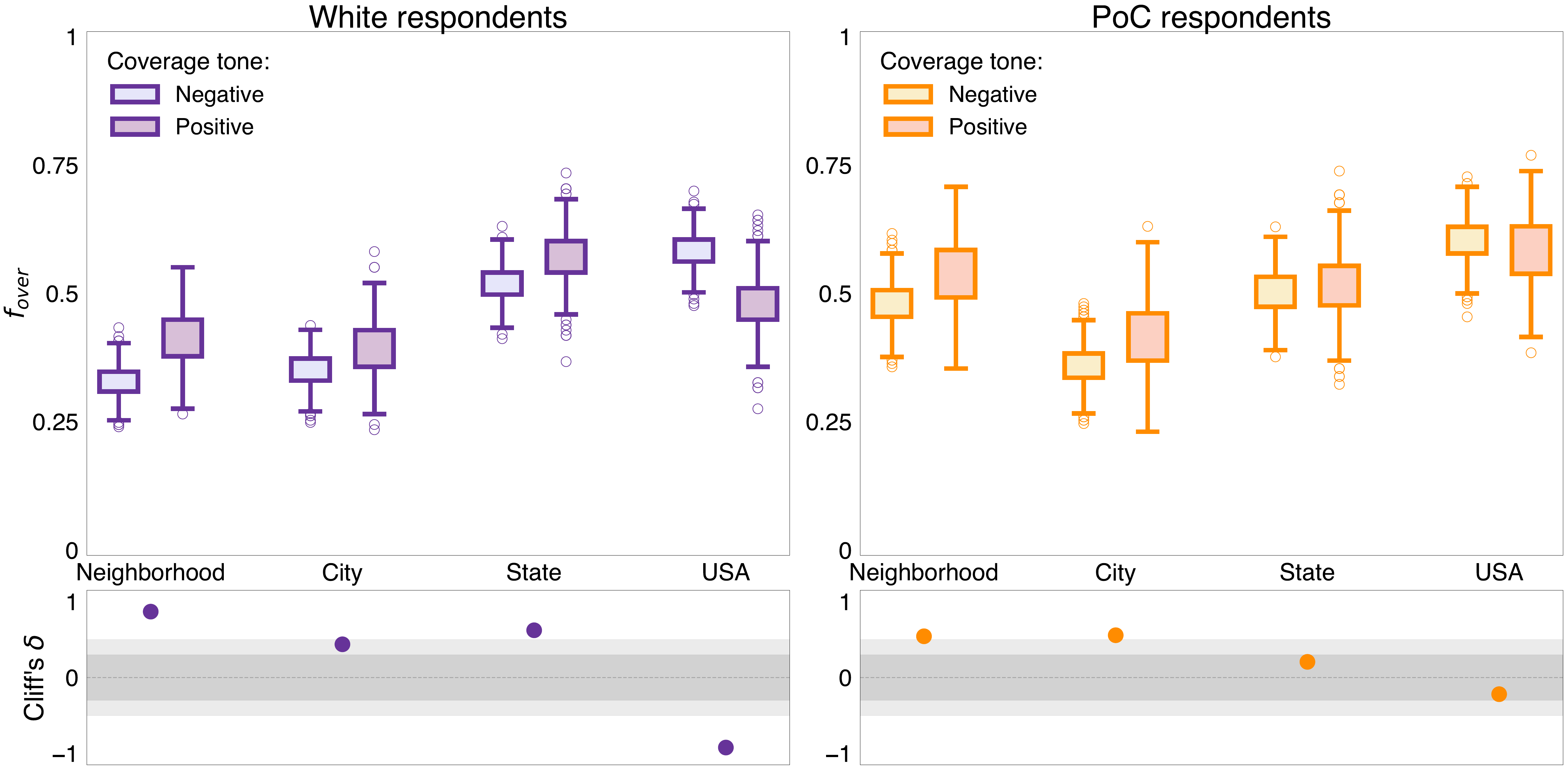}
        \caption{Proportional buffer at 10\% (as in the main text).}
    \end{subfigure}\hfill
    \begin{subfigure}{0.45\textwidth}
        \includegraphics[width=\linewidth]{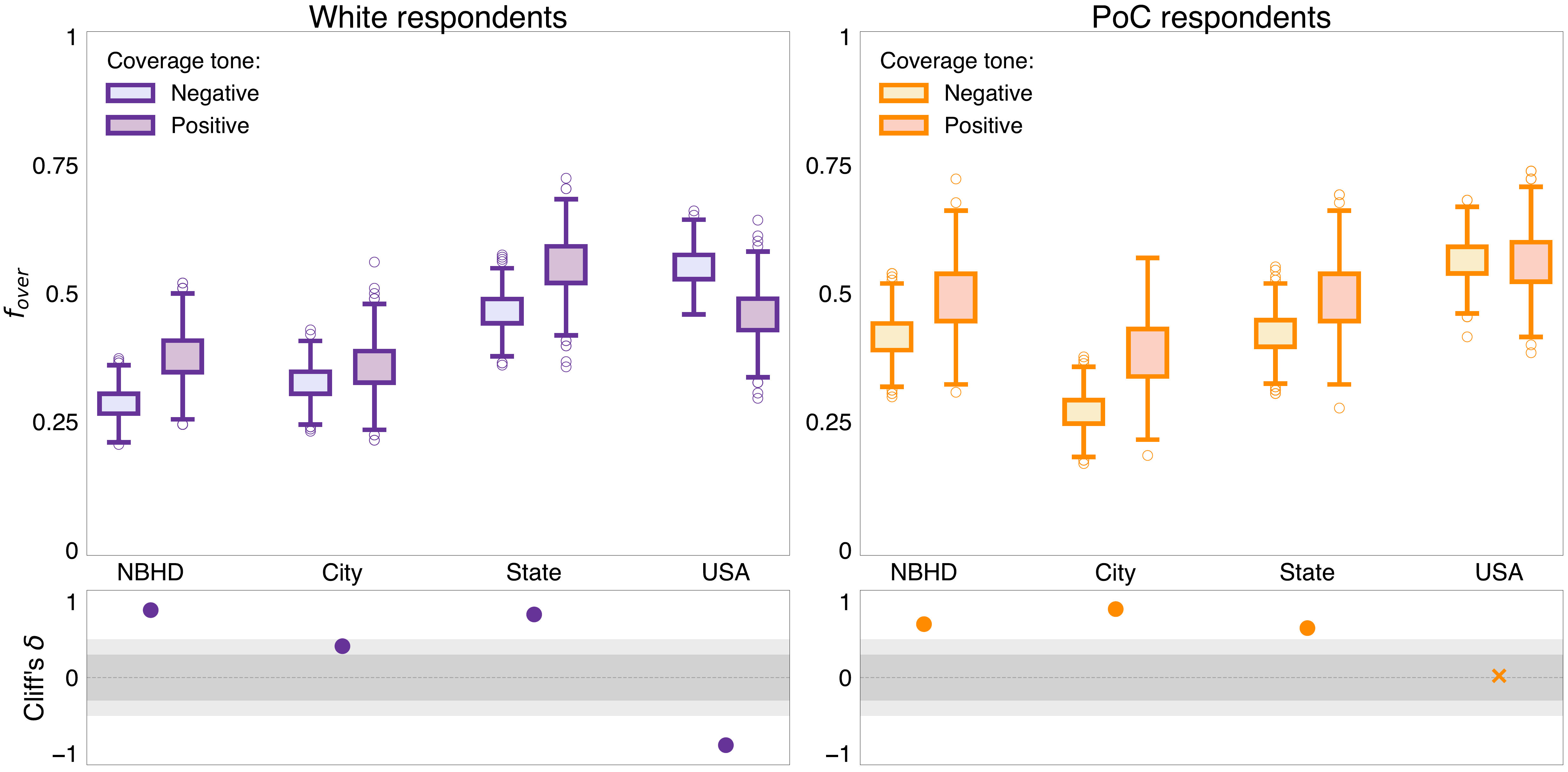}
        \caption{Proportional buffer at 15\%.}
    \end{subfigure}
    \caption{Figure~\ref{fig:coverage}c-d reproduced for different types and values of the buffer.}
\end{figure}

\begin{figure}[ht!]
    \centering
    \begin{subfigure}{0.45\textwidth}
        \includegraphics[width=\linewidth]{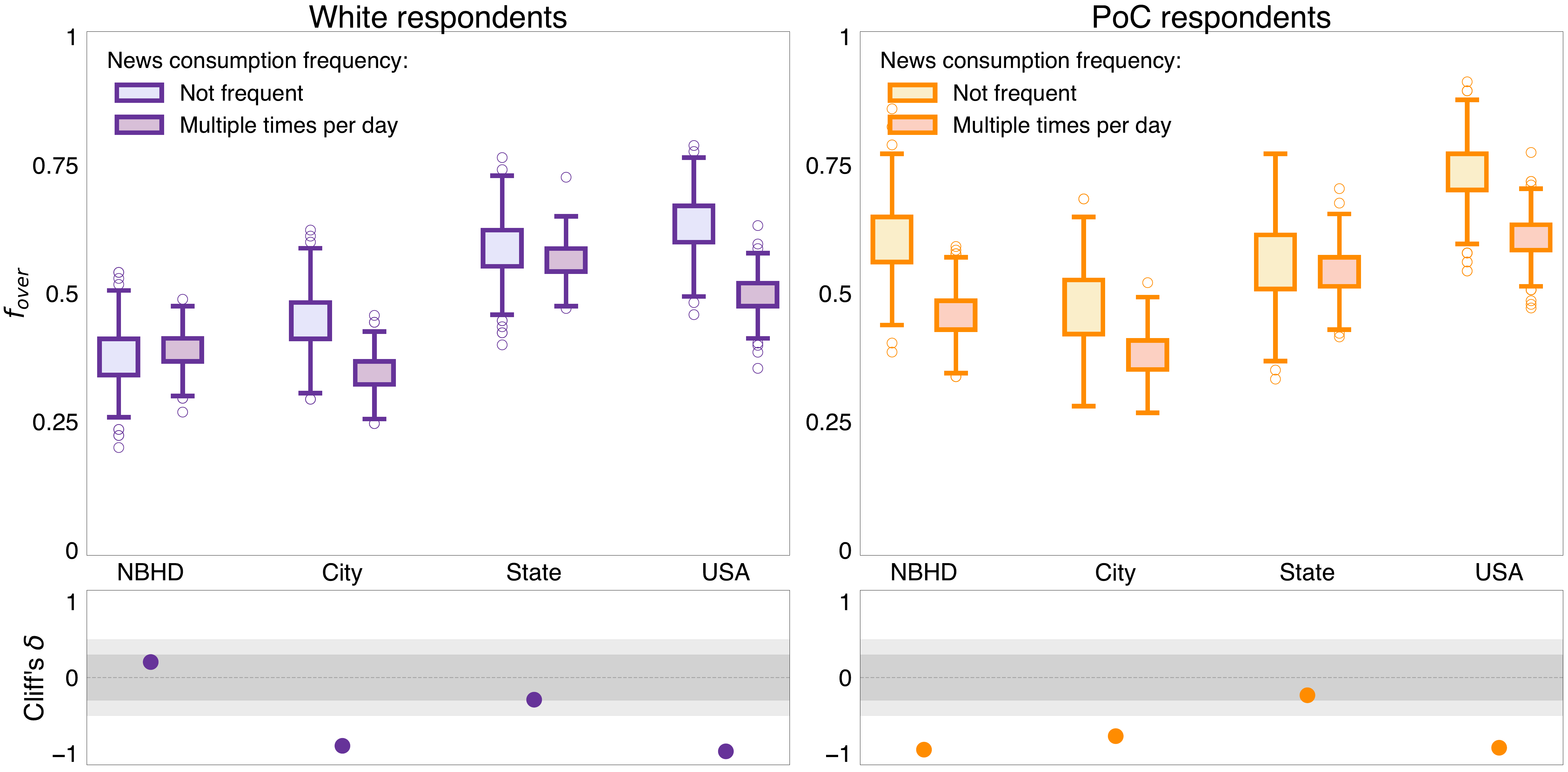}
        \caption{Fixed buffer at 3\%.}
    \end{subfigure}\hfill
    \begin{subfigure}{0.45\textwidth}
        \includegraphics[width=\linewidth]{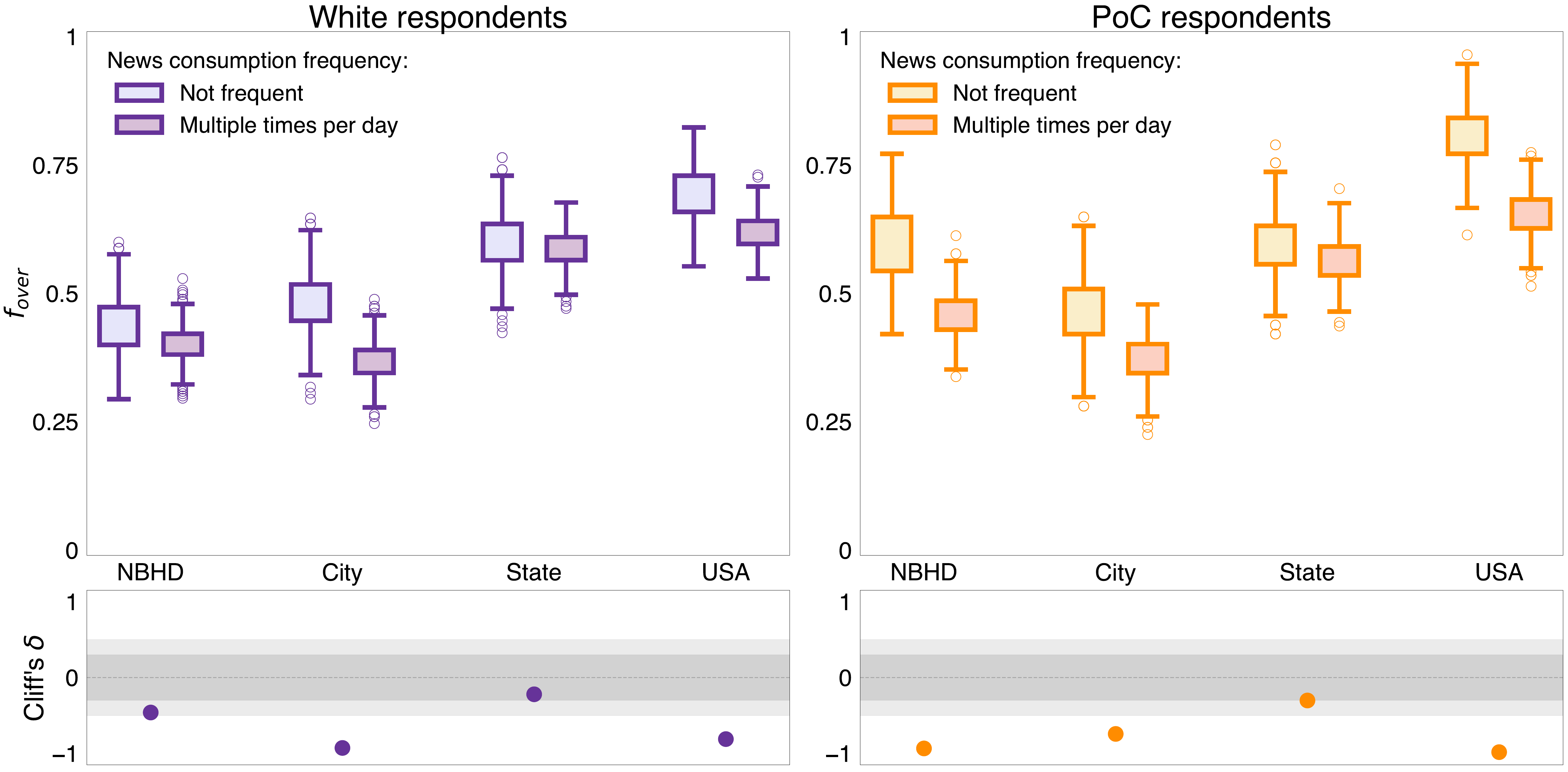}
        \caption{Proportional buffer at 5\%.}
    \end{subfigure}
    \vspace{0.5em}
    \begin{subfigure}{0.45\textwidth}
        \includegraphics[width=\linewidth]{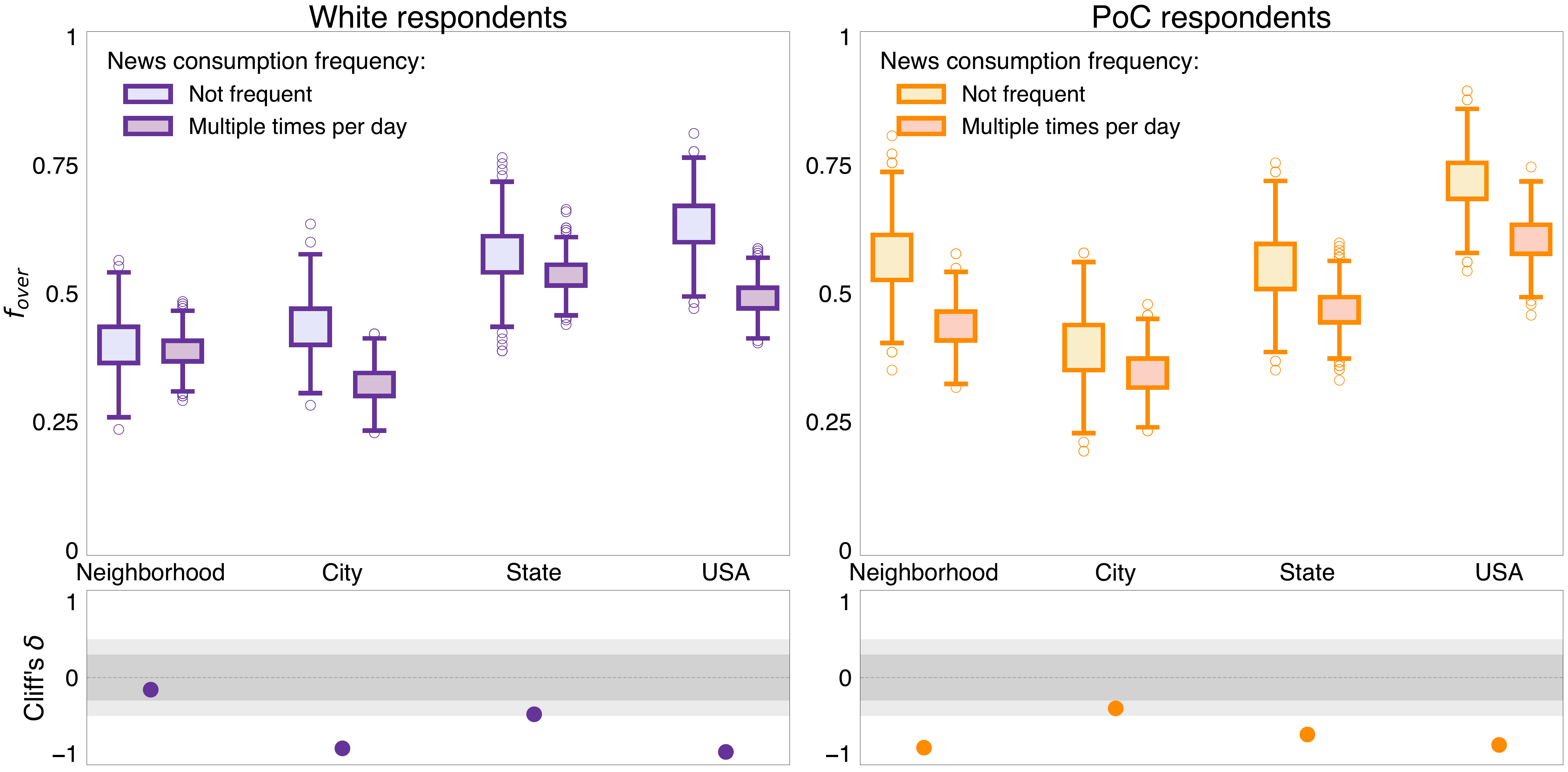}
        \caption{Proportional buffer at 10\%(as in the main text).}
    \end{subfigure}\hfill
    \begin{subfigure}{0.45\textwidth}
        \includegraphics[width=\linewidth]{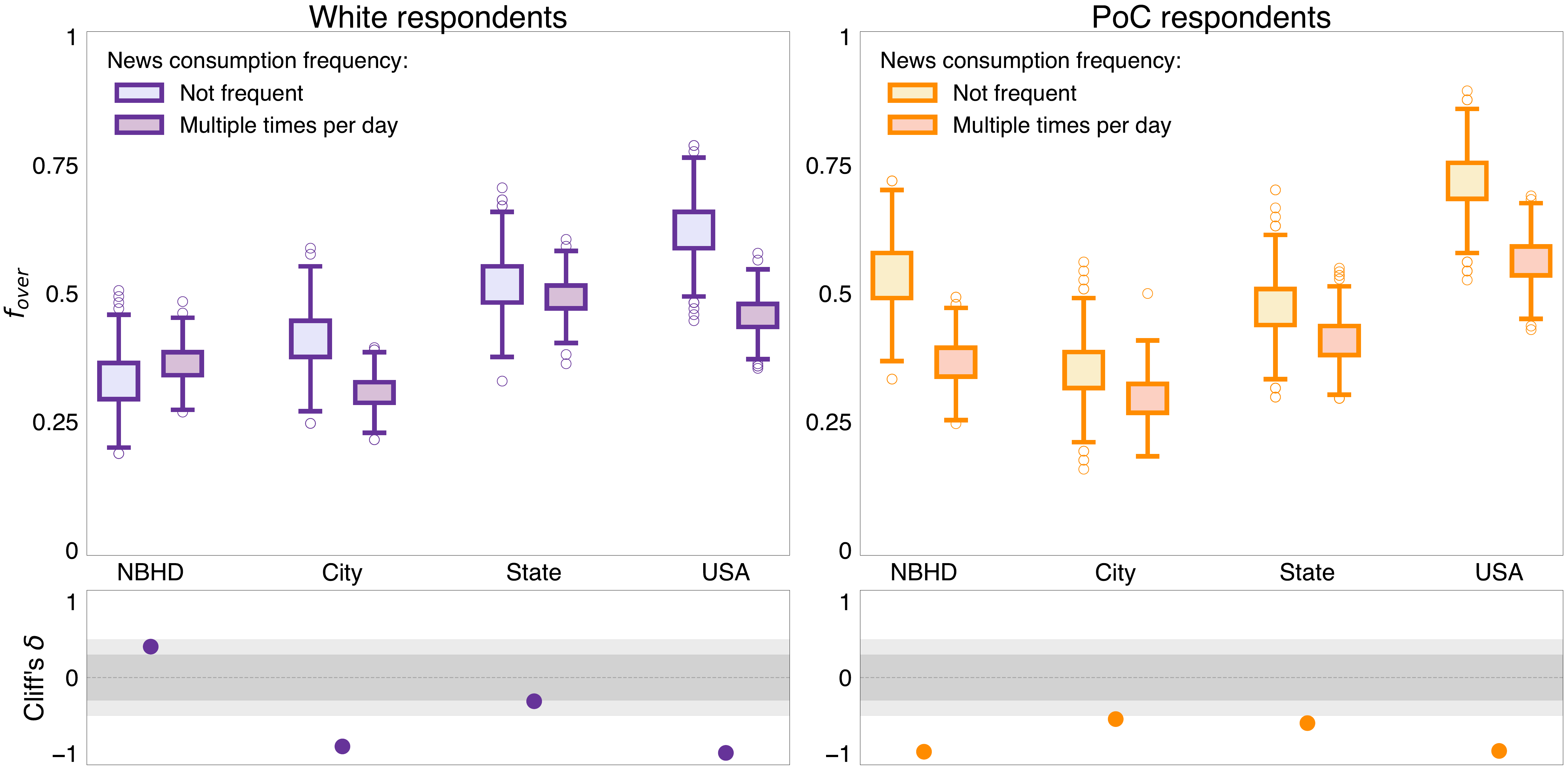}
        \caption{Proportional buffer at 15\%.}
    \end{subfigure}
    \caption{Figure~\ref{fig:frequency}a-b reproduced for different types and values of the buffer. }
\end{figure}

\begin{figure}[ht!]
    \centering
    \begin{subfigure}{0.45\textwidth}
        \includegraphics[width=\linewidth]{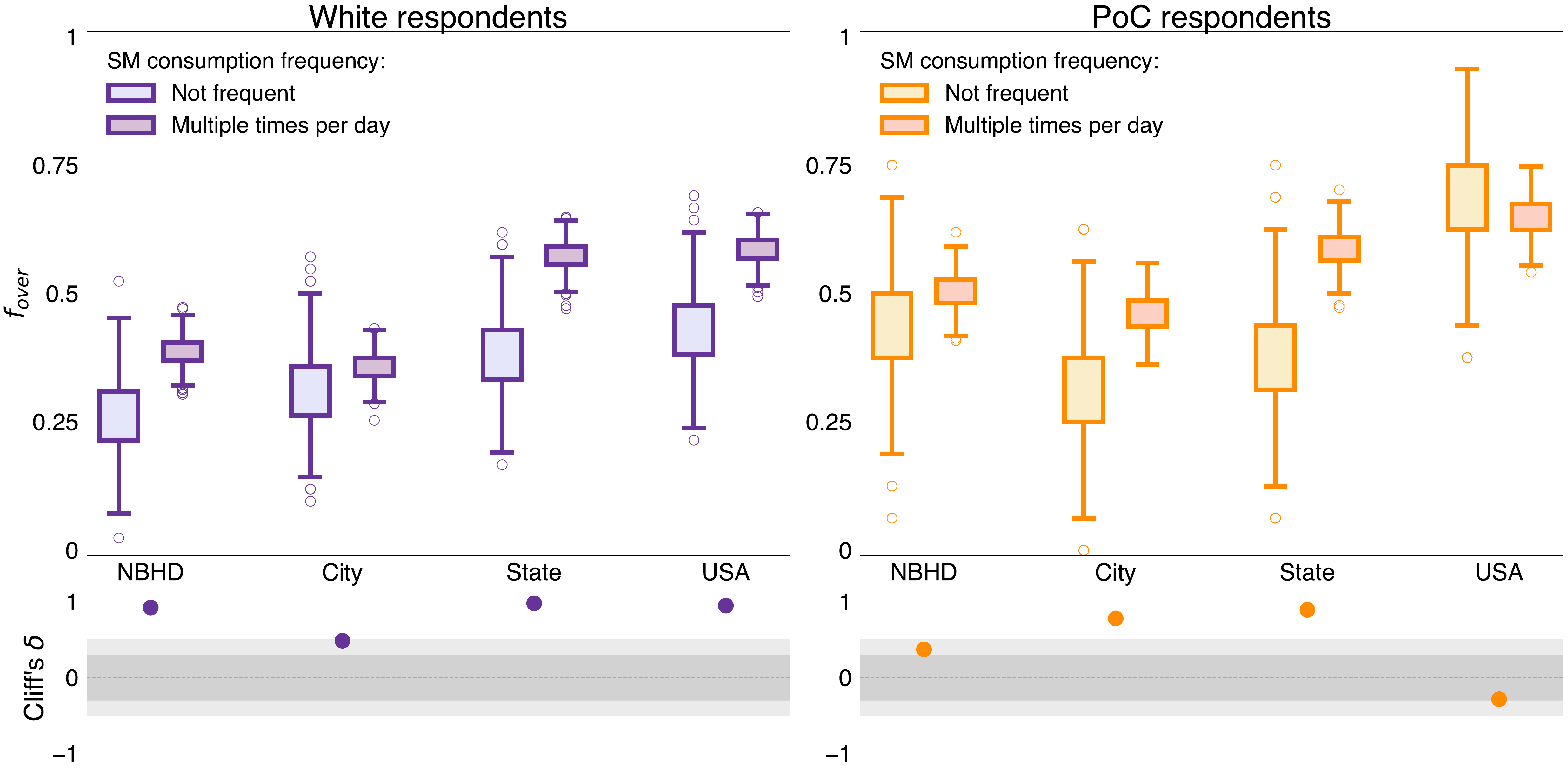}
        \caption{Fixed buffer at 3\%.}
    \end{subfigure}\hfill
    \begin{subfigure}{0.45\textwidth}
        \includegraphics[width=\linewidth]{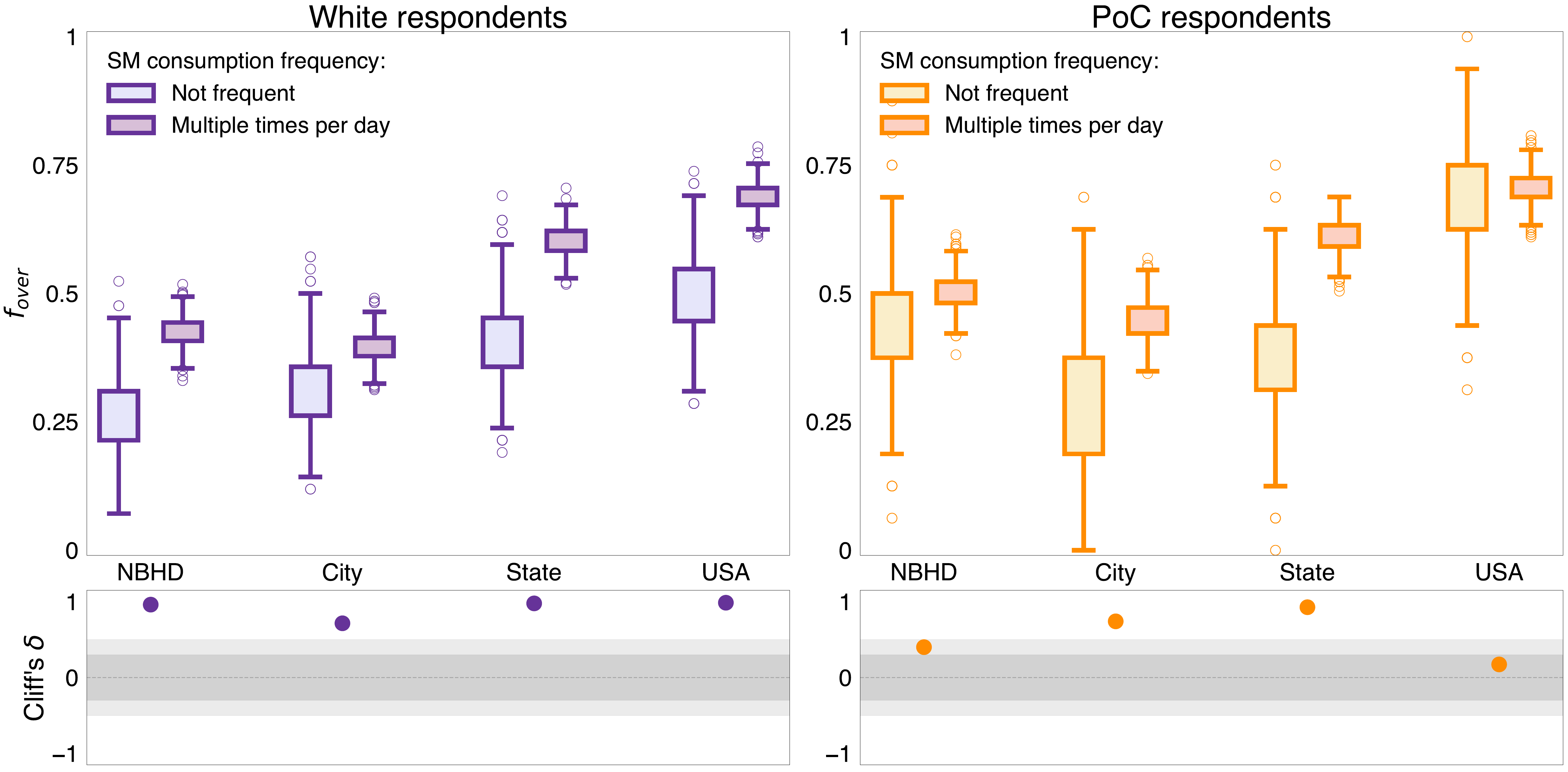}
        \caption{Proportional buffer at 5\%.}
    \end{subfigure}
    \vspace{0.5em}
    \begin{subfigure}{0.45\textwidth}
        \includegraphics[width=\linewidth]{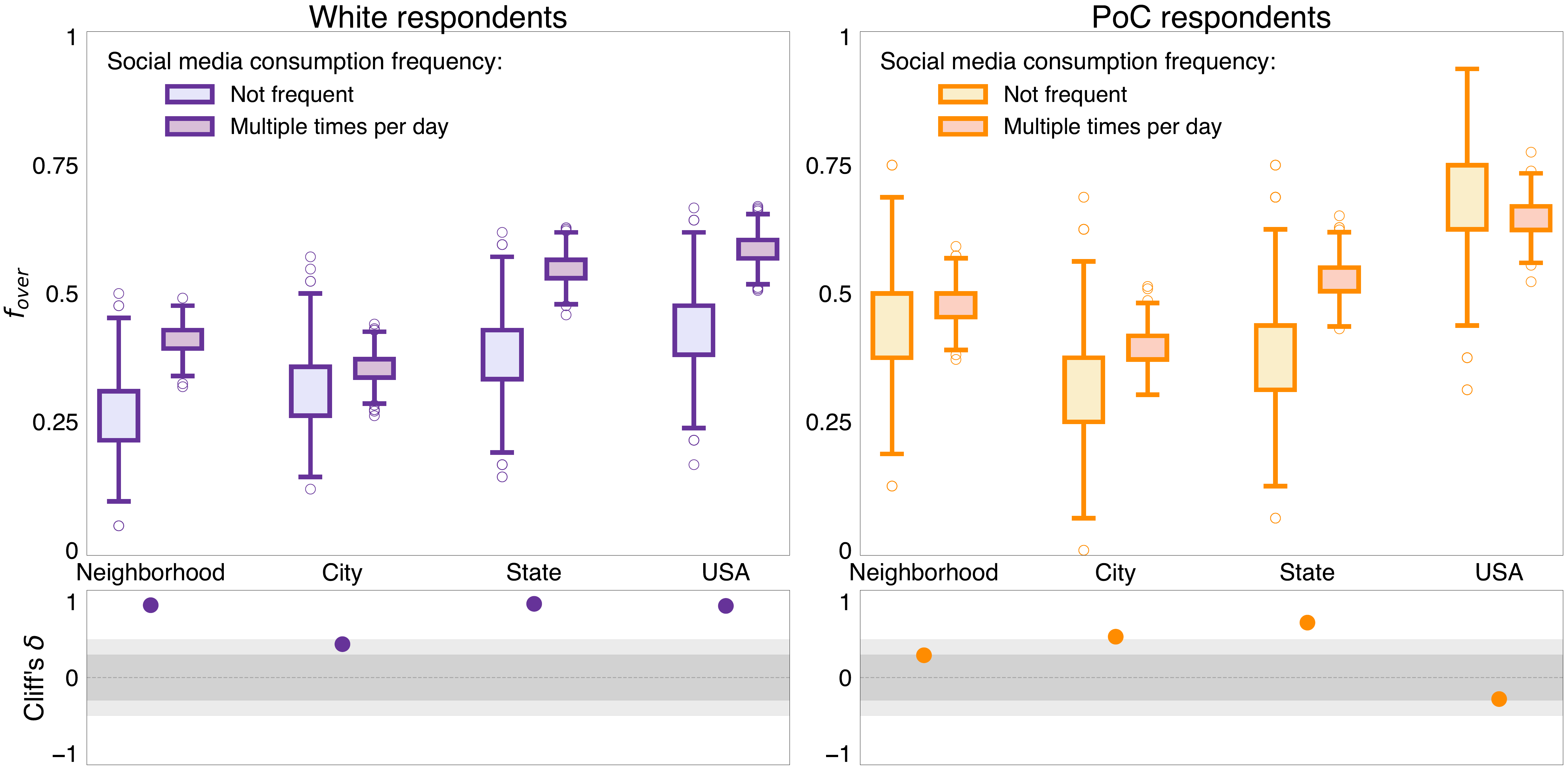}
        \caption{Proportional buffer at 10\%(as in the main text).}
    \end{subfigure}\hfill
    \begin{subfigure}{0.45\textwidth}
        \includegraphics[width=\linewidth]{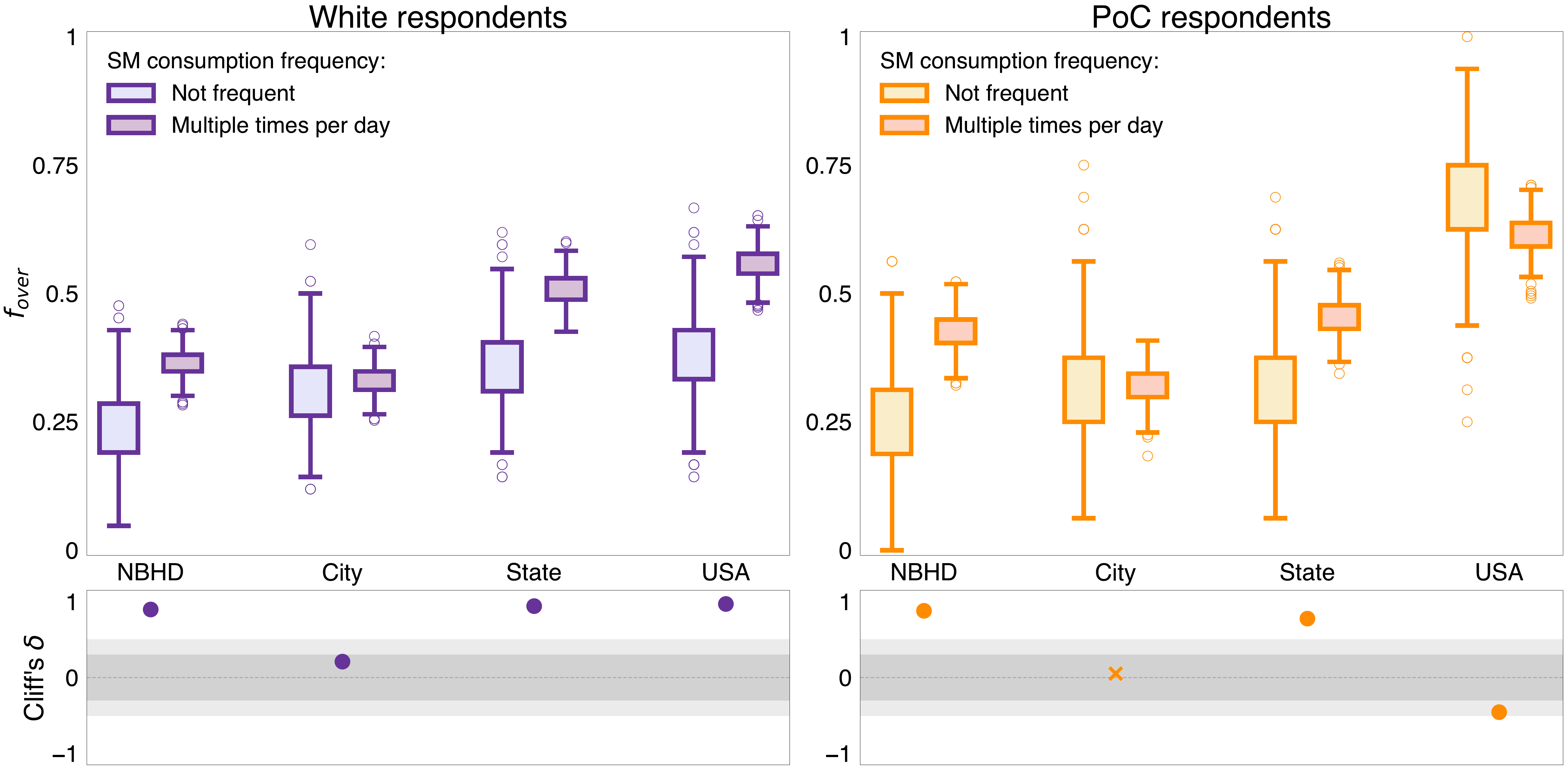}
        \caption{Proportional buffer at 15\%.}
    \end{subfigure}
    \caption{Figure~\ref{fig:frequency}a-b reproduced for different types and values of the buffer. }
\end{figure}

\FloatBarrier

\section{Analysis of under and correct estimations} \label{sec:supmat:undercorrect}
Examining overestimation alone does not provide a complete picture of the results. The fact that respondents with characteristic $A$ are more likely to overestimate than those with characteristic $B$ does not imply that respondents with characteristic $B$ are more likely to be correct, as they may be more likely to underestimate. While it is still correct to state that a group of respondents is more likely to overestimate relative to another group, it would not be correct to conclude that the second group is more likely to provide accurate estimates. For this reason, we also provide for completeness the patterns of correct and underestimation. 

\begin{figure}[ht!]
    \centering
    \begin{subfigure}{0.5\textwidth}
        \centering
        \includegraphics[width=\linewidth]{img/geores_focus_PoC_proportional_10_over.pdf}
        \caption{Over (as in the text)}
    \end{subfigure}
    \begin{subfigure}{0.5\textwidth}
        \centering
        \includegraphics[width=\linewidth]{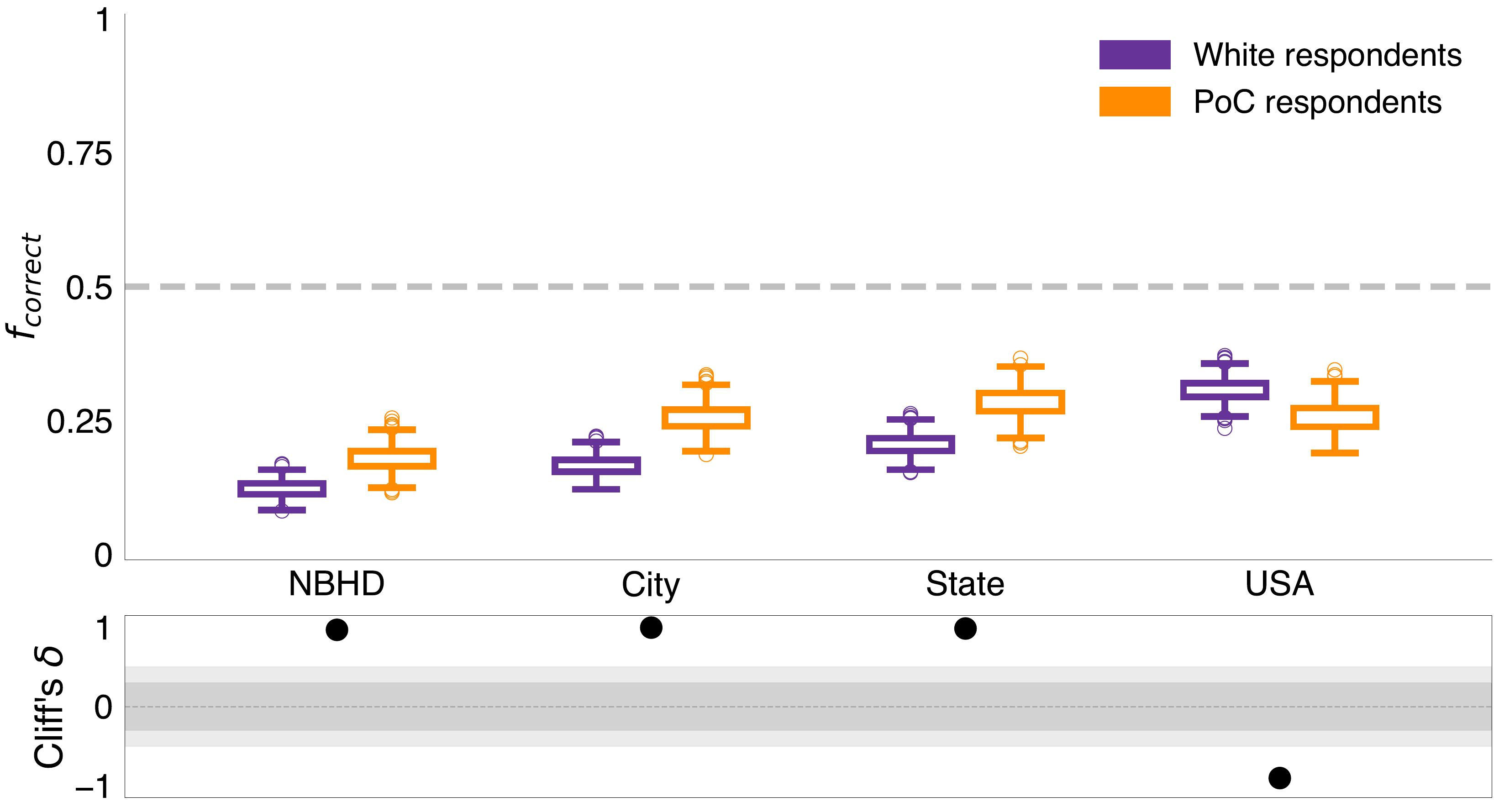}
        \caption{Correct}
    \end{subfigure}
    \begin{subfigure}{0.5\textwidth}
        \centering
        \includegraphics[width=\linewidth]{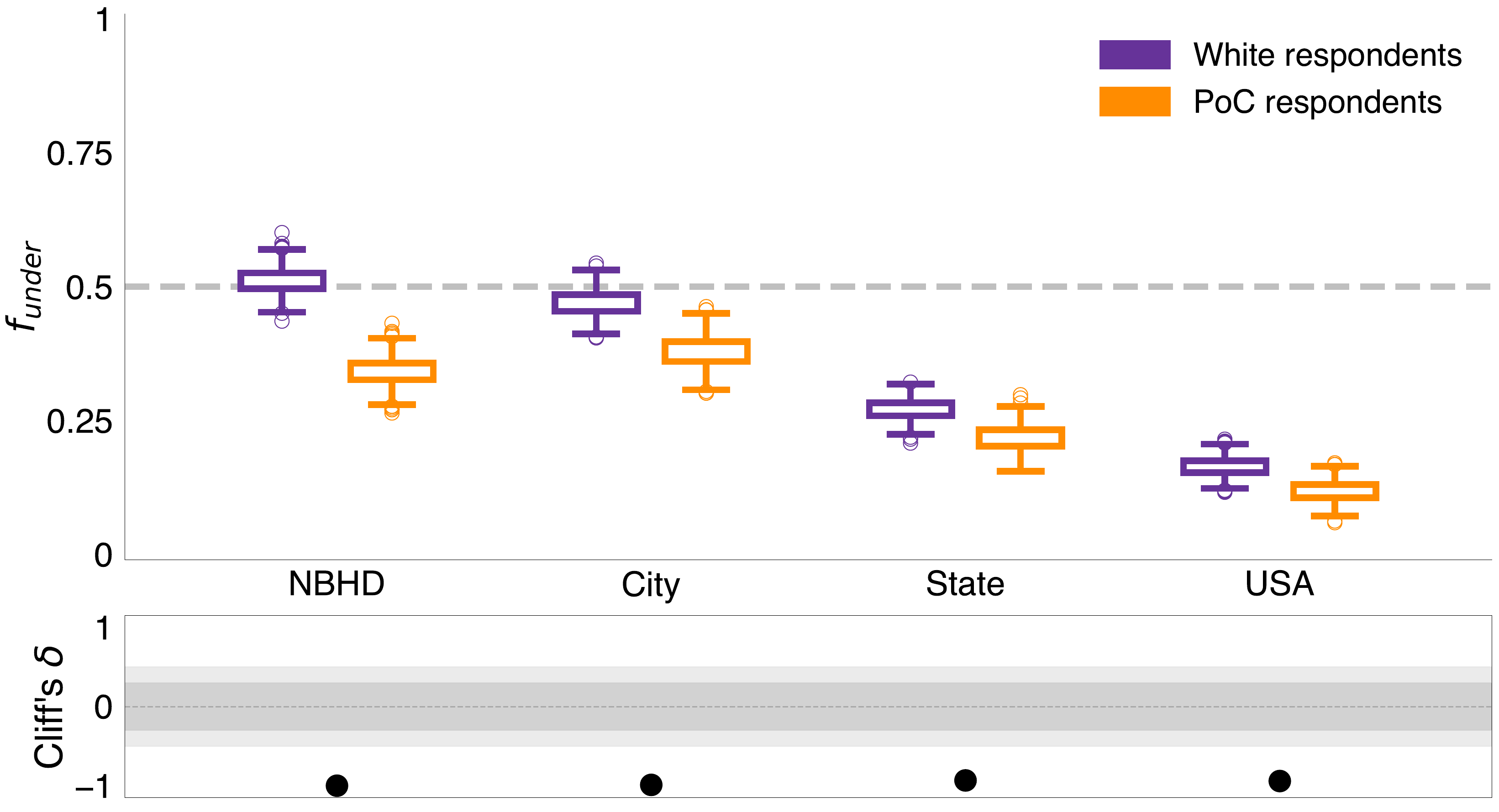}
        \caption{Under}
    \end{subfigure}
    \caption{Figure~\ref{fig:methods_geores}a reproduced for the fraction of correct and underestimations}
    \label{fig:supmat:overcorrect:methods_geores}
\end{figure}

\begin{figure}[ht!]
    \centering
    \begin{subfigure}{0.65\textwidth}
        \centering
        \includegraphics[width=\linewidth]{img/soccircle_focus_PoC_proportional_10_over.pdf}
        \caption{Over (as in the text)}
    \end{subfigure}
    \begin{subfigure}{0.65\textwidth}
        \centering
        \includegraphics[width=\linewidth]{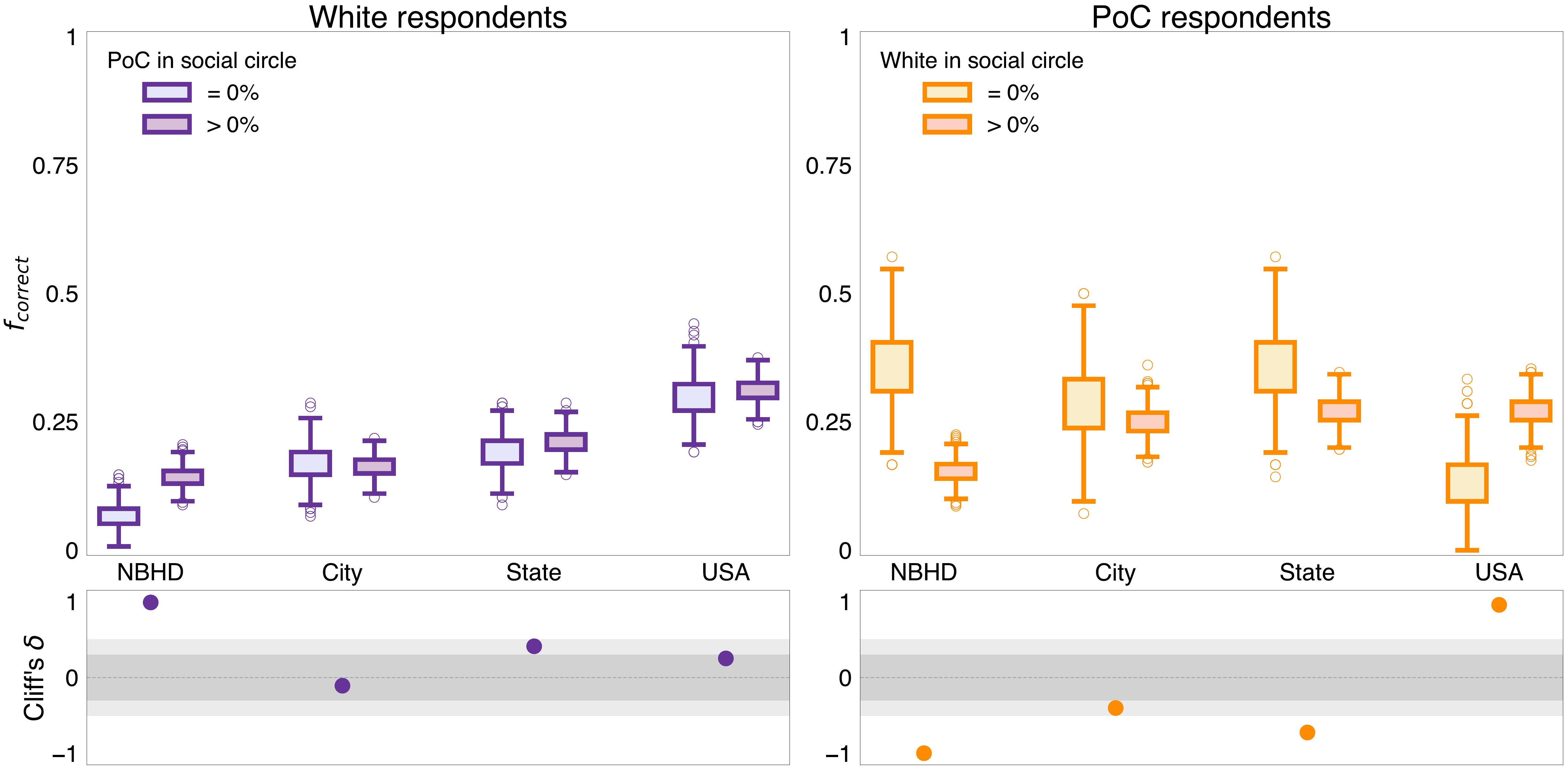}
        \caption{Correct}
    \end{subfigure}
    \begin{subfigure}{0.65\textwidth}
        \centering
        \includegraphics[width=\linewidth]{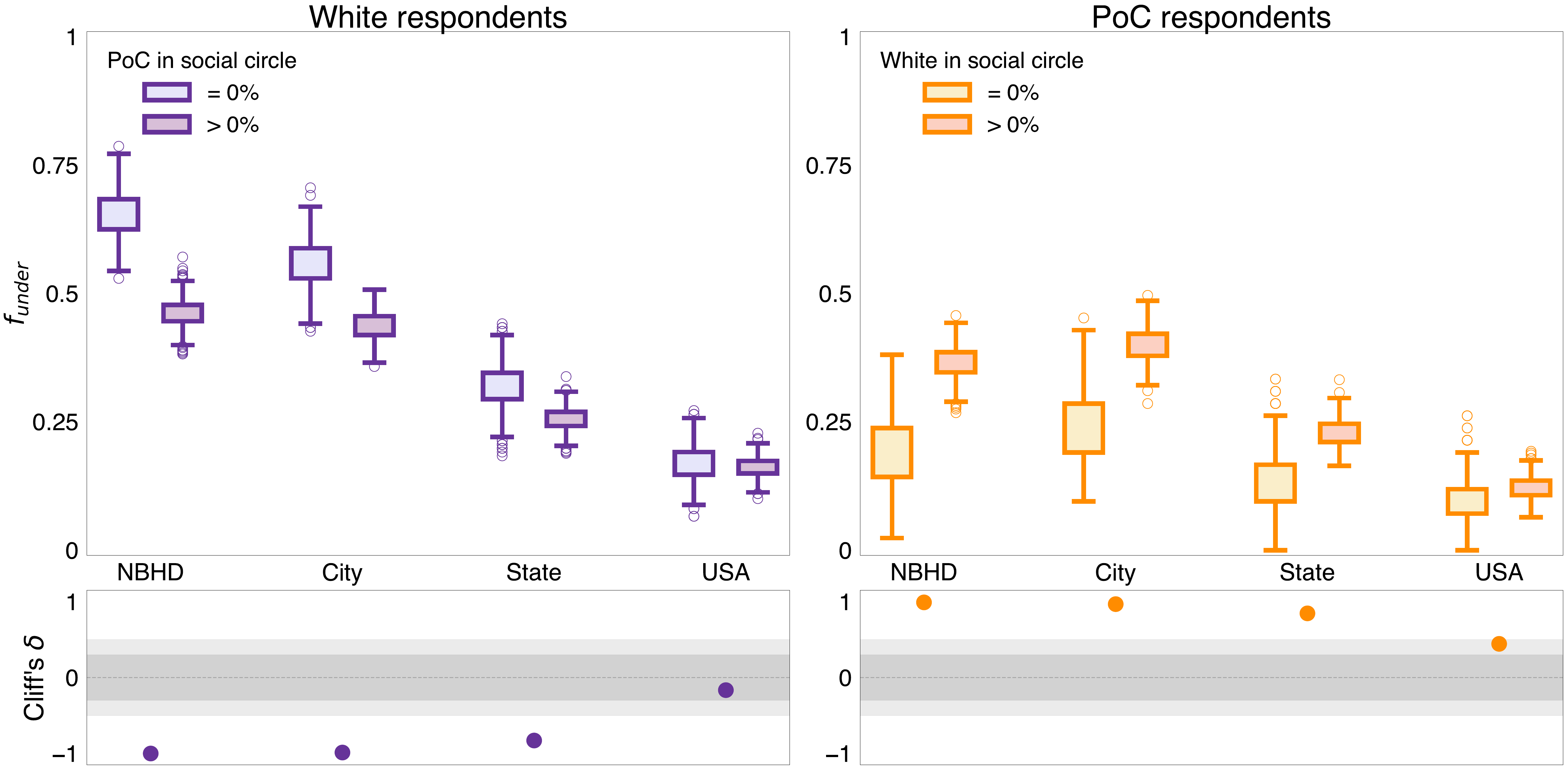}
        \caption{Under}
    \end{subfigure}
    \caption{Figure~\ref{fig:soccirc} reproduced for the fraction of correct and under estimations}
\end{figure}

\begin{figure}[ht!]
    \centering
    \begin{subfigure}{0.65\textwidth}
        \centering
        \includegraphics[width=\linewidth]{img/covfreq_focus_PoC_proportional_10_over.pdf}
        \caption{Over (as in the text)}
    \end{subfigure}
    \begin{subfigure}{0.65\textwidth}
        \centering
        \includegraphics[width=\linewidth]{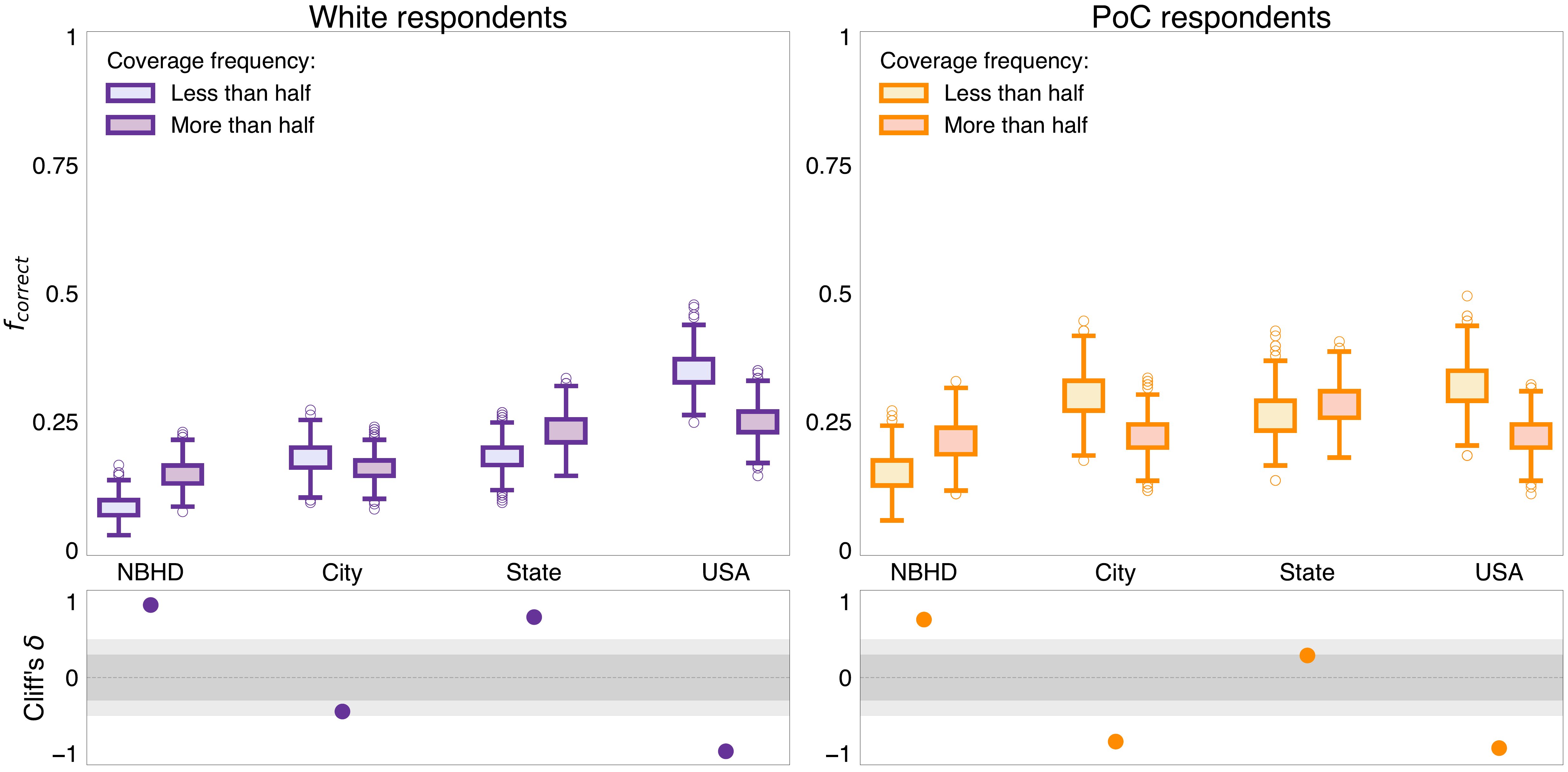}
        \caption{Correct}
    \end{subfigure}
    \begin{subfigure}{0.65\textwidth}
        \centering
        \includegraphics[width=\linewidth]{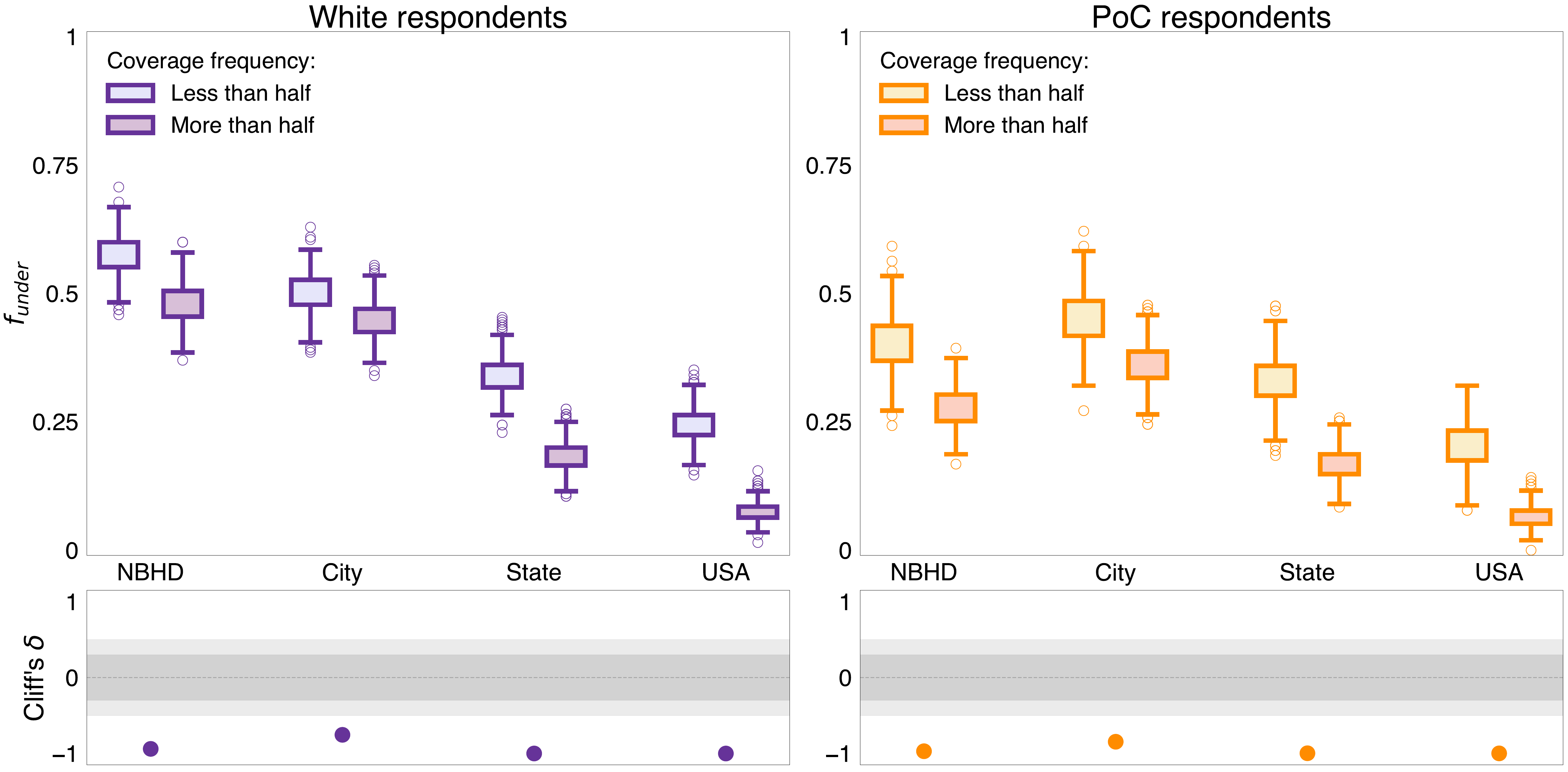}
        \caption{Under}
    \end{subfigure}
    \caption{Figure~\ref{fig:coverage}a-b reproduced for the fraction of correct and underestimations}
\end{figure}

\begin{figure}[ht!]
    \centering
    \begin{subfigure}{0.65\textwidth}
        \centering
        \includegraphics[width=\linewidth]{img/covtone_focus_PoC_proportional_10_over.pdf}
        \caption{Over (as in the text)}
    \end{subfigure}
    \begin{subfigure}{0.65\textwidth}
        \centering
        \includegraphics[width=\linewidth]{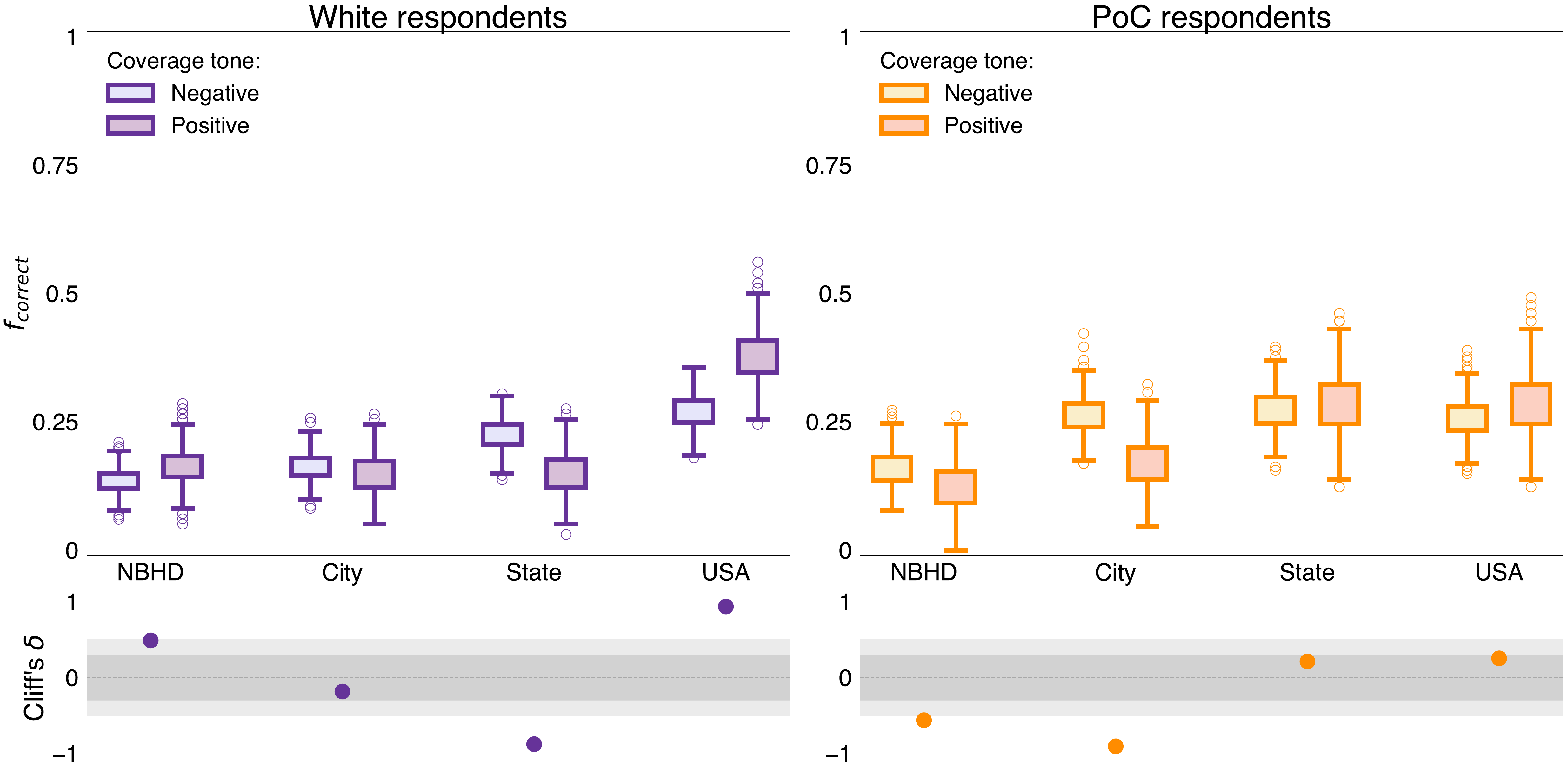}
        \caption{Correct}
    \end{subfigure}
    \begin{subfigure}{0.65\textwidth}
        \centering
        \includegraphics[width=\linewidth]{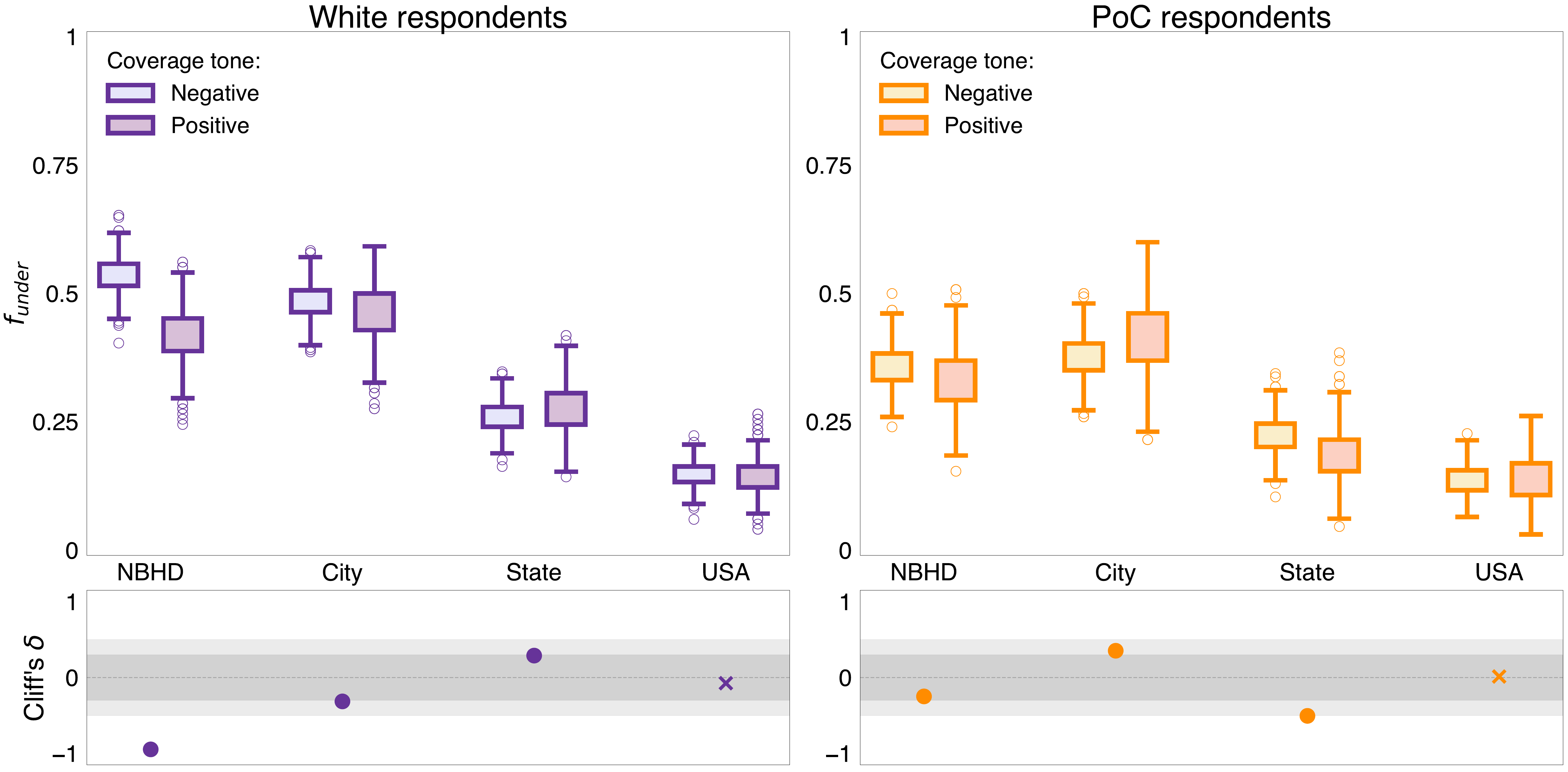}
        \caption{Under}
    \end{subfigure}
    \caption{Figure~\ref{fig:coverage}c-d reproduced for the fraction of correct and underestimations}
\end{figure}

\begin{figure}[ht!]
    \centering
    \begin{subfigure}{0.65\textwidth}
        \centering
        \includegraphics[width=\linewidth]{img/Newsuse_focus_PoC_proportional_10_over.pdf}
        \caption{Over (as in the text)}
    \end{subfigure}
    \begin{subfigure}{0.65\textwidth}
        \centering
        \includegraphics[width=\linewidth]{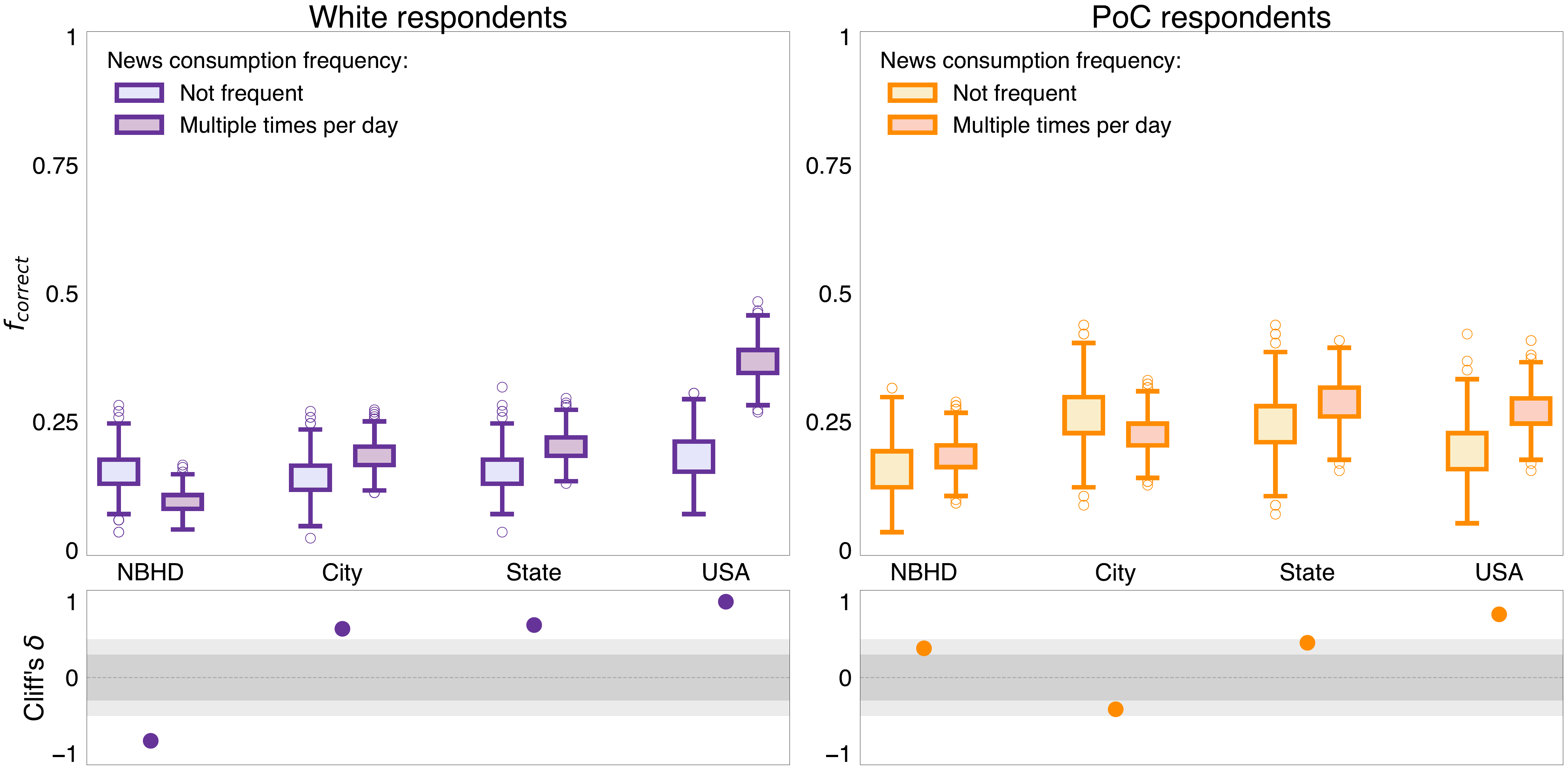}
        \caption{Correct}
    \end{subfigure}
    \begin{subfigure}{0.65\textwidth}
        \centering
        \includegraphics[width=\linewidth]{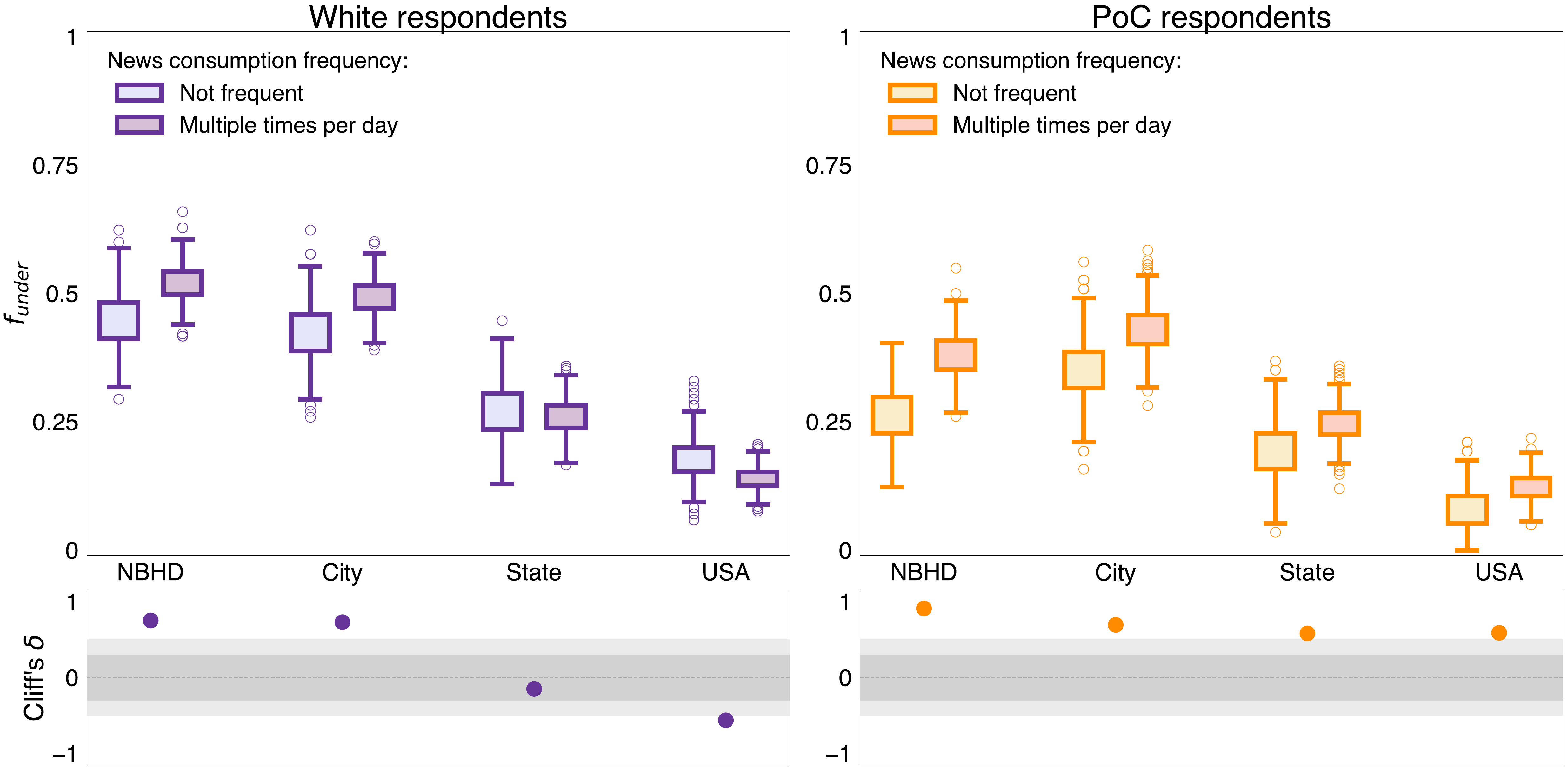}
        \caption{Under}
    \end{subfigure}
    \caption{Figure~\ref{fig:frequency}a-b reproduced for the fraction of correct and underestimations}
\end{figure}

\begin{figure}[ht!]
    \centering
    \begin{subfigure}{0.65\textwidth}
        \centering
        \includegraphics[width=\linewidth]{img/SMuse_focus_PoC_proportional_10_over.pdf}
        \caption{Over (as in the text)}
    \end{subfigure}
    \begin{subfigure}{0.65\textwidth}
        \centering
        \includegraphics[width=\linewidth]{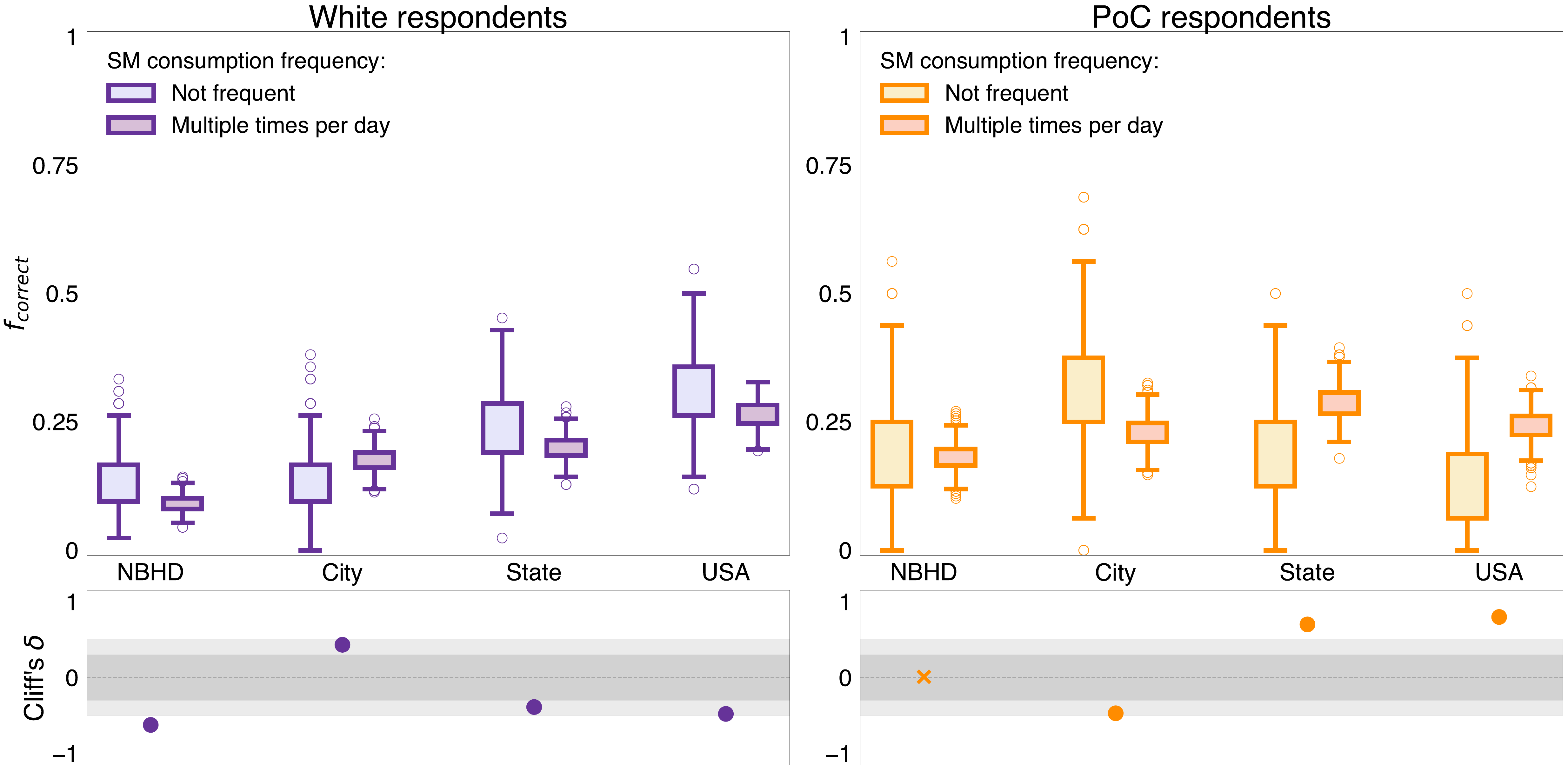}
        \caption{Correct}
    \end{subfigure}
    \begin{subfigure}{0.65\textwidth}
        \centering
        \includegraphics[width=\linewidth]{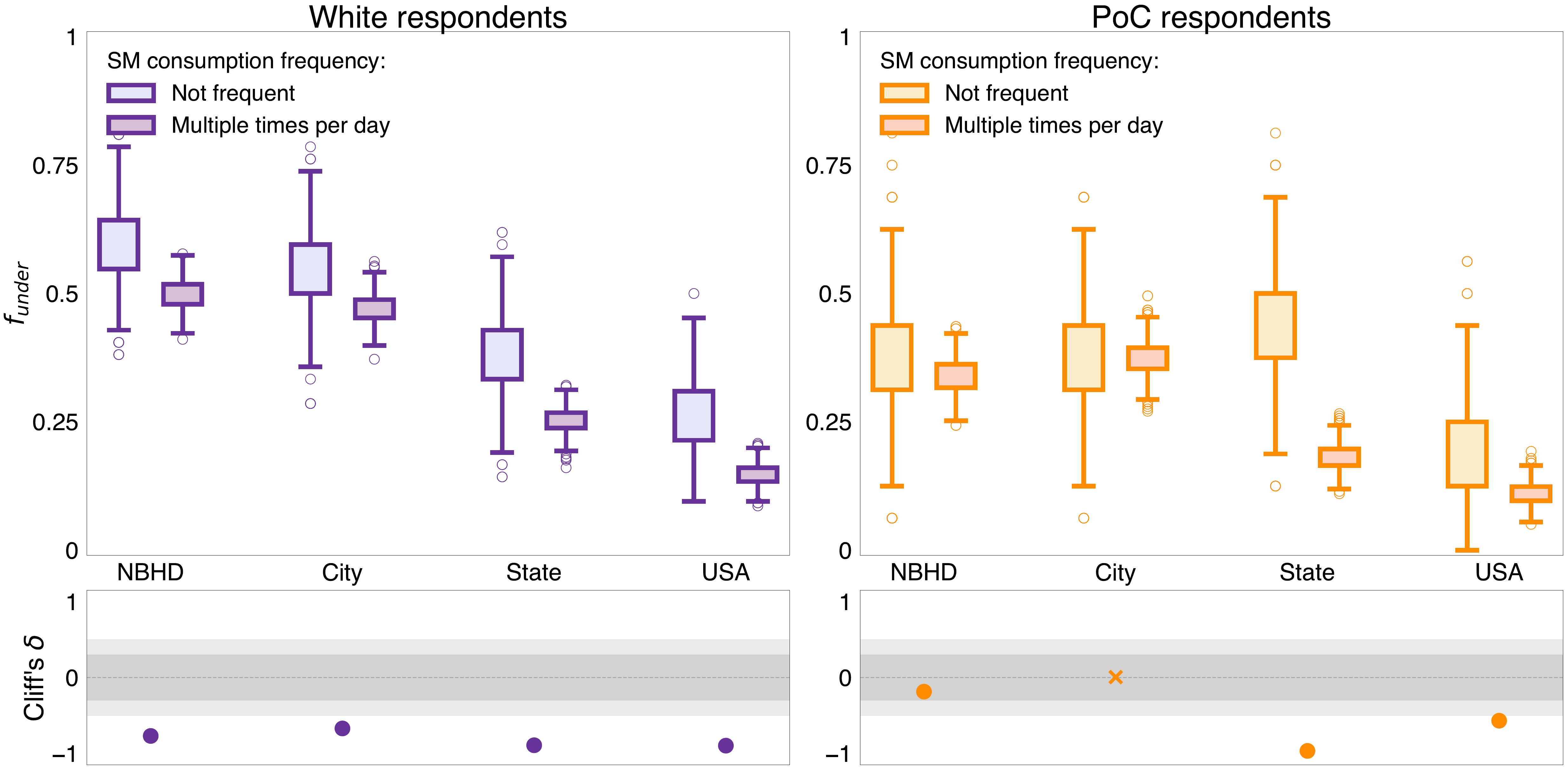}
        \caption{Under}
    \end{subfigure}
    \caption{Figure~\ref{fig:frequency}c-d reproduced for the fraction of correct and underestimations}
\end{figure}

\clearpage

\section{Independent variables used in the Random Forest classifiers}
\begin{table}[h]
\begin{center}
\renewcommand{\arraystretch}{0.45}
\caption{Definition of the independent variables used in the Random Forest classifiers.}
\label{tab:rfvars}
\begin{tabular}{p{4cm} p{12cm}}
\hline
\textbf{Variable} &  \textbf{Definition} \\
\hline
\\
Gender &  Categorical variable describing the gender of the respondents. Only Female/Male options were retained during data cleaning, as other values were dropped due to the small sample size. The reference category is `Female'. \\
\\
Age &  Age retrieved through the data shared by Prolific regarding the collected sample. The provided age bins were subsequently mapped to a corresponding ordinal numerical scale. \\
\\
Education &  The original survey responses (ranging from `Some high school or less' to `Graduate or professional degree') were mapped to an ordinal numerical scale. \\
\\
Income &  Yearly income. The original survey responses (ranging from `Less than \$25,000' to `\$150,000 or more') were mapped to an ordinal numerical scale. \\
\\
Occupation &  The original survey responses were collapsed into a binary variable corresponding to `Employed'/`Non-employed'. \\
\\
Political leaning &  After dropping non-informative values during data cleaning, we retain the distinction between Democrat, Republican, and Independent (the latter being the reference category). \\
\\
\makecell[l]{Social media use \\ frequency}&  Numerical scale where 0 means social media is used less than daily, 1 means social media is used daily but not more than once, and 2 means social media is used multiple times per day. \\
\\
\makecell[l]{News consumption\\frequency}&  Numerical scale where 0 means news are consumed less than daily, 1 that they are consumed daily but not more than once, and 2 that they are consumed multiple times per day. \\
\\
Coverage frequency &  Perceived frequency of coverage of PoC expressed as a percentage of the total coverage. \\
\\
Coverage tone &  Perceived tone of the news towards people of color. It is expressed as a numerical scale were lower values correspond to negative tones and higher values to positive tones. \\
\\
Social circle (percentage)  &  Percentage of a respondent's social circle they reported being made of people of color. \\
\\
Confidence &  Score reflecting the respondents' confidence in their guess about the population composition at a certain geographical level. Expressed on a scale of 1 (very low confidence) to 6 (high confidence). \\

\hline
\end{tabular}
\end{center}
\end{table}

\clearpage

\section{Hyperparameters used for tuning the Random Forest classifiers}
\begin{table}[h]
\begin{center}
\renewcommand{\arraystretch}{0.5}
\caption{Hyperparameter grid used for tuning the Random Forest classifiers.}
\label{tab:hyperpars}
\begin{tabular}{{p{4cm} p{4cm} p{8cm}}}
\hline
\textbf{Parameter} & \textbf{Values} & \textbf{Definition} \\
\hline
\\
n\_estimators & 100, 300 & Number of trees in the forest. \\
\\
max\_depth & 3, 5, 7, 10 & Maximum depth allowed for each decision tree. \\
\\
max\_features & sqrt, 0.8 & Number/fraction of features considered when splitting a node. \\
\\
min\_samples\_split & 2, 5, 10 & Minimum number of samples required to split an internal node. \\
\\
min\_samples\_leaf & 1, 2, 4 & Minimum number of samples required at a leaf node. \\
\\
class\_weight & balanced, balanced\_subsample, None & Weights assigned to classes to handle class imbalance. \\
\\
\hline
\end{tabular}
\end{center}
\end{table}